\documentclass[conference, draftclsnofoot,onecolumn]{IEEEtran}
\usepackage[english]{babel} 
\usepackage{amsmath,amsfonts, amssymb, latexsym}
\usepackage{dsfont}
\usepackage{bbold}
\usepackage{cite}
\usepackage{float}
\usepackage{graphicx}
\usepackage{subfigure}
\usepackage{colortbl}
\usepackage{multirow}
\usepackage{hhline}
\usepackage{footnote}
\usepackage{booktabs}
\usepackage{enumitem}

\renewcommand{\thefootnote}{}

\newtheorem{theorem}{Theorem}
\newtheorem{lemma}{Lemma}

\newtheorem{definition}{Definition}
\newtheorem{remark}{Remark}

\newcommand{\etal}{\emph{et al. }}
\newcommand{\vect}[1]{\mathbf{#1}} 
\newcommand{\dmat}[1]{\mathsf{#1}} 
\newcommand{\mat}[1]{\boldsymbol{#1}}
\newcommand{\eqdef}{\triangleq} 
\newcommand{\set}[1]{\mathcal{#1}}   
\newcommand{\Hsum}{H_{\text{sum}}}
\newcommand{\infr}{\textnormal{i}} 
\newcommand{\Ee}{\mathcal{E}}
\newcommand{\Pehat}[1]{\hat{P}_{e_{#1}}^{(N)}}
\newcommand{\E}[2][]{\textnormal{\textsf{E}}_{#1}\!\left[#2\right]} 

\usepackage{xcolor}

  \setitemize{parsep=6pt,itemsep=0pt,leftmargin=*,labelsep=5.5mm}
\setenumerate{parsep=6pt,itemsep=0pt,leftmargin=*,labelsep=5.5mm}
\setlist[description]{itemsep=0mm}   

\usepackage{environ}
\NewEnviron{myequation}{%
\begin{equation}
\scalebox{0.88}{$\BODY$}
\end{equation}}
\begin{document}

\IEEEoverridecommandlockouts
\title{Robust Signaling for Bursty Interference}
\author{
\IEEEauthorblockN{Grace Villacr\'es\IEEEauthorrefmark{2}, Tobias Koch\IEEEauthorrefmark{2}, Aydin Sezgin\IEEEauthorrefmark{1} and Gonzalo Vazquez-Vilar\IEEEauthorrefmark{2}\\}
\IEEEauthorblockA{\IEEEauthorrefmark{2}%
Universidad Carlos III de Madrid, Legan\'es, Spain, and Gregorio Mara\~n\'on Health Research Institute, Madrid, Spain.\\
\IEEEauthorrefmark{1}Ruhr-Universit\"at Bochum, Bochum, Germany\\Emails: \{gvillacres, koch\}@tsc.uc3m.es, aydin.sezgin@rub.de, gvazquez@ieee.org} 
\thanks{This work has been funded in part by the European Research Council (ERC) under the European Union's Horizon 2020 research and innovation programme (grant agreement number 714161), from the Spanish Ministerio de Econom\'ia y Competitividad under Grants TEC2013-41718-R, RYC-2014-16332, IJCI-2015-27020, TEC2016-78434-C3-3-R (AEI/FEDER, EU), from the Comunidad de Madrid under Grant S2103/ICE-2845, and from `Ayudas para la Movilidad del Programa Propio de Investigaci\'on UC3M 2016''.}}
\maketitle

\begin{abstract}
This paper studies a bursty interference channel, where the presence/absence of interference is modeled by a block-i.i.d.\ Bernoulli process that stays constant for a duration of $T$ symbols (referred to as coherence block) and then changes independently to a new state. We consider both a quasi-static setup, where the interference state remains constant during the whole transmission of the codeword, and an ergodic setup, where a codeword spans several coherence blocks. For the quasi-static setup, we study the largest rate of a coding strategy that provides reliable communication at a basic rate and allows an increased (opportunistic) rate when there is no interference. For the ergodic setup, we study the largest achievable rate. We study how non-causal knowledge of the interference state, referred to as \emph{channel-state information (CSI)}, affects the achievable rates. We derive converse and achievability bounds for (i) local CSI at the receiver-side only; (ii) local CSI at the transmitter- and receiver-side, and (iii) global CSI at all nodes. Our bounds allow us to identify when interference burstiness is beneficial and in which scenarios global CSI outperforms local CSI. The joint treatment of the quasi-static and ergodic setup further allows for a thorough comparison of these two setups.
\end{abstract}
\setcounter{footnote}{0}
\renewcommand{\thefootnote}{\arabic{footnote}}


\setcounter{footnote}{0}
\renewcommand{\thefootnote}{\arabic{footnote}}
\section{Introduction}
Interference is a key limiting factor for the efficient use of the spectrum in modern wireless networks. It is, therefore, not surprising that the \emph{interference channel (IC)} has been studied extensively in the past; see, e.g., \cite[Ch.~6]{elgamal2011book} and references therein. Most of the information-theoretic work developed for the IC assumes that interference is always present. However, certain physical phenomena, such as shadowing, can make the presence of interference intermittent or bursty. Interference can also be bursty due to the bursty nature of data traffic, distributed medium access control mechanisms, and decentralized networking protocols. For this reason, there has been an increasing interest in understanding and exploring the effects of burstiness of interference.

Seminal works in this area were performed by Khude \etal in \cite{khude09_1} for the Gaussian channel and in \cite{Khude09} by using a model which corresponds to an approximation to the two-user Gaussian IC. \color{black} They tried to harness the burstiness of the interference by taking advantage of the time instants when the interference is not present to send opportunistic data. Specifically, \cite{Khude09,khude09_1} considered a channel model where the interference state stays constant during the transmission of the entire codeword, which corresponds to a quasi-static channel. Motivated by the idea of degraded message sets by K\"{o}rner and Marton \cite{Korner77}, Khude \etal studied the largest rate of a coding strategy that provides reliable communication at a basic rate $R$ and allows an increased (opportunistic) rate $R+\Delta R$ when there is no interference. The idea of opportunism was also used by Diggavi and Tse \cite{Diggavi06} for the quasi-static flat fading channel and, recently, by Yi and Sun \cite{yi18} for the $K$-user IC with states.

Wang \etal \cite{Wang13} modeled the presence of interference using an \emph{independent and identically distributed (i.i.d.)} Bernoulli process that indicates whether interference is present or not, which corresponds to an ergodic channel. They further assume that the interference links are fully correlated.  Wang \etal mainly studied the effect of causal feedback under this model, but also presented converse bounds for the non-feedback case. Mishra \etal considered the generalization of this model to multicarrier systems, modeled as parallel two-user bursty ICs, for the feedback \cite{mishra17} and non-feedback case \cite{mishra13}.

The bursty IC is related to the binary fading IC, for which the four channel coefficients are in the binary field $\{0,1\}$ according to some Bernoulli distribution. Note, however, that neither of the two models is a special case of the other. While a zero channel coefficient of the cross link corresponds to intermittence of interference, the bursty IC allows for non-binary signals. Conversely, in contrast to the binary fading IC, the direct links in the bursty IC cannot be zero, since only the interference can be intermittent.  Vahid \etal \cite{Vahid11,Vahid14,Vahid14_dCSIT, Calderbank, Vahid17} studied the capacity region of the binary fading IC. \mbox{Specifically,~\cite{Vahid14,Vahid17}} study the capacity region of the binary fading IC when the transmitters do not have access to the channel coefficients, and \cite{Vahid14_dCSIT} study the capacity region when the transmitters have access to the past channel coefficients. Vahid and Calderbank additionally study the effect on the capacity region when certain correlation is available to all nodes as side information \cite{Calderbank}. 

\color{black} The focus of the works by Khude \etal \cite{Khude09} and Wang \etal \cite{Wang13} was on the \emph{linear deterministic model (LDM)}, which was first  introduced by Avestimehr {\cite{Avestimehr11}, but falls within the class of more general deterministic channels whose capacity was obtained by El Gamal and Costa in \cite{gamal82}. The LDM maps the Gaussian IC to a channel whose outputs are deterministic functions of their inputs. 
Bresler and Tse demonstrated in~\cite{bresler08} that the generalized degrees of freedom (first-order capacity approximation) of the two-user Gaussian IC coincides with the normalized capacity of the corresponding deterministic channel. The LDM thus offers insights on the Gaussian IC.

\subsection{Contributions}
In this work, we consider the LDM of a bursty IC. We study how interference burstiness and the knowledge of the interference states (throughout referred to as \emph{channel-state information (CSI)}) affects the capacity of this channel. We point out that this CSI is different from the one sometimes considered in the analysis of ICs (see, e.g., \cite{Kao13}), where CSI refers to knowledge of the channel coefficients. (In this regard, we assume that all transmitters and receivers have access to the channel coefficients.) For the sake of compactness, we focus on non-causal CSI and leave other CSI scenarios, such as causal or delayed CSI, for future work.

We consider the following cases: (i) only the receivers know the corresponding interference state (local CSIR); (ii) transmitters and receivers know their corresponding interference states (local CSIRT); and  (iii) both transmitters and receivers know all interference states (global CSIRT).  For each CSI level we consider both (i) the quasi-static channel and (ii) the ergodic channel. Specifically, in the quasi-static channel the interference is present or absent during the whole message transmission and we harness the realizations when the channel experiences better conditions (no presence of interference) to send extra messages. In the ergodic channel the presence/absence of interference is modeled as a Bernoulli random variable which determines the interference state. The interference state stays constant for a certain coherence time $T$ and then changes independently to a new state. This model includes the i.i.d.\ model by Wang \etal as a special case, but also allows for scenarios where the interference state changes more slowly. Note, however, that when the receivers know the interference state (as we shall assume in this work), then the capacity of this model becomes independent of $T$ and coincides with that of the i.i.d. model. The proposed analysis is performed for the two extreme cases where the states of each of the interfering links are independent, and where states of the interfering links are fully correlated. Hence we unify the scenarios already treated in the literature \cite{khude09_1,Khude09,Wang13}. Nevertheless, some of our presented results can be extended to consider an arbitrary correlation between the interfering states. The works by Vahid and Calderbank~\cite{Calderbank} and Yeh and Wang~\cite{Yeh-arXiv1} characterize the capacity region of the two-user binary IC and the MIMO X-channel, respectively. While \cite{Calderbank,Yeh-arXiv1} consider a general spatial correlation between communication and interfering links, they do not consider the correlation between interfering links. \color{black}

Our analysis shows that, for both the quasi-static and ergodic channels,  for all interference regions except the very strong interference region, global CSIRT outperforms local CSIR/CSIRT. This result does not depend on the correlation between the states of the interfering links. For local CSIR/CSIRT and the quasi-static scenario, the burstiness of the channel is of benefit only in the very weak and weak interference regions. For the ergodic case and local CSIR, interference burstiness is only of clear benefit if the interference is either weak or very weak, or if it is present at most half of the time. This is in contrast to local CSIRT, where interference burstiness is beneficial in all interference regions.

Specific contributions of our paper include:
\begin{itemize}
\item A joint treatment of the quasi-static and the ergodic model: Previous literature on the bursty IC considers either the quasi-static model or the ergodic model. Furthermore, due to space constraints, the proofs of some of the existing results were either omitted or contain little details. In contrast, our paper discusses both models, allowing for a thorough comparison between the two.
\item Novel achievability and converse bounds: For the ergodic model, the achievability bounds for local CSIRT, and the achievability and converse bounds for global CSIRT, are novel. In particular, novel achievability strategies are proposed that exploit certain synchronization between the users. To keep the paper self-contained, we further present the proof of the achievability bound for local CSIR that has appeared in the literature without proof.
\item Novel converse proofs for the quasi-static model: In contrast to existing converse bounds, which are based on Fano's inequality, our proofs of the converse bounds for the rates of the worst-case and opportunistic messages are based on an information density approach (more precise, they are based on the Verd\'u-Han lemma). This approach does not only allow for rigorous yet clear proofs, but it would also enable a more refined analysis of the probabilities that worst-case and opportunistic messages can be decoded correctly.
\item A thorough comparison of the sum capacity of various scenarios: \emph{Inter alia}, the obtained results are used to study the advantage of featuring different levels of CSI, the impact of the burstiness of the interference, and the effect of the correlation between the channel states of both users.
\end{itemize}

The rest of this paper is organized as follows. Section~\ref{Sec: ChannelModel} introduces the system model, where we define the bursty IC quasi-static setup, the ergodic setup, and briefly summarize previous results on the non-bursty IC. In Sections~\ref{Sec: local_CSIR}--\ref{Sec: global_CSIRT} we present our results for local CSIR, local CSIRT and global CSIRT, respectively. Section~\ref{Sec: Exploiting_CSI} studies the impact of featuring different CSI levels. Section~\ref{Sec: Exploiting_Burstiness} analyzes in which scenarios exploiting burstiness of interference is beneficial. Section~\ref{Sec: Summary} concludes the paper with a summary of the results. Most proofs of the presented results are deferred to the appendix.

\subsection{Notation}\label{Notation}
To differentiate between scalars, vectors, and matrices we use different fonts:
scalar random variables and their realizations are denoted by upper and lower case letters, respectively, e.g., $B$, $b$; vectors are denoted using bold face, e.g., $\vect{X}$, $\vect{x}$; random matrices are denoted via a special font, e.g., $\mat{X}$; and for deterministic matrices we shall use yet another font, e.g.,~$\dmat{S}$. For sets we use the calligraphic font, e.g., $\mathcal{S}$. We denote sequences such as $A_{i,1},\ldots,A_{i,M}$ by $A_i^{M}$. We define $\max\{0, x\}$ as $(x)^+$.

 We use $\mathbb{F}_2$ to denote the binary Galois field and $\oplus$ to denote the modulo 2 addition. Let the down-shift matrix $\dmat{S}_u \in \mathbb{F}_2^{q\times q}$, a matrix of dimension $q \times q$, be defined as
 
  \[
   \dmat{S}_u=
   \begin{bmatrix}
        \mathsf{0}^T_{u\times (q-u)} & 0  \\
        \mathsf{I}_{u} & \mathsf{0}_{u\times (q-u)}
      \end{bmatrix}_{q\times q}
    \]
with $0_{q-1} \in \mathbb{F}_2^{q-1}$ the all-zero vector and $\mathsf{I}_{u} \in \mathbb{F}_2^{u \times u}$ the identity matrix. 

Similarly, we define the matrix $\dmat{L}_d\in \mathbb{F}_2^{q\times q}$ of dimension $q \times q$ that selects the $d$ lowest components of a vector of dimension $q$:

\[
 \dmat{L}_d=
  \begin{bmatrix}
        0 &  \mathsf{0}^T_{d\times (q-d)}  \\
       \mathsf{0}_{d\times (q-d)} & \mathsf{I}_{d}
     \end{bmatrix}_{q\times q}.
 \]

We shall denote by $H_b(p)$ the entropy of a binary random variable $X$ with probability mass function ($p, 1-p$), i.e.,
 \begin{IEEEeqnarray}{lCl}
H_b (p)
\triangleq -p\log p-(1-p)\log(1-p).
 \end{IEEEeqnarray}
Similarly, we denote by $H_{\text{sum}}(p,q)$ the entropy $H(X\oplus  \tilde{X})$ where $X$ and $\tilde{X}$ are two independent binary random variables with probability mass functions $(p,1-p)$ and $(q,1-q)$, respectively:
 \begin{IEEEeqnarray}{lCl}
H_{\text{sum}}(p,q)
\triangleq H_b(p(1-q)+(1-p)q)
 \end{IEEEeqnarray}
For this function it holds that $H_{\text{sum}}(p,q) = H_{\text{sum}}(1-p,q) = H_{\text{sum}}(p,1-q) = H_{\text{sum}}(1-p,1-q)$. Finally, $\mathds{1}(\cdot)$ denotes the indicator function, i.e., $\mathds{1}(\textnormal{statement})$ is $1$ if the statement is true and $0$ if it is~false.

\section{System Model} \label{Sec: ChannelModel}
Our analysis is based on the LDM, introduced by Avestimehr \etal \cite{Avestimehr11} for some relay network. This model is, on the one hand, simple to analyze and, on the other hand, captures the essential structure of the Gaussian channel in the high signal-to-noise ratio regime.

We consider a bursty IC where i) the interference state remains constant during the whole transmission of the codeword of length $N$ (quasi-static setup) or ii) the interference state remains constant for a duration of $T$ consecutive symbols and then changes independently to a new state (ergodic setup). 
For one coherence block, the two-user bursty IC is depicted in Figure~\ref{Fig:Ch_Model_det}, where $n_{d}$ and $n_{c}$ are the channel gains of the direct and cross links, respectively. We assume that $n_{d}$ and $n_{c}$ are known to both the transmitter and receiver and remain constant during the whole transmission of the codeword. For simplicity, we shall assume that $n_d$ and $n_c$ are equal for both users. Nevertheless, most of our results generalize to the asymmetric case. More precisely, all converse and achievability bounds generalize to the asymmetric case, while the direct generalization of the proposed achievability schemes may be loose in some asymmetric  regions. 
\begin{figure}[H]
		\centering
		\includegraphics[width=0.4\linewidth]{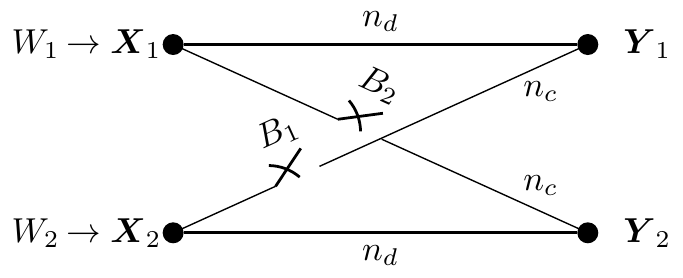}
		\caption{Channel model of the bursty interference channel.}
		\label{Fig:Ch_Model_det}
	\end{figure}

For the $k$-th block, the input-output relation of the channel is given~by
\begin{eqnarray}\label{Ch_Model}
\mat{Y}_{1,k}&=&\dmat{S}_{n_{d}}\mat{X}_{1,k}\oplus B_{1,k}\dmat{S}_{n_{c}}\mat{X}_{2,k},\label{Eq: Y1_det_int}\\
\mat{Y}_{2,k}&=&\dmat{S}_{n_{d}}\mat{X}_{2,k}\oplus B_{2,k}\dmat{S}_{n_{c}}\mat{X}_{1,k}.\label{Eq: Y2_det_int}
\end{eqnarray}

Let $q \triangleq \max \{n_{d}, n_{c}\}$. In \eqref{Eq: Y1_det_int} and \eqref{Eq: Y2_det_int}, $\mat{X}_{i,k}\in\mathbb{F}_2^{q\times T}$ and $\mat{Y}_{i,k}\in\mathbb{F}_2^{q\times T}$, $i=1,2$. The interference states $B_{i,k}$, $i=1,2$,  $k=1,\ldots, K$, are sequences of i.i.d.\ Bernoulli random variables with activation probability~$p$. 

Regarding the sequences $B_1^K$ and $B_2^K$, we consider two cases: (i) $B_1^K$ and $B_2^K$ are independent of each other and (ii) $B_1^K$ and $B_2^K$ are fully correlated sequences, i.e., $B_1^K=B_2^K$. For both cases we assume that the sequences are independent of the messages $W_1$ and $W_2$. 

We shall define the normalized interference level as $\alpha\eqdef\frac{n_c}{n_d}$, based on which we can divide the interference into the following regions (a similar division was used by Jafar and Vishwanath \cite{jafar10}): 
\begin{itemize}
\item \emph{very weak interference (VWI)} for $\alpha\leq\frac{1}{2}$,
\item \emph{weak interference (WI)}  for $\frac{1}{2}<\alpha\leq\frac{2}{3}$, 
\item \emph{moderate interference (MI)} for $\frac{2}{3}<\alpha\leq 1$,
\item \emph{strong interference (SI)} for $1<\alpha\leq 2$,
\item  \emph{very strong interference (VSI)} for $2<\alpha$. 
\end{itemize}

\subsection{Quasi-Static Channel}\label{QS_Ch}
The channel defined in \eqref{Eq: Y1_det_int} and \eqref{Eq: Y2_det_int}  may experience a slowly-varying change on the interference state. In this case, the duration of each of the transmitted codewords of length $N=KT$ is smaller than the coherence time $T$ of the channel and the interference state stays constant over the duration of each codeword, i.e., $K=1$, $T=N$. In the wireless communications literature such a channel is usually referred to as a quasi-static channel \cite[Sec.~5.4.1]{tse2005book}.
In this scenario, the rate pair of achievable rates $(R_1, R_2)$ is dominated by the worst case, which corresponds to the presence of interference at both receivers. However, in absence of interference, it is possible to communicate at a higher date rate, so planning a system for the worst case may be too pessimistic. Assuming that the receivers have access to the interference states, the transmitters could send opportunistic messages that are decoded only if the interference is absent, in addition to the regular messages that are decoded irrespective of the interference state.  We make the notion of opportunistic messages and rates precise in the subsequent~paragraphs.

Let  $U_{i,k}$ indicate the level of CSI available at the transmitter side in coherence block $k$, and let $V_{i,k}$ indicate the level of CSI at the receiver side in coherence block $k$:
\begin{enumerate}
\item local CSIR: $\qquad U_{i,k} =\emptyset\ \text{and}\  V_{i,k}=B_{i,k},\ i=1,2, \ k=1,\ldots, K$,
\item local CSIRT: $\quad\ \;U_{i,k} = V_{i,k}=  B_{i,k} ,\ i=1,2,  \ k=1,\ldots, K$,
\item global CSIRT: $\quad U_{i,k} = V_{i,k}=  (B_{1,k}, B_{2,k} ),\ i=1,2, \ k=1,\ldots, K$.
\end{enumerate}
We define the set of opportunistic messages according to the level of CSI at the receiver as $\{\Delta {W_i}{(\cdot)}\}\eqdef \{\Delta {W_i}(v_i), v_i\in\set{V}_i\}$, where $\set {V}_i$ denotes the set of  possible interference states $V_i$. Specifically, 
\begin{enumerate}
\item for local CSIR: $\{\Delta {W_i}(\cdot)\}=\{\Delta {W_i}(1),\Delta {W_i}(0)\},\ i=1,2$,
\item for local CSIRT: $\{\Delta {W_i}(\cdot)\}=\{\Delta {W_i}(1),\Delta {W_i}(0)\},\ i=1,2$,
\item for global CSIRT: $\{\Delta {W_i}(\cdot)\}=\{\Delta {W_i}(00),\Delta {W_i}(01),\Delta {W_i}(10),\Delta {W_i}(11)\}, \ i=1,2$.
\end{enumerate}
Then, we define an opportunistic code as follows.
\begin{definition}[Opportunistic code for the bursty IC]\label{Def:opp_code}
An $\bigl(N,R_1,R_2,\{\Delta {R_1}(\cdot)\},\{\Delta {R_2}(\cdot)\}\bigr)$ opportunistic code for the bursty IC is defined as:
\begin{enumerate}
\item two independent messages $W_1$ and $W_2$ uniformly distributed over the message sets $\mathcal{W}_i\triangleq \{1,2,\ldots,2^{N R_i}\},\ i=1,2$;
\item two independent sets of opportunistic messages $\{\Delta {W_1}(\cdot)\}$ and $\{\Delta {W_2}(\cdot)\}$  uniformly distributed over the message sets $\Delta {\mathcal{W}_i}(v_i)\eqdef \{1,2,\ldots,2^{N \Delta {R_i}(v_i)}\}, v_i\in\mathcal{V}_i$, $\ i=1,2$, 
\item two encoders: $f_i: (W_i,\{\Delta {W_i}(\cdot)\},U_i) \mapsto \mat{X}_i,\ i=1,2,$
\item two decoders: $g_i: (\mat{Y}_i,V_i) \mapsto (\hat{W}_i, \Delta {\hat{W}_i}(V_i)),\ i=1,2$.
\end{enumerate}
Here $\hat{W}_i$ and $\Delta {\hat{W}_i}(V_i)$ denote the decoded message and the decoded opportunistic message, respectively. We set $\Delta R_i(1)=0$, $i=1,2$ (for local CSIR/CSIRT) and $\Delta R_i(11)=0$ (for global CSIRT).
\end{definition}

To better distinguish the rates $(R_1,R_2)$ from the opportunistic rates $\{\Delta {R_i}(\cdot)\}$, $i=1,2$, we shall refer to $(R_1,R_2)$ as worst-case rates, because the corresponding messages can be decoded even if the channel is in its worst state (see also Definition~\ref{Def: Delta R_i+R_i}).

\begin{definition}[Achievable opportunistic rates] \label{Def: Delta R_i+R_i}
A rate tuple $\bigl(R_1, R_2, \{\Delta {R_1}(\cdot)\}, \{\Delta {R_2}(\cdot)\}\bigr)$ is achievable if there exists a sequence of codes $\bigl(N,R_1,R_2,\{\Delta  {R_1}(\cdot)\}, \{\Delta {R_2}(\cdot)\}\bigr)$ such that
\begin{IEEEeqnarray}{lCl}\label{QS_IC_Pr}
\Pr\bigl\{\hat{W}_1 \neq W_1 \cup \hat{W}_2\neq W_2\bigr\}\to 0 \quad\text{as}\quad N\to\infty
\end{IEEEeqnarray} 
and
\begin{IEEEeqnarray}{lCl}\label{QS_Pr_1}
\Pr\bigl\{(\hat{W}_1,\Delta {\hat{W}_1}(V_1))\neq(W_1,\Delta {W_1}(V_1))|V_1=v_1\bigr\}\to 0 \quad\text{as}\quad N\to\infty,\ v_1\in\set {V}_1,
\end{IEEEeqnarray}
\begin{IEEEeqnarray}{lCl}\label{QS_Pr_2}
\Pr\bigl\{(\hat{W}_2,\Delta {\hat{W}_2}(V_2))\neq(W_2,\Delta  {W_2}(V_2))|V_2=v_2\bigr\}\to 0 \quad\text{as}\quad N\to\infty, \ v_2\in\set {V}_2.
\end{IEEEeqnarray}
The capacity region is the closure of the set of achievable rate tuples \cite[Sec.~6.1]{elgamal2011book}. We define the worst-case sum rate as $R\eqdef R_1+R_2$ and the opportunistic sum rate as $\Delta {R}(V_1,V_2) \eqdef \Delta {R_1}(V_1)+\Delta {R_2}(V_2)$. The worst-case sum capacity $C$ is the supremum of all achievable worst-case sum rates, the opportunistic sum capacity $\Delta C(V_1,V_2)$ is the supremum of all opportunistic sum rates, and the total sum capacity is defined as $C+\Delta {C}(V_1,V_2)$. Note that the opportunistic sum capacity depends on the worst-case sum rate. 
\end{definition}

\begin{remark}
The worst-case sum rate and opportunistic sum rates in the quasi-static setting depend only on the collection of possible interference states: for independent interference states we have $\displaystyle{\vect{B} \in\{00,01,10,11\}}$, and for fully correlated interference states we have $\displaystyle{\vect{B}\in\{00,11\}}$. In principle, our proof techniques could also be applied to analyze other collections of interference states.
\end{remark}

\begin{remark}
In the CSIRT setting the transmitters have access to the interference state. Therefore, in this setting the messages are strictly speaking not opportunistic. Instead, transmitters can adapt their rate based on the state of the interference links, which is sometimes referred to as rate adaptation in the literature.
\end{remark}

\subsection{Ergodic Channel} \label{Ergo_Ch}
In this setup, we shall restrict ourselves to codes whose blocklength $N$ is an integer multiple of the coherence time $T$. A codeword of length $N=KT$ thus spans $K$ independent channel realizations. 

\begin{definition}[Code for the bursty IC]\label{Def_code_BIC}
A $\bigl(K,T,R_1,R_2\bigr)$ code for the bursty IC is defined as:
\begin{enumerate}
\item two independent messages $W_1$ and $W_2$ uniformly distributed over the message sets $\mathcal{W}_i\triangleq \{1,2,\ldots,2^{KTR_i}\},\ i=1,2;$
\item two encoders: $f_i: (W_i,U_i^K) \mapsto \mat{X}_i^K,\ i=1,2;$
\item two decoders: $g_i: (\mat{Y}_i^K,V_i^K) \mapsto \hat{W}_i,\ i=1,2.$
\end{enumerate}
Here $\hat{W}_i$ denotes the decoded message, and $U_i^K$ and $V_i^K$ indicate the level of CSI at the transmitter and receiver side, respectively, which are defined as for the quasi-static channel in Section~\ref{QS_Ch}.
\end{definition}

\begin{definition}[Ergodic achievable rates]\label{Def: R1_R2}
A rate pair $(R_1,R_2)$ is achievable for a fixed $T$ if there exists a sequence of codes $\bigl(K,T,{ R_1},{ R_2}\bigr)$ (parametrized by $K$) such that

\begin{equation}
\Pr\bigl\{\hat{W}_1 \neq W_1 \cup \hat{W}_2\neq W_2\bigr\}\to 0 \quad\text{as}  \quad K\to\infty.
\end{equation}
The capacity region is the closure of the set of achievable rate pairs. 
We define the sum rate as $R\eqdef R_1+R_2$, the sum capacity $C$ is the supremum of all achievable sum rates.
\end{definition}

\subsection{The Sum Capacities of the Non-Bursty and the Quasi-Static Bursty IC}\label{IC}
When the activation probability $p$ is $1$, we recover in both the ergodic and quasi-static scenarios the deterministic IC. For a general deterministic IC the capacity region was  obtained in \cite[Th.~1]{gamal82} and then by Bresler and Tse in \cite{bresler08} for a specific deterministic IC. For completeness, we present the sum capacity region for the deterministic non-bursty IC in the following theorem. 
\begin{theorem}\label{Thm: non-bursty_IC}
The sum capacity region of the two-user deterministic IC is equal to the union of the set of all sum rates $R$ satisfying
\begin{IEEEeqnarray}{lCl}
R\leq 2n_d \label{IC1}\\
R\leq (n_{d}-n_{c})^++\max(n_{d},n_{c})\label{IC2}\\
R\leq 2\max\{(n_{d}-n_{c})^+,n_{c}\}.\label{IC4}
\end{IEEEeqnarray}
\end{theorem}
\begin{IEEEproof}
The proof is given in \cite[Sec.~II]{gamal82}. For the achievability bounds,  El Gamal and Costa \cite[Th.~1]{gamal82} use the Han-Kobayashi scheme \cite{han81} for a general IC. Bresler and Tse \cite[Section~4]{bresler08} use a specific Han-Kobayashi strategy for the special case of the LDM. Jafar and Vishwanath \cite{jafar10} present an alternative achievability scheme for the $K$-user IC, which particularized for the two-user IC will be referenced in this work.
\end{IEEEproof}
We can achieve the sum rates \eqref{IC1} and \eqref{IC4} over the quasi-static channel by treating the bursty IC as a non-bursty IC. The following theorem demonstrates that this is the largest achievable worst-case sum rate irrespective of the availability of CSI and the correlation between $B_1$ and $B_2$.

\begin{theorem}[Sum capacity for the quasi-static bursty IC] \label{Thm:conv_realiable_rates}
For $0 \leq p\leq 1$, the worst-case sum capacity of the bursty IC is equal to the supremum of the set of sum rates $R$ satisfying
\begin{itemize}
\item For $p=0$, 
\begin{IEEEeqnarray}{lCl}
 R &\leq& 2n_d \label{Eq: IC1}.
 \end{IEEEeqnarray} 
\item For $0 < p\leq 1$
\begin{IEEEeqnarray}{lCl}
 R&\leq& (n_{d}-n_{c})^++\max(n_{d},n_{c})\label{Eq: IC3}\\
  R &\leq& 2\max\{(n_{d}-n_{c})^+,n_{c}\}.\label{Eq: IC2}
 \end{IEEEeqnarray} 
 
\end{itemize}
\end{theorem}

\begin{IEEEproof}
The converse bounds are proved in Appendix~\ref{IC_Verdu-Han}. Achievability follows directly from Theorem~\ref{Thm: non-bursty_IC} by treating the bursty IC as a non-bursty IC.
\end{IEEEproof}

Theorem~\ref{Thm:conv_realiable_rates} shows that the worst-case sum capacity does not depend on the level of CSI available at the transmitter and receiver side. However, this is not the case for the opportunistic rates as we will see in the next sections.

\begin{remark}\label{Remark_worst-case}
In principle, one could reduce the worst-case rates in order to increase the opportunistic rates. However, it turns out that such a strategy is not beneficial in terms of total rates $R_i+\Delta R_i(V_i)$, $i=1,2$. In other words, setting $\Delta R_i(1)=0$, $i=1,2$ (for local CSIR/CSIRT) and $\Delta R_i(11)=0$ (for global CSIRT), as we have done in Definition~\ref{Def: Delta R_i+R_i}, incurs no loss in total rate. Furthermore, in most cases it is preferable to maximize the worst-case rate, since it can be guaranteed irrespective of the interference state.
\end{remark}

\section{Local CSIR}\label{Sec: local_CSIR}
For the quasi-static and ergodic setups, described in Sections~\ref{QS_Ch} and \ref{Ergo_Ch}, respectively, we derive converse and achievability bounds for the independent and fully correlated scenarios when the interference state is only available at the receiver side.

\subsection{Quasi-Static Channel}
\unskip
\subsubsection{Independent Case}\label{QS: Local_CSIR_uncorr}
We present converse and achievability bounds for local CSIR when $B_1$ and $B_2$ are independent. The converse bounds are derived for local CSIRT, hence they also apply to this case. Since converse and achievability bounds coincide, this implies that local CSI at the transmitter is not beneficial in the quasi-static setup.
\begin{theorem}[Opportunistic sum capacity for local CSIR/CSIRT]\label{Thm:convBIC_local}
Assume that $B_1$ and $B_2$ are independent of each other. For $0< p <1$, the opportunistic sum capacity region is the union of the set of rate tuples $(R,\{\Delta R_1(b_1)+\Delta R_2(b_2), b_i\in\{0,1\}\})$, where $\Delta R_1(1)=\Delta R_2(1)=0$, and $R$, $\Delta R_1(0)$ and $\Delta R_2(0)$ satisfy \eqref{Eq: IC1}--\eqref{Eq: IC2} and
\begin{IEEEeqnarray}{lCl}
R+\Delta {R_1}{(0)}+\Delta {R_2}{(0)} &\leq& 2n_{d}\label{Eq: local_Delta0}\\
R+\Delta {R_1}{(0)} &\leq& (n_{d}-n_{c})^++\max(n_{d},n_{c})\label{Eq: local_Delta1}\\
R+\Delta {R_2}{(0)}  &\leq& (n_{d}-n_{c})^++\max(n_{d},n_{c})\label{Eq: local_Delta2}.
\end{IEEEeqnarray}
\end{theorem}
\begin{IEEEproof}
The converse bounds are proved in Appendix~\ref{Proof:convBIC_local} and the achievability bounds are proved in Appendix~\ref{Ach. apportunistic}. 
\end{IEEEproof}
\begin{remark}
 The converse bounds in Theorem~\ref{Thm:convBIC_local} coincide with those in \cite[Th.~2.1]{Khude09}, particularized for the symmetric setting. Theorem~\ref{Thm:convBIC_local}, however, is proven for local CSIRT, which is not considered in the model from~\cite{Khude09}. The proof included in Appendix~\ref{Proof:convBIC_local}  is based on an information density approach and provides a unified framework for treating local CSIR, local CSIRT and global CSIRT, as will be shown in Section~\ref{Sec: global_CSIRT}.
\end{remark}

As discussed in Remark~\ref{Remark_worst-case},  one could reduce the worst-case sum rate $R$ and increase the opportunistic rates $\Delta R(V_1,V_2)$. However, in the case of one-shot transmission this is not desirable, since the worst-case sum rate is the only rate that can be guaranteed irrespective of the interference state. (With one-shot transmission we refer to the case where we transmit one codeword of length $N$ over the quasi-static channel. This is in contrast to the case discussed, e.g., in Section~\ref{CSIR_QS_ER}, where we are interested in transmitting many codewords, each over $N$ channel uses of independent quasi-static channels.) Thus, one is typically interested in the opportunistic sum capacity when the worst-case rate $R$ is maximized. For this case, the results of Theorem~\ref{Thm:convBIC_local} are summarized in Table~\ref{Tab:qs_local_uncorr} for the VWI, WI, MI and SI regions. 
\begin{table}[tbp]
\centering \small
\caption{Opportunistic sum capacity for local CSIR when the worst-case sum rate is maximized.}
\label{Tab:qs_local_uncorr}
\begin{tabular}{ccccc}
\toprule
\textbf{Rates}&\textbf{VWI}&\textbf{WI}&\textbf{MI}&\textbf{SI}\\\midrule
{\textbf{$C$}}& $2(n_d-n_c)$&$2n_c$&{$2n_d-n_c$}&{$n_c$}\\\midrule
{\textbf{${\Delta C}(00)$}}& $2n_c$&$2(2n_d-3n_c)$&{$0$}&{$0$}\\ \midrule
{\textbf{${\Delta C}(01)/{\Delta C}(10)$}}& $n_c$&$2n_d-3n_c$&{$0$}&{$0$}\\ \bottomrule
\end{tabular}
\end{table}
Observe that converse and achievability bounds coincide. Further observe that opportunistic messages can only be transmitted reliably for VWI or WI. In the other interference regions, the opportunistic sum capacity is zero.

\subsubsection{Fully Correlated Case} \label{ER_corr}
Assume now that  the sequences $B_1$ and $B_2$ are fully correlated ($B_1=B_2$).
For local CSIR, the correlation between $B_1$ and $B_2$ has no influence on the opportunistic sum capacity region. Indeed, in this case the channel inputs are independent of $(B_1,B_2)$ and the opportunistic sum capacity region of the quasi-static bursty IC depends on $(B_1,B_2)$ only via the marginal distributions of $B_i$, $i=1,2$. Hence, it follows that Theorem~\ref{Thm:convBIC_local} as well as Table~\ref{Tab:qs_local_uncorr} apply also to the fully correlated case and  local CSIR scenario. For completeness, a proof of the converse part is given in Appendix~\ref{Sec:Local_00_corr}. The achievability part is included in Appendix~\ref{Ach. apportunistic}.

\subsection{Ergodic Channel}
\unskip
\subsubsection{Independent Case}\label{ER_uncorr}
For the case where the sequences $B_1^K$ and $B_2^K$ are independent of each other, we have the following~theorems.
\begin{theorem}[Converse bounds for local CSIR]\label{Thm:converse-indCSI}
Assume that $B_1^K$ and $B_2^K$ are independent of each other. The sum rate $R$ for the bursty IC is upper-bounded by
\begin{align}\label{UB_TK}
R &\leq2\frac{1-p}{1+p}n_d+2\frac{p}{1+p}\left[(n_d-n_c)^++\max(n_d,n_c)\right]
\end{align}
and
\begin{equation}\label{UB_Wang}
R \leq
\begin{cases} 
      2(1-2p)n_d+2p\bigl[(n_d-n_c)^++\max(n_d,n_c)\bigr] & \ p\leq\tfrac{1}{2}, \\
      2(1-p)\left[(n_d-n_c)^++\max(n_d,n_c)\right]+2(2p-1)\left[\max\{(n_d-n_c)^+,n_c\}\right]& \ p>\tfrac{1}{2}.
   \end{cases}
\end{equation}

\end{theorem}
\begin{IEEEproof}\label{UBP}
Bound \eqref{UB_TK} coincides with \cite[Eq.~(3)]{Wang13}. Specifically, \cite[Eq.~(3)]{Wang13} derives \eqref{UB_TK} for the considered channel model with $T=1$ and feedback. The proof for this bound under local CSIRT (without feedback) is given in Appendix~\ref{Ap: UB_local_CSI}. Bound \eqref{UB_Wang} coincides with \cite[Lemma~A.1]{Wang13_pre}. Specifically,~\cite[Lemma~A.1]{Wang13_pre} derives \eqref{UB_Wang} for the model considered with $T=1$. The proof of \cite[Lemma~A.1]{Wang13_pre} directly generalizes to arbitrary $T$.
\end{IEEEproof}

\begin{theorem}[Achievability bounds for local CSIR]\label{Thm:achiev-noCSI}
Assume that $B_1^K$ and $B_2^K$ are independent of each other. The following sum rate $R$ is achievable over the bursty IC:

\begin{equation}\label{Ach. CSIR2}
R =
\begin{cases} 
      2(1-2p)n_d+2p\bigl[(n_d-n_c)^++\max(n_d,n_c)\bigr], & p\leq\tfrac{1}{2}, \\
      \min\left\{(n_d-n_c)^++\max(n_d,n_c),\right.& \\
       \left.2(1-p)\left[(n_d-n_c)^++\max(n_d,n_c)\right]+2(2p-1)\left[\max\{(n_d-n_c)^+,n_c\}\right]\right\},& p>\tfrac{1}{2}.\\
\end{cases}
\end{equation}

\end{theorem}
\begin{IEEEproof}
The achievability scheme for VWI for all values of $p$, and for WI and MI  when $0\leq p\leq\tfrac{1}{2}$, is described in Appendix~\ref{Ap:Ach-NoCSI-1}. The achievability scheme for WI and $\tfrac{1}{2}< p\leq 1$ is described in Appendix~\ref{Ap:Ach-NoCSI-2}. The scheme for SI and $0\leq p\leq \tfrac{1}{2}$ is summarized in Appendix~\ref{Ap:Ach-NoCSI-3}. 
For MI and SI when $\tfrac{1}{2}<p\leq {1}$, the achievability bound in the theorem corresponds to the one of the non-bursty IC~\cite{jafar10}. This also implies that in this sub-region we do not exploit the burstiness of the IC.
\end{IEEEproof}
\begin{remark}
 The achievability schemes presented in Theorem~\ref{Thm:achiev-noCSI} are similar to those described in \cite{Vahid14,Vahid17}. They achieve the capacity region by applying point-to-point erasure codes with appropriate rates at each transmitter and using either treating-interference-as-erasure or interference-decoding at each receiver. Specifically, we apply treating-interference-as-erasure in the VWI region and for all values of $p$, and for all interference regions, except VSI, and $p\leq\tfrac{1}{2}$. Interference-decoding at each receiver is applied in the MI and SI regions for $p>\tfrac{1}{2}$.
\end{remark}
\begin{remark}
Wang et al. claim in \cite[Lemma~A.1]{Wang13_pre} that the converse bound (18) is tight for $0\leq p \leq\tfrac{1}{2}$ without providing an achievability bound. Instead, they refer to Khude et al. \cite{Khude09} for the inner bound which, alas, does not apply to the ergodic setup. While it is possible to adapt the achievability schemes considered in \cite{Khude09} to prove \eqref{Ach. CSIR2}, a number of steps are required. For completeness, we include the achievability schemes for the ergodic setup and $0 \leq p\leq\frac{1}{2}$ in Appendix~\ref{Ap:Ach-NoCSI-1}.
\end{remark}

Table~\ref{Tab:local_uncorr} summarizes the results of Theorems \ref{Thm:converse-indCSI} and \ref{Thm:achiev-noCSI}. We write the sum capacities in bold face when the converse and achievability bounds match. In Table~\ref{Tab:local_uncorr}, we define 
\begin{IEEEeqnarray}{lCl}
\mathbf{C_{\text{LMI}}}&\eqdef& \min\left\{2[2(n_d-n_c)+p(3n_c-2n_d)],2\left[\frac{1-p}{1+p}n_d+\frac{p}{1+p}(2n_d-n_c)\right]\right\}\label{Eq: C_MI}\\
\mathbf{C_{\text{LSI}}}&\eqdef& \min\left\{2pn_c,2\left[\frac{1-p}{1+p}n_d+\frac{p}{1+p}n_c\right]\right\}\label{Eq: C_SI}
\end{IEEEeqnarray}

where ``{L}'' stands for ``local CSIR''.
\begin{table}[tbp]
\centering \scriptsize
\caption{Sum capacity for local CSIR.}\label{Tab:local_uncorr}
\begin{tabular}{cccc}
\toprule
\textbf{Regions}&$p\leq\tfrac{1}{2}$&\multicolumn{2}{c}{$p>\tfrac{1}{2}$}\\\midrule
\textbf{VWI}& $\mathbf{2(n_d-pn_c)}$ &\multicolumn{2}{c}{$\mathbf{2(n_d-pn_c)}$}\\ \midrule
{\textbf{WI}}&$\mathbf{2(n_d-pn_c)}$ &\multicolumn{2}{c}{$\mathbf{4(n_d-n_c)+2p(3n_c-2n_d)}$}\\\midrule
 {\textbf{MI}}& {$\mathbf{2(n_d-pn_c)}$} &\multicolumn{2}{c}{${2n_d-n_c}\leq R \leq\mathbf{C_{\text{LMI}}}$} \\ \midrule
 {\textbf{SI}}& {$\mathbf{2(1-2p)n_d+2pn_c}$}  &\multicolumn{2}{c}{${n_c}\leq R \leq \mathbf{C_{\text{LSI}}}$}\\ \bottomrule
\end{tabular}
\end{table} 
\subsubsection{Fully Correlated Case}
For local CSIR, the dependence  between $B_1^K$ and $B_2^K$ has no influence on the capacity region. Indeed, in this case the channel inputs are independent of $(B_1^K,B_2^K)$ and decoder $i$ has only access to $B_{i,k}$ and $(\dmat{S}_{n_d}\mat{X}_{i,k}\oplus B_{i,k}\dmat{S}_{n_c}\mat{X}_{j,k})$, $k=1,\ldots,K$, $j=3-i$ and $i=1,2$. Furthermore, $\Pr\{\hat{W}_1\neq W_1\cup\hat{W}_2\neq W_2\}$ vanishes as $K\to\infty$ if, and only if, $\Pr\{\hat{W}_i\neq W_i\}$, $i=1,2$, vanishes as $K\to \infty$. Since $\Pr (\hat{W}_i\neq W_i)$ depends only on $B_i^K$, the capacity region of the bursty IC depends on $(B_1^K,B_2^K)$ only via the marginal distributions of $B_1^K$ and $B_2^K$. Hence, Theorems~\ref{Thm:converse-indCSI} and \ref{Thm:achiev-noCSI} as well as Table~\ref{Tab:local_uncorr} apply also to the case where $B_1^K=B_2^K$. This is consistent with the observation by Sato \cite{Sato77} that ``the capacity region is the same for all two-user channels that have the same marginal probabilities.''

\subsection{Quasi-Static vs. Ergodic Setup}\label{CSIR_QS_ER}
In general, the sum capacities of the quasi-static and ergodic channels cannot be compared, because in the former case we have a set of sum capacities (worst case and opportunistic), whereas in the latter case only one is defined. To allow for a comparison, we introduce for the quasi-static channel the average sum capacity as
\begin{IEEEeqnarray}{lCl}\label{R_LocalCSIR}
\bar{C} \eqdef \sup_{(R,\Delta R_1(0),\Delta R_2(0))}\{R+(1-p)(\Delta R_1(0)+\Delta R_2(0))\}
\end{IEEEeqnarray}
where the suprema is over all tuples $(R,\Delta R_1(0), \Delta R_2(0))$ that satisfy \eqref{Eq: IC1}--\eqref{Eq: local_Delta2}. Intuitively, the average rate corresponds to the case where we send many messages over independent quasi-static fading channels. By the law of large numbers, a fraction of $p$ transmissions will be affected by interference, the remaining transmissions will be interference-free. Table~\ref{Tab:local_av_Qs} summarizes the average sum capacity for the different interference regions.

By comparing Tables~\ref{Tab:local_uncorr} and \ref{Tab:local_av_Qs}, we can observe that for $p\leq\tfrac{1}{2}$ and all interference regions, and for $p>\tfrac{1}{2}$ and VWI/WI, the average sum capacity in the quasi-static setup coincides with the sum capacity in the ergodic setup. For $p>\tfrac{1}{2}$, and MI/SI (where converse and achievability bounds do not coincide), the average sum capacities in the quasi-static setup coincide with the achievability bounds of the ergodic setup.

\begin{table}[tbp]
\centering \scriptsize
\caption{Average sum capacities for local CSIR.}\label{Tab:local_av_Qs}
\begin{tabular}{cccc}
\toprule
\textbf{Regions}&$p\leq\tfrac{1}{2}$&\multicolumn{2}{c}{$p>\tfrac{1}{2}$}\\\midrule
\textbf{VWI}&$2(n_d-pn_c)$ & \multicolumn{2}{c}{$2(n_d-pn_c)$}\\ \midrule
{\textbf{WI}}&$2(n_d-pn_c)$ &\multicolumn{2}{c}{$4(n_d-n_c)+2p(3n_c-2n_d)$}\\\midrule
 {\textbf{MI}}& {$2(n_d-pn_c)$} &\multicolumn{2}{c}{${2n_d-n_c}$} \\ \midrule
 {\textbf{SI}}& {$2(1-2p)n_d+2pn_c$}  &\multicolumn{2}{c}{${n_c}$}\\ \bottomrule
\end{tabular}
\end{table}

\section{Local CSIRT}\label{Sec: local_CSIRT}
For the quasi-static and ergodic setups,  we present converse and achievability bounds when transmitters and receivers have access to their corresponding interference states. We shall only consider the independent case here, because when $B_1^K=B_2^K$ local CSIRT coincides with global CSIRT, which will be discussed in Section~\ref{Sec: global_CSIRT}.

\subsection{Quasi-Static Channel}
For the quasi-static channel, the converse and achievability bounds were already presented in Theorem~\ref{Thm:convBIC_local} in Section~\ref{QS: Local_CSIR_uncorr}. Indeed, the converse bounds were derived for local CSIRT, whereas the achievability bounds in that theorem were derived for local CSIR. Since these bounds coincide for all interference regions and all probabilities of $0<p<1$ it follows that, for the quasi-static channel, availability of local CSI at the transmitter in addition to local CSI at the receiver is not beneficial. The converse and achievability bounds are  then given in Theorem~\ref{Thm:convBIC_local}.
 
\subsection{Ergodic Channel}
The converse bound \eqref{UB_TK} presented in Theorem~\ref{Thm:converse-indCSI} was derived for local CSIRT, so it applies to the case at hand. We next present achievability bounds for this setup that improve upon those for CSIR. The aim of these bounds is to provide computable expressions showing that local CSIRT outperforms local CSIR in the whole range of the $\alpha$ parameter. While the particular achievability schemes are sometimes involved, the intuition behind these schemes can be explained with the following toy~example.

\vspace{6pt}
\noindent\textbf{Example:} 
{\it Let us assume that $n_d=n_c=T=1$, and suppose that at time $k$ the transmitters send the bits $(B_{1,k},B_{2,k})\in\{0,1\}^2$. If there is no interference, then receiver $i$ receives $X_{i,k}$. If there is interference, then receiver $i$ receives $X_{1,k}\oplus X_{2,k}$. Consequently, the channel flips $X_{1,k}$ if $B_{1,k}=X_{2,k}=1$, and it flips $X_{2,k}$ if $B_{2,k}=X_{1,k}=1$. It follows that each transmitter-receiver pair experiences a \emph{binary symmetric channel (BSC)} with a given crossover probability that depends on $p$ and on the probabilities that $(X_1,X_2)$ are one.  Specifically, let}
\begin{IEEEeqnarray}{lCl}
P_{X_1|B_1}(X_1=1|B_1=0)\eqdef p_1\label{p1}\\
P_{X_1|B_1}(X_1=1|B_1=1)\eqdef p_2\label{p2}\\
P_{X_2|B_2}(X_2=1|{B_2}=0)\eqdef q_1\label{q1}\\
P_{X_2|B_2}(X_2=1|{B_2}=1)\eqdef q_2\label{q2}
\end{IEEEeqnarray}
{\it and define $p_3\triangleq (1-p)p_1+pp_2$ and $q_3\triangleq (1-p)q_1+pq_2$, which are the crossover probabilities of the BSCs experienced by receivers $1$ and $2$, respectively, when they are affected by interference. By drawing for each user two codebooks (one for $B_{i,k}=0$ and one for $B_{i,k}=1$) i.i.d.\ at random according to the probabilities $p_1$, $p_2$, $q_1$, and $q_2$, and by following a random-coding argument, it can be shown that this scheme achieves the sum rate}
\begin{IEEEeqnarray}{lCl}\label{Bounda1}
R&=& (1-p)[H_b(p_1)+H_b(q_1)]+p[\Hsum(p_2,q_3)-H_b(q_3)]+p[\Hsum(q_2,p_3)-H_b(p_3)].
\end{IEEEeqnarray}
{\it This expression holds for any set of parameters $(p_1, p_2, q_1,q_2)$, and the largest sum rate achieved by this scheme is obtained by maximizing over $(p_1,p_2,q_1,q_2)\in \bigl[0,\tfrac{1}{2}\bigl]^4$.}
 
 \vspace{6pt}
In the following, we present the achievable sum rates that can be obtained by generalizing the above achievability scheme to general $n_d$ and $n_c$. The achievability schemes that achieve these rates are presented in Appendix~\ref{App: Local_CSIRT}. The largest achievable sum rates can then be obtained by numerically maximizing over the parameters $(p_1,p_2,q_1,q_2,\ldots)$ (which depend on the interference region).

\begin{enumerate}
\item For the VWI region, we achieve the sum rate
\begin{IEEEeqnarray}{lCl}\label{VWI_local_CSIRT}
R=2(n_d-pn_c).
\end{IEEEeqnarray}

\item For the WI region, we can achieve for any $(p_1,p_2, q_1,q_2) \in \bigl[0,\tfrac{1}{2}\bigl]^4$
\begin{IEEEeqnarray}{lCl}\label{WI_local_CSIRT}
 R_1&=&(n_d-n_c)+(1-p)[(n_d-n_c)+(2n_c-n_d)H_b(p_1)]+p(2n_c-n_d)(1-H_b(q_3))\label{WI_local_CSIRT_R1}\\
 R_2&=&(n_d-n_c)+(1-p)[(n_d-n_c)+(2n_c-n_d)H_b(q_1)]+p(2n_c-n_d)(1-H_b(p_3))\label{WI_local_CSIRT_R2}
\end{IEEEeqnarray}
where $p_3 = (1-p){p}_1+p{p}_2$ and $q_3 = (1-p){q}_1+p{q}_2$.
\item To present the achievable rates for MI, we need to divide the region into the following four subregions:
\begin{enumerate}
\item For $\tfrac{2}{3} \leq \alpha \leq \tfrac{3}{4}$, we can achieve for any $(p_1,p_2,\tilde{p}_1,\tilde{p}_2,\hat{p}_1,q_1,q_2,\tilde{q}_1,\tilde{q}_2,\hat{q}_1)\in \bigl[0,\tfrac{1}{2}\bigl]^{10}$ and $(\eta_1,\gamma_1)\in \bigl[\frac{1}{2},1\bigr]^2$ 
 \begin{myequation}
  \begin{aligned}
R_1=&(n_d-n_c)\\
&{}+(1-p)\left[\left(\tfrac{3n_c-2n_d}{2}\right)\left(H_b(\eta_1)+H_b(\hat{p}_1)+H_b({p}_1)\right)+\left(\tfrac{4n_d-5n_c}{2}\right)H_b(\tilde{p}_1)+(n_d-n_c)\right]\\
&{}+ p\Big[\left(\tfrac{3n_c-2n_d}{2}\right)\left(1+\Hsum(p_2,\tilde{\gamma})-H_b(\tilde{\gamma})+\Hsum(\tilde{p}_2,q_3)-H_b(q_3)-H_b(\hat{q}_3)\right)\\
& {}+\left(\tfrac{4n_d-5n_c}{2}\right)\left(1-H_b(\tilde{q}_3)\right)\Big]\label{MI_local_CSIRT_sr1_R1}
\end{aligned}
\end{myequation}
where $q_3 = (1-p){q}_1+p{q}_2$, $\tilde{q}_3 = (1-p)\tilde{q}_1+p\tilde{q}_2$,  $\hat{q}_3 = (1-p)\hat{q}_1$, and $\tilde{\gamma}=p+\gamma_1(1-p)$, and
\begin{myequation}
  \begin{aligned}
R_2=&(n_d-n_c)\\
&{}+(1-p)\left[\left(\tfrac{3n_c-2n_d}{2}\right)\left(H_b(\gamma_1)+H_b(\hat{q}_1)+H_b({q}_1)\right)+\left(\tfrac{4n_d-5n_c}{2}\right)H_b(\tilde{q}_1)+(n_d-n_c)\right]\\
&{}+ p\left[\left(\tfrac{3n_c-2n_d}{2}\right)\left(1+\Hsum(q_2,\tilde{\eta})-H_b(\tilde{\eta})+\Hsum(\tilde{q}_2,p_3)-H_b(p_3)-H_b(\hat{p}_3)\right)\right.\big.\\
& {}+\left(\tfrac{4n_d-5n_c}{2}\right)\left(1-H_b(\tilde{p}_3)\right)\big]\label{MI_local_CSIRT_sr1_R2}
\end{aligned}
\end{myequation}
where  $p_3 = (1-p){p}_1+p{p}_2$, $\tilde{p}_3 = (1-p)\tilde{p}_1+p\tilde{p}_2$,  $\hat{p}_3 = (1-p)\hat{p}_1$, and $\tilde{\eta}=p+\eta_1(1-p)$. 
\begin{remark}\label{Remark_eta_gamma}
After combining \eqref{MI_local_CSIRT_sr1_R1} and \eqref{MI_local_CSIRT_sr1_R2}, $\eta_1$ and $\gamma_1$ appear only through the functions $H_b(\eta_1)-H_b(p+\eta_1(1-p))$ and $H_b(\gamma_1)-H_b(p+\gamma_1(1-p))$, respectively. Hence, $\eta_1$ and $\gamma_1$ can be optimized separately from the remaining terms.
\end{remark}
\item For $\tfrac{3}{4} \leq \alpha \leq \tfrac{4}{5}$, we can achieve for any $(p_1,p_2,\tilde{p}_1,\tilde{p}_2,\hat{p}_1,q_1,q_2,\tilde{q}_1,\tilde{q}_2,\hat{q}_1)\in \bigl[0,\tfrac{1}{2}\bigl]^{10}$ and $(\eta_1,\gamma_1)\in \bigl[\frac{1}{2},1\bigr]^2$ 
\begin{myequation}
  \begin{aligned}
R_1=&(n_d-n_c)\\
&{}+(1-p)\Big[\left(\tfrac{3n_c-2n_d}{2}\right)\left(H_b({p}_1)+H_b(\eta_1)+H_b(\hat{p}_1)\right)+(\tfrac{4n_d-5n_c}{2})H_b(\tilde{p}_1)+(n_d-n_c)\Big]\\
&{}+p\Big[\left(\tfrac{3n_c-2n_d}{2}\right)\left(\Hsum({p}_2,\tilde{\gamma})-H_b(\tilde{\gamma})+1-H_b(\hat{q}_3)\right)\\
&  {}+\left(\tfrac{4n_d-5n_c}{2}\right)\left(\Hsum(\tilde{p}_2,q_3)-H_b({q}_3)+1-H_b(\tilde{q}_3)\right)\Big]\label{MI_local_CSIRT_sr2_R1}
\end{aligned}
\end{myequation}
where $q_3 = (1-p){q}_1+p{q}_2$, $\tilde{q}_3 = (1-p)\tilde{q}_1+p\tilde{q}_2$,  $\hat{q}_3 = (1-p)\hat{q}_1$, and $\tilde{\gamma}=p+\gamma_1(1-p)$,~and
\begin{myequation}
  \begin{aligned}
R_2=&(n_d-n_c)\\
&{}+(1-p)\Big[\left(\tfrac{3n_c-2n_d}{2}\right)\left(H_b({q}_1)+H_b(\gamma_1)+H_b(\hat{q}_1)\right)+(\tfrac{4n_d-5n_c}{2})H_b(\tilde{q}_1)+(n_d-n_c)\Big]\\
&{}+p\Big[\left(\tfrac{3n_c-2n_d}{2}\right)\left(\Hsum({q}_2,\tilde{\eta})-H_b(\tilde{\eta})+1-H_b(\hat{p}_3)\right)\\
& {}+\left(\tfrac{4n_d-5n_c}{2}\right)\left(\Hsum(\tilde{q}_2,p_3)-H_b({p}_3)+1-H_b(\tilde{p}_3)\right)\Big] \label{MI_local_CSIRT_sr2_R2}
\end{aligned}
\end{myequation}

where  $p_3 = (1-p){p}_1+p{p}_2$, $\tilde{p}_3 = (1-p)\tilde{p}_1+p\tilde{p}_2$,  $\hat{p}_3 = (1-p)\hat{p}_1$, and $\tilde{\eta}=p+\eta_1(1-p)$. Remark~\ref{Remark_eta_gamma} also applies to the parameters $\eta_1$ and $\gamma_1$ in \eqref{MI_local_CSIRT_sr2_R1} and \eqref{MI_local_CSIRT_sr2_R2}.
\item For $\tfrac{4}{5} \leq \alpha \leq \tfrac{6}{7}$, we can achieve for any $(p_1,p_2,\hat{p}_1,q_1,q_2,\hat{q}_1)\in \bigl[0,\tfrac{1}{2}\bigl]^{6}$ and $(\eta_1,\eta',\gamma_1,\gamma') \in \bigl[\tfrac{1}{2},1\bigl]^{4}$
\begin{myequation}
  \begin{aligned}
R_1=&(n_d-n_c)\\
&{}+(1-p)\Big[\left(\tfrac{5n_c-4n_d}{2}\right)(1+H_b(\eta'))+(n_d-n_c)\left(1+H_b({p}_1)+H_b(\eta_1)+H_b(\hat{p}_1)\right)\Big]\\
&{}+p\Big[\left(\tfrac{5n_c-4n_d}{2}\right)\left(1-H_b(\tilde{\gamma})+\Hsum({p}_2,\gamma')-H_b(\gamma')\right.\\
&{}+\Hsum(\eta'(1-\tilde{\gamma})+(1-\eta')\tilde{\gamma},q_3)-H_b(q_3)\big)\\
& {}+\left(\tfrac{6n_d-7n_c}{2}\right)\left(\Hsum({p}_2,\tilde{\gamma})-H_b(\tilde{\gamma})\right)+(n_d-n_c)(1-H_b(\hat{q}_3))\Big]\label{MI_local_CSIRT_sr3_R1}
\end{aligned}
\end{myequation}
where $q_3 = (1-p){q}_1+p{q}_2$,  $\hat{q}_3 = (1-p)\hat{q}_1$, and $\tilde{\gamma}=p+\gamma_1(1-p)$, and
\begin{myequation}
  \begin{aligned}
R_2=&(n_d-n_c)\\
&{}+(1-p)\Big[\left(\tfrac{5n_c-4n_d}{2}\right)(1+H_b(\gamma'))+(n_d-n_c)\left(1+H_b({q}_1)+H_b(\gamma_1)+H_b(\hat{q}_1)\right)\Big]\\
&{}+p\Big[\left(\tfrac{5n_c-4n_d}{2}\right)\left(1-H_b(\tilde{\eta})+\Hsum({q}_2,\eta')-H_b(\eta')\right.\\
&{}+\left.\Hsum(\gamma'(1-\tilde{\eta})+(1-\gamma')\tilde{\eta},p_3)-H_b(p_3)\right)\\
&{}+\left(\tfrac{6n_d-7n_c}{2}\right)\left(\Hsum({q}_2,\tilde{\eta})-H_b(\tilde{\eta})\right)+(n_d-n_c)(1-H_b(\hat{p}_3))\Big]\label{MI_local_CSIRT_sr3_R2}
\end{aligned}
\end{myequation}
where  $p_3 = (1-p){p}_1+p{p}_2$, $\hat{p}_3 = (1-p)\hat{p}_1$, and $\tilde{\eta}=p+\eta_1(1-p)$.
\item For $\tfrac{6}{7} \leq \alpha \leq 1$ we can achieve for any $(p_1,p_2,\hat{p}_1,q_1,q_2,\hat{q}_1)\in \bigl[0,\tfrac{1}{2}\bigl]^{6}$ and $(\eta_1,\eta',\gamma_1,\gamma') \in \bigl[\tfrac{1}{2},1\bigl]^{4}$
\begin{myequation}
  \begin{aligned}
R_1
=&(n_d-n_c)\\
&{}+(1-p)\big[(6n_c-5n_d)H_b(p_1)+(n_d-n_c)\left(2+H_b(\eta_1)+H_b(\eta')+H_b(\hat{p}_1)\right)\big]\\
&{}+ p\big[(n_d-n_c)\left(2-H_b(\tilde{\gamma})-H_b(\hat{q}_3)+\Hsum(\eta'(1-\tilde{\gamma})+(1-\eta')\tilde{\gamma},q_3)-H_b(q_3)\right)\\
&{}+(n_d-n_c)\left(\Hsum(p_2,\gamma')-H_b(\gamma')\right)\\
&{}+(7n_c-6n_d)\left(\Hsum(p_2,q_3)-H_b(q_3)\right)\big]\label{MI_local_CSIRT_sr4_R1}
\end{aligned}
\end{myequation}
where $q_3 = (1-p){q}_1+p{q}_2$,  $\hat{q}_3 = (1-p)\hat{q}_1$, and $\tilde{\gamma}=p+\gamma_1(1-p)$, and
\begin{myequation}
  \begin{aligned}
R_2
=&(n_d-n_c)\\
&{}+(1-p)\big[(6n_c-5n_d)H_b(q_1)+(n_d-n_c)\left(2+H_b(\gamma_1)+H_b(\gamma')+H_b(\hat{q}_1)\right)\big]\\
&{}+ p\big[(n_d-n_c)\left(2-H_b(\tilde{\eta})-H_b(\hat{p}_3)+\Hsum(\gamma'(1-\tilde{\eta})+(1-\gamma')\tilde{\eta},p_3)-H_b(p_3)\right)\\
&{}+(n_d-n_c)\left(\Hsum(q_2,\eta')-H_b(\eta')\right)\\
&{}+(7n_c-6n_d)\left(\Hsum(q_2,p_3)-H_b(p_3)\right)\big]\label{MI_local_CSIRT_sr4_R2}
\end{aligned}
\end{myequation}
where  $p_3 = (1-p){p}_1+p{p}_2$, $\hat{p}_3 = (1-p)\hat{p}_1$, and $\tilde{\eta}=p+\eta_1(1-p)$.
\end{enumerate}
\item To present the achievable rates for SI, we divide the region into the following four subregions:
\begin{enumerate}
\item For $1 \leq \alpha \leq \tfrac{6}{5}$, we can achieve for any $(p_1,p_2,q_1,q_2) \in \bigl[0,\frac{1}{2}\bigr]^4$ and $(\eta_1,\eta',\gamma_1,\gamma' )\in \bigl[\frac{1}{2},1\bigr]^4$
\begin{myequation}
  \begin{aligned}
R_1
=&(n_c-n_d)+(1-p)\big[(5n_d-4n_c)H_b(p_1)+(n_c-n_d)\left(1+H_b(\eta_1)+H_b(\eta')\right)\big]\\
&{}+ p\big[(n_c-n_d)\left(1-H_b(\tilde{\gamma})+\Hsum(\eta'(1-\tilde{\gamma})+(1-\eta')\tilde{\gamma},q_3)-H_b(q_3)\right)\\
&{}+(n_c-n_d)\left(\Hsum(p_2,\gamma')-H_b(\gamma')\right)\\
&{}+(6n_d-5n_c)\left(\Hsum(p_2,q_3)-H_b(q_3)\right)\big]\label{SI_local_CSIRT_sr1_R1}
\end{aligned}
\end{myequation}
where $q_3 = (1-p){q}_1+p{q}_2$ and $\tilde{\gamma}=p+\gamma_1(1-p)$, and
\begin{myequation}
  \begin{aligned}
R_2
=&(n_c-n_d)+(1-p)\big[(5n_d-4n_c)H_b(q_1)+(n_c-n_d)\left(1+H_b(\gamma_1)+H_b(\gamma')\right)\big]\\
&{}+ p\big[(n_c-n_d)\left(1-H_b(\tilde{\eta})+\Hsum(\gamma'(1-\tilde{\eta})+(1-\gamma')\tilde{\eta},q_3)-H_b(p_3)\right)\\
&{}+(n_c-n_d)\left(\Hsum(q_2,\eta')-H_b(\eta')\right)\\
&{}+(6n_d-5n_c)\left(\Hsum(q_2,p_3)-H_b(p_3)\right)\big]\label{SI_local_CSIRT_sr1_R2}
\end{aligned}
\end{myequation}
where  $p_3 = (1-p){p}_1+p{p}_2$ and $\tilde{\eta}=p+\eta_1(1-p)$.
\item For $\tfrac{6}{5} \leq \alpha \leq \tfrac{4}{3}$, we can achieve for any $(p_1,p_2,q_1,q_2) \in \bigl[0,\frac{1}{2}\bigr]^4$ and $(\eta_1,\gamma_1)\in \bigl[\frac{1}{2},1\bigl]^2$
\begin{myequation}
  \begin{aligned}
R_1=&\left(2n_d-\tfrac{3n_c}{2}\right)+(1-p)\big[\left(2n_d-\tfrac{3n_c}{2}\right)H_b(\eta_1)+2(n_c-n_d)+(3n_d-2n_c)H_b({p}_1)\big ]\\
&{}+p\big[(n_c-n_d)\left(1-H_b({q}_3)\right)+(2n_d-\tfrac{3n_c}{2})\left(1-H_b(\tilde{\gamma})\right)+(\tfrac{5n_c}{2}-3n_d)\big ]\label{SI_local_CSIRT_sr2_R1}
\end{aligned}
\end{myequation}
where $q_3 = (1-p){q}_1+p{q}_2$, and $\tilde{\gamma}=p+\gamma_1(1-p)$, and
\begin{myequation}
  \begin{aligned}
R_2=&\left(2n_d-\tfrac{3n_c}{2}\right)+(1-p)\big[\left(2n_d-\tfrac{3n_c}{2}\right)H_b(\gamma_1)+2(n_c-n_d)+(3n_d-2n_c)H_b({q}_1)\big ]\\
&{}+p\big[(n_c-n_d)\left(1-H_b({p}_3)\right)+(2n_d-\tfrac{3n_c}{2})\left(1-H_b(\tilde{\eta})\right)+(\tfrac{5n_c}{2}-3n_d)\big ] \label{SI_local_CSIRT_sr2_R2}
\end{aligned}
\end{myequation}
where  $p_3 = (1-p){p}_1+p{p}_2$,  and $\tilde{\eta}=p+\eta_1(1-p)$. Remark~\ref{Remark_eta_gamma} also applies to the parameters $\eta_1$ and $\gamma_1$ in \eqref{SI_local_CSIRT_sr2_R1} and \eqref{SI_local_CSIRT_sr2_R2}.
\item For $\tfrac{4}{3} \leq \alpha \leq \tfrac{3}{2}$, we can achieve for any $(p_1,p_2,q_1,q_2)\in \bigl[0,\frac{1}{2}\bigl]^4$ and $(\eta_1,\gamma_1) \in \bigl[\frac{1}{2},1\bigl]^2$,
\begin{myequation}
  \begin{aligned}
R_1=&(n_d-\tfrac{n_c}{2})+(1-p)\big [(3n_d-2n_c)(1+H_b(p_1))+\left(\tfrac{3n_c}{2}-2n_d\right)(1+H_b(\eta_1))\big ]\\
&{}+p\big [(3n_d-2n_c)(1- H_b({q}_3))+(\tfrac{3n_c}{2}-2n_d)(1-H_b(\tilde{\gamma})\big]\label{SI_local_CSIRT_sr3_R1}
\end{aligned}
\end{myequation}
\begin{myequation}
  \begin{aligned}
R_2=&(n_d-\tfrac{n_c}{2})+(1-p)\big [(3n_d-2n_c)(1+H_b(q_1))+\left(\tfrac{3n_c}{2}-2n_d\right)(1+H_b(\gamma_1))\big ]\\
&{}+p\big [(3n_d-2n_c)(1- H_b({p}_3))+(\tfrac{3n_c}{2}-2n_d)(1-H_b(\tilde{\eta})\big]\label{SI_local_CSIRT_sr3_R2}
\end{aligned}
\end{myequation}
where $q_3 = (1-p){q}_1+p{q}_2$, $\tilde{\gamma}=p+\gamma_1(1-p)$, $p_3 = (1-p){p}_1+p{p}_2$ and $\tilde{\eta}=p+\eta_1(1-p)$. Remark~\ref{Remark_eta_gamma} also applies to the parameters $\eta_1$ and $\gamma_1$ in \eqref{SI_local_CSIRT_sr3_R1} and \eqref{SI_local_CSIRT_sr3_R2}.
\item For $\tfrac{3}{2} \leq \alpha \leq 2$, we can achieve for any $\eta_1,\gamma_1 \in \bigl[\frac{1}{2},1\bigr]$
\begin{equation}
R_1=(n_c-n_d)+(1-p)\big [(n_d-\tfrac{n_c}{2})(1+H_b(\eta_1))\big ] + p(n_d-\tfrac{n_c}{2})(1-H_b(\tilde{\gamma}))\label{SI_local_CSIRT_sr4_R1}
\end{equation}
\begin{equation}
R_2=
\big (n_c-n_d)+(1-p)\big [(n_d-\tfrac{n_c}{2})(1+H_b(\gamma_1))\big ] + p(n_d-\tfrac{n_c}{2})(1-H_b(\tilde{\eta}))\label{SI_local_CSIRT_sr4_R2}
\end{equation}
where $\tilde{\gamma}=p+\gamma_1(1-p)$ and  $\tilde{\eta}=p+\eta_1(1-p)$. Remark~\ref{Remark_eta_gamma} also applies to the parameters $\eta_1$ and $\gamma_1$ in \eqref{SI_local_CSIRT_sr4_R1} and \eqref{SI_local_CSIRT_sr4_R2}.
\end{enumerate}
\end{enumerate}

In each region, we optimize numerically over the set of parameters, exploiting in some cases that there is symmetry (except for $\alpha=1$ ) between the corresponding parameters of both users.

\subsection{Local CSIRT vs. Local CSIR}
To evaluate the effect of exploiting local CSI at the transmitter side, we plot in Figures~\ref{Fig:testwi}--\ref{Fig:testsi_1} the converse and achievability bounds for local CSIR and local CSIRT. For each interference region, we choose one value of $\alpha$. We omit the VWI region because in this region both local CSIR and local CISRT coincide. We observe that for all interference regions, except in the VWI region, local CSIRT outperforms local CSIR. We further observe that the largest improvement is obtained for $p=\frac{1}{2}$. This is not surprising, since in this case the uncertainty about the interference states is the largest.

\begin{figure}[tbp]
\centering
\includegraphics[width=0.48\linewidth]{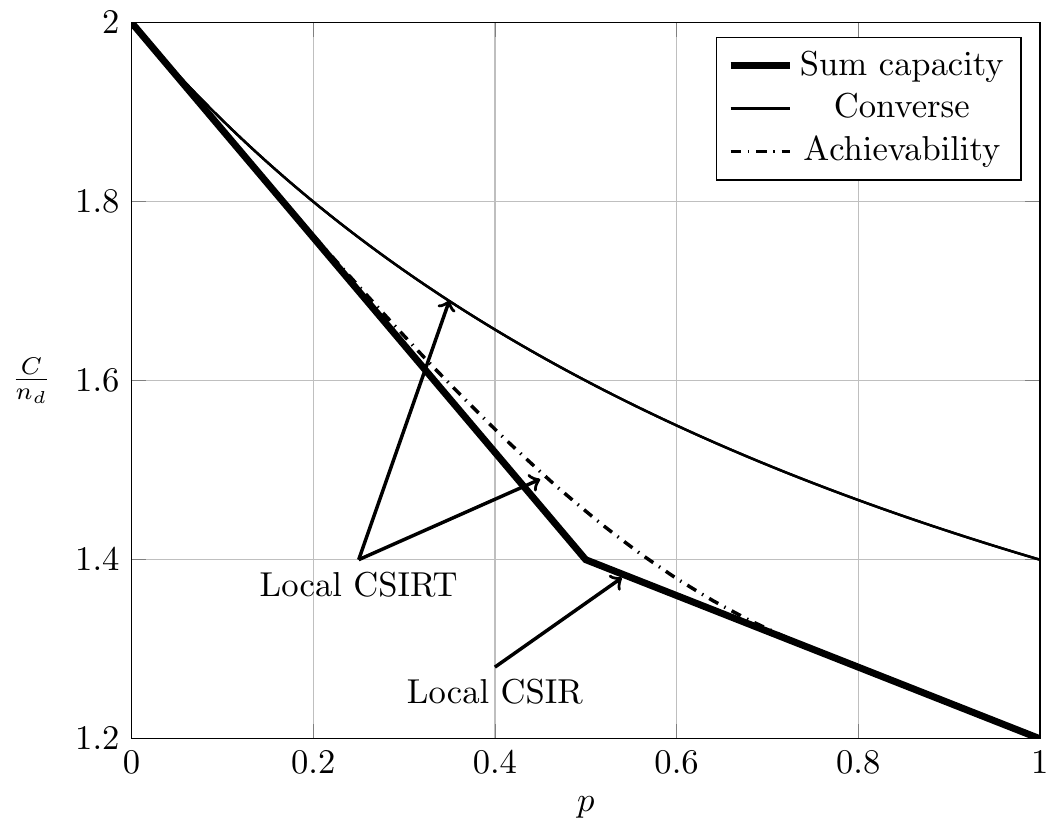}
\caption{Local CSIRT vs. local CSIR for $\alpha=\tfrac{3}{5}$ (WI).}
\label{Fig:testwi}
\end{figure}
\unskip
\begin{figure}[tbp]
\centering
\includegraphics[width=0.48\linewidth]{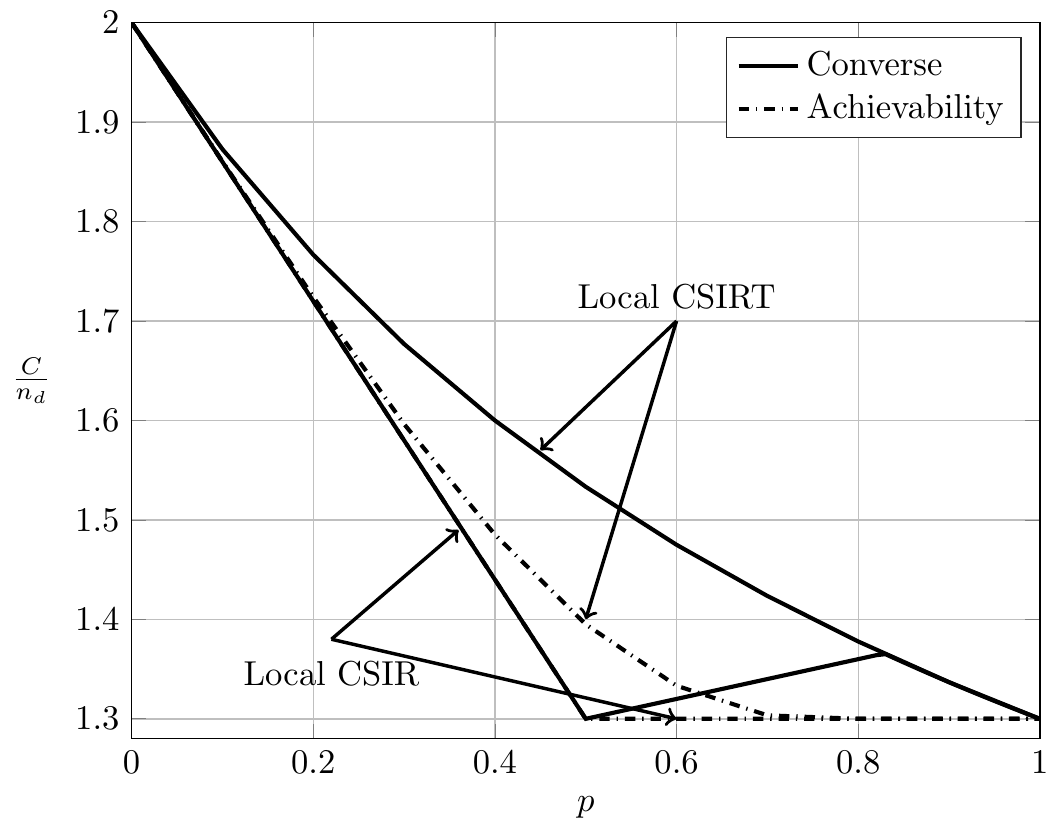}
\caption{Local CSIRT vs. local CSIR for $\alpha=\tfrac{7}{10}$ (MI).}
\label{Fig:testmi_1}
\end{figure}
\unskip
\begin{figure}[tbp]
		\centering
		\includegraphics[width=0.48\linewidth]{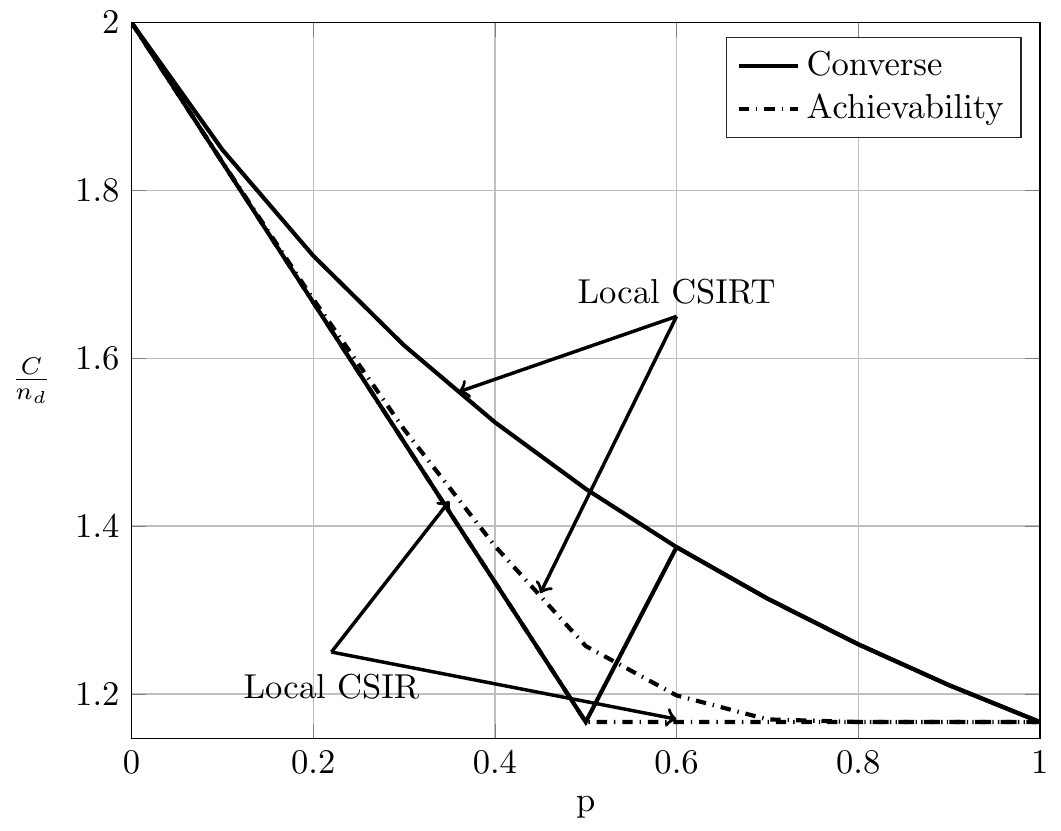}
		\caption{Local CSIRT vs. local CSIR for $\alpha=\tfrac{7}{6}$ (SI).}
	    \label{Fig:testsi_1}
\end{figure}

\subsection{Quasi-Static vs. Ergodic Setup}
As observed in the previous subsection, for the ergodic setup local CSIRT outperforms local CSIR in all interference regions (except VWI). In contrast, the opportunistic rates achievable in the quasi-static setup for local CSIRT coincide with those achievable for local CSIR. In other words, the availability of local CSI at the transmitter is only beneficial in the ergodic setup but not in the quasi-static one. This remains to be true even if we consider the average sum capacity rather than the sum rate region. Intuitively, in the coherent setup, the achievable rates depend on the input distributions of $X_1^K$ and $X_2^K$, and adapting these distributions to the interference state yields a rate gain. In contrast, in the quasi-static setup, we treat the two interference states separately: the worst-case rates are designed for the worst case (where both receivers experience interference), and the opportunistic rates are designed for the best case (where the corresponding receiver is interference-free). 

Given that the opportunistic rate region $(R,\Delta R(V_1,V_2))$ is not enhanced by the availability of local CSI at the transmitter, it follows directly that the same is true for the average sum capacity, defined in \eqref{R_LocalCSIR}.  Note, however, that it is unclear whether \eqref{R_LocalCSIR} corresponds to the best strategy to transmit several messages over independent uses of a quasi-static channel when the transmitters have access to local CSI. Indeed, in this case transmitter $i$ may choose the values for $R_i$ and $\Delta R_i(0)$ as a function of the interference state $B_i$, potentially giving rise to a larger average sum capacity. Yet, the set of achievable rate pairs $(R_i,\Delta R_i(0))$ depends on the choice of $(R_j,\Delta R_j(0))$ of transmitter $j\neq i$, which transmitter $i$ may not deduce since it has no access to the other transmitter's CSI. How the transmitters should adapt their rates to the interference state remains therefore an open question.

\section{Global CSIRT}\label{Sec: global_CSIRT}
We next present converse and achievability bounds for global CSIRT. In this scenario, the transmitters may agree on a specific coding scheme that depends on the realization of $(B_1^K,B_2^K)$. This allows for a more elaborated cooperation between the transmitters and strictly increases the sum capacity compared to the local CSIR/CSIRT scenarios.
\subsection{Quasi-Static Channel}
In the quasi-static scenario with global CSIRT, the messages are, strictly speaking, not opportunistic. Instead, transmitters can choose the message depending on the true state of the interference links, so the strategy is perhaps better described as \emph{rate adaptation}. Nevertheless, the definitions of worst-case sum rate and opportunistic sum rate in Section~\ref{QS_Ch} still apply in this case. To keep notation consistent, we use the definition of ``opportunism'' also for global CSIRT. 

\subsubsection{Independent Case}
Assume first that the sequences $B_1$ and $B_2$ are independent of each other.
\begin{theorem}[Opportunistic sum capacity for global CSIRT]\label{Thm:conv_BIC_global}
Assume that $B_1$ and $B_2$ are independent of each other. For $0 < p < 1$, the opportunistic sum capacity region is the union of the set of rate tuples $(R, \Delta R(00), \Delta R(01), \Delta R(10))$ satisfying \eqref{Eq: IC1}--\eqref{Eq: IC2} and
\begin{IEEEeqnarray}{lCl}
R+\Delta {R}(00) &\leq& 2n_d \label{Eq: local_Delta00}\\
R+\Delta {R}(01) &\leq& (n_{d}-n_{c})^++\max(n_{d},n_{c})\\
R+\Delta {R}(10) &\leq& (n_{d}-n_{c})^++\max(n_{d},n_{c}).
\end{IEEEeqnarray} 

\end{theorem}

\begin{IEEEproof}
The converse bounds are proved in Appendix~\ref{Proof:convBIC_global}. The achievability bounds are achieved by the following achievability scheme: For $\vect{B}=[0,0]$ we use all the $n_d$ sub-channels of both parallel channels. For $\vect{B}=[0,1]$ and $\vect{B}=[1,0]$ and the VWI/WI regions, we use all $n_d$ sub-channels and the receivers decode them only if they are not affected by interference.  For the MI/SI regions, we treat the bursty IC as a non-bursty IC and use the achievability schemes of the IC proposed in \cite{jafar10}. The details can be found in Appendix~\ref{Ach. apportunistic_global_uncorr}.
\end{IEEEproof}
\begin{remark}\label{Remark_corr}
The proofs of Theorems~\ref{Thm:convBIC_local} and \ref{Thm:conv_BIC_global} merely require that the  joint distribution $p_{b_1b_2}\eqdef\Pr\{\vect{B}=[b_1,b_2]\}$ satisfies $p_{00}<1$, $p_{01}>0$, $p_{10}>0$ and $p_{11}>0$. Thus, these theorems also apply to the case where $B_1$ and $B_2$ are dependent, as long as they are not fully correlated.
\end{remark}

Table \ref{Tab:QS_global_uncorr} summarizes the results of Theorem~\ref{Thm:conv_BIC_global}. Observe that for VWI and WI opportunistic messages can be transmitted reliably at a positive rate, while for MI and SI this is only the case if both links are interference-free.
\begin{table}[tbp]
\centering \scriptsize
\caption{Opportunistic sum capacity for global CSIRT when the worst-case sum rate is maximized and $B_1$ and $B_2$ are independent.}
\label{Tab:QS_global_uncorr}
\begin{tabular}{ccccc}
\toprule
\textbf{Rates}&\textbf{VWI}&\textbf{WI}&\textbf{MI}&\textbf{SI}\\\midrule
{\textbf{$C$}}& {$2(n_d-n_c)$}&{$2n_c$}&$2n_d-n_c$&$n_c$\\\midrule
{\textbf{${\Delta C}(00)$}}& {$2n_c$}&{$2(n_d-n_c)$}&$n_c$&$2n_d-n_c$\\\midrule
{\textbf{${\Delta C}(01)/{\Delta C}(10)$}}& {$n_c$}&{$2n_d-3n_c$}&$0$&$0$\\\bottomrule
\end{tabular}
\end{table}

\subsubsection{Fully Correlated Case}
Next, we consider the case in which the interference states are fully correlated. In this scenario, local CSIRT coincides with global CSIRT.

\begin{theorem}[Opportunistic sum capacity for global CSIRT]\label{Thm:convBIC_global_corr}
Assume that $B_1$ and $B_2$ are fully correlated. For $0 \leq p < 1$, the opportunistic sum capacity region is the union of the set of rate pairs $(R, \Delta R(00))$ satisfying~\eqref{Eq: IC1}--\eqref{Eq: IC2} and

\begin{IEEEeqnarray}{lCl}
R+\Delta {R}(00) &\leq& 2n_{d}.
\end{IEEEeqnarray} 

\end{theorem}
\begin{IEEEproof}
For the converse bound, we note that the analysis in Appendix~\ref{Proof:convBIC_global} applies directly to the case where the states $B_1$ and $B_2$ are fully correlated, with the only difference that there are only two possible cases  $\vect {B}=[0,0]$ and $\vect {B}=[1,1]$. The result follows then from \eqref{IC_Global1}, \eqref{IC_Global3} and \eqref{P2P_global}. For the achievability bound, we use an achievability scheme where the opportunistic messages are only decoded in absence of interference at the intended receiver. In this case, we have two parallel interference-free channels, for which the optimal strategy consists of transmitting uncoded bits in the $n_d$ sub-channels.
\end{IEEEproof}

Table \ref{Tab:QS_global_corr} summarizes the results of Theorem~\ref{Thm:convBIC_global_corr}. Observe that the worst-case sum capacity $C$ and the opportunistic sum capacity $\Delta C(00)$ when the channel is interference-free do not depend on the correlation between $B_1$ and $B_2$. The only difference between the independent and fully correlated case is that the interference states $[0,1]$ and $[1,0]$ are impossible if $B_1=B_2$.
\begin{table}[tbp]
\centering \scriptsize
\caption{Opportunistic sum capacity for global CSIRT when the worst-case sum rate is maximized and $B_1$ and $B_2$ are fully correlated.}
\label{Tab:QS_global_corr}
\begin{tabular}{ccccc}
\toprule
\textbf{Rates}&\textbf{VWI}&\textbf{WI}&\textbf{MI}&\textbf{SI}\\\midrule
{\textbf{$C$}}& {$2(n_d-n_c)$}&{$2n_c$}&$2n_d-n_c$&$n_c$\\\midrule
{\textbf{${\Delta C}(00)$}}& {$2n_c$}&{$2(n_d-n_c)$}&$n_c$&$2n_d-n_c$\\\bottomrule
\end{tabular}
\end{table}

\subsection{Ergodic Channel}
\unskip
\subsubsection{Independent Case}
When the sequences $B_1^K$ and $B_2^K$ are independent of each other, we have the following theorems.
\begin{theorem}[Converse bounds for global CSIRT]
\label{Thm:converse-fullCSI}
Assume that $B_1^K$ and $B_2^K$ are independent of each other. The sum rate $R$ for the bursty IC is upper-bounded by
\begin{align}
R&\leq  2(1-p)n_d+p\left[(n_d-n_c)^++\max(n_d,n_c)\right]\label{UBF1}
\end{align}
and
\begin{align}
R&\leq 2 \left[p(1-p)\{(n_d-n_c)^++\max(n_d,n_c)\} +(1-p)^2n_d+p^2\max\{(n_d-n_c)^+,n_c\}\right].\label{UBF2}
\end{align}

\end{theorem}

\begin{IEEEproof}\label{C_Full}
The proof of \eqref{UBF1} follows along similar lines as \eqref{UB_TK} but noting that, for global CSIRT, $\mat{X}_i^K$ depends on both $B_1^K$ and $B_2^K$. The proof of \eqref{UBF2} is based on pairing the interference states according the four possible combinations of ($B_{1,k},B_{2,k}$). See Appendix~\ref{Proof_UBF} for details. 
\end{IEEEproof}
\begin{remark}\label{Remark_spatial_corr_global}
 The proof of Theorem~\ref{Thm:converse-fullCSI} can be extended to consider an arbitrary joint distribution $\displaystyle{p_{b_1b_2}\triangleq\Pr\{\vect{B}_k=[b_{1},b_{2}]\}}$. In this case \eqref{UBF1} is replaced by
\begin{align*}
R&\leq  2(p_{00}+p_{01})n_d+(p_{10}+p_{11})\left[(n_d-n_c)^++\max(n_d,n_c)\right]\nonumber\\
R&\leq  2(p_{00}+p_{10})n_d+(p_{01}+p_{11})\left[(n_d-n_c)^++\max(n_d,n_c)\right]
\end{align*}
 and \eqref{UBF2} becomes
\begin{equation*}
R\leq  (p_{01}+p_{10})[(n_d-n_c)^++\max(n_d,n_c)] +2\left[p_{00}n_d+p_{11}\max\{(n_d-n_c)^+,n_c\}\right].
\end{equation*}
\end{remark}
\begin{theorem}[Achievability bounds for global CSIRT]\label{Thm:achiev-fullCSI}
Assume that $B_1^K$ and $B_2^K$ are independent of each other. The following sum rates $R$ are achievable over the bursty IC:
\begin{align}
  R&= 2 \left[p(1-p)(2n_d-n_c)+(1-p)^2n_d + p^2\max\{(n_d-n_c)^+,n_c\}\right],\quad\text{(VWI, WI)}\label{Eq: LB_1}\\
  R & = 4 n_d p_{\min}  +  2n_d (1-p)^2  +  \bigl(2n_d-n_c\bigr)\bigl(2p-p^2-3 p_{\min}\bigr), \quad\text{(MI)}\label{Eq: LB_2}\\
  R& = 2(n_d+n_c) p_{\min} + 2n_d (1-p)^2 + n_c \bigl(2p-p^2 - 3  p_{\min}\bigr), \quad\text{(SI)}\label{Eq: LB_3}
\end{align}
where $p_{\min}\eqdef\min(p^2,p(1-p))$.
\end{theorem}

\begin{IEEEproof}
 The sum rate \eqref{Eq: LB_1} is achieved by using the optimal scheme for the non-bursty IC when any of the two receivers is affected by interference~\cite{jafar10}, and by using uncoded transmission when there is no interference. The sum rates \eqref{Eq: LB_2} and \eqref{Eq: LB_3} are novel. 
 See Appendix~\ref{Ap: Ach_Full_CSI} for details.
\end{IEEEproof}

\begin{remark}
In contrast to the local CSIR scenario, the achievability schemes presented in Theorem~\ref{Thm:achiev-fullCSI} differ noticeably from those in \cite{Vahid14_dCSIT} for the binary IC. Indeed, while both works exploit global CSIRT to enable cooperation between users, \cite{Vahid14_dCSIT} assumes that only delayed CSI is present. The achievability schemes presented in Theorem~\ref{Thm:achiev-fullCSI} thus cannot be applied directly to the model considered in~\cite{Vahid14_dCSIT}.
\end{remark}

Table~\ref{Tab:global_uncorr} summarizes the results of Theorems \ref{Thm:converse-fullCSI} and \ref{Thm:achiev-fullCSI}. We write the sum capacity in bold face when converse and achievability bounds coincide. In Table~\ref{Tab:global_uncorr}, we define
\begin{IEEEeqnarray}{lCl}
\mathbf{C_{\text{GMI}}}&\eqdef&\min\left\{2n_d-pn_c,2\left[(1-p^2)-(1-2p)\alpha p\right]\right\}\label{Eq: CMIG}\\
\mathbf{C_{\text{GSI}}}&\eqdef&\min\left[n_cp+2(1-p)n_d,2n_d(1-p)^2+2n_c p\right]\label{Eq: CSIG}
\end{IEEEeqnarray}

where ``G'' stands for ``global CSIRT''.

\begin{table}[tbp]
\centering \scriptsize
\caption{Bounds on the sum capacity $C$ for global CSIRT when $B_1^K$ and $B_2^K$ are independent.}
\label{Tab:global_uncorr}
\begin{tabular}{ccc}
\toprule
\textbf{Regions}&\textbf{Achievability}&\textbf{Converse}\\\midrule
\textbf{VWI}& \multicolumn{2}{c}{$\mathbf{2(n_d-pn_c)}$}\\ \midrule
\textbf{WI}& \multicolumn{2}{c}{$\mathbf{2[(1-p^2)n_d+(1-2p)pn_c}]$}\\ \midrule
\textbf{MI}& $4n_dp_{\min}+2n_d(1-p)^2+(2n_d-n_c)(2p-p^2-3p_{\min})$&$\mathbf{C_{\text{GMI}}}$\\ \midrule
\textbf{SI}& {$2(n_d+n_c)p_{\min}+2n_d(1-p)^2+n_c(2p-p^2-3p_{\min})$}&$\mathbf{C_{\text{GSI}}}$\\ \bottomrule
\end{tabular}
\end{table}

\subsubsection{Fully Correlated Case} \label{ER_global_corr}
We next discuss the case where the sequences $B_1^K$ and $B_2^K$ are fully correlated, i.e., $B_1^K=B_2^K$.
 \begin{theorem}[Converse bounds for global CSIRT]
\label{Thm:converse-globalCSI_corr}
Assume that $B_1^K$ and $B_2^K$ are fully correlated. The sum rate $R$ for the bursty IC is upper-bounded by
\begin{align}
R&\leq 2(1-p)n_d+p\{(n_d-n_c)^++\max(n_d,n_c)\}\label{UBF_corr1}
\end{align}
\begin{align}
R&\leq 2 \left[(1-p)n_d+p\max\{(n_d-n_c)^+,n_c\}\right].\label{UBF_corr2}
\end{align}

\end{theorem}
\begin{IEEEproof}\label{Converse_GlobalCSI_corr}
The proof of \eqref{UBF_corr1} follows similar steps as in Appendix~\ref{Proof_UBF1}  but considering $\displaystyle{B_1^K=B_2^K=B^K}$. The proof of \eqref{UBF_corr2} is given in  Appendix~\ref{Proof_Global_corr}. See also Remark~\ref{Remark_spatial_corr_global}.
\end{IEEEproof}

\begin{theorem}[Achievability bounds for global CSIRT]\label{Thm:achiev-globalCSI_corr}
Assume that $B_1^K$ and $B_2^K$ are fully correlated. The following sum rates $R$ are achievable over the bursty IC:
\begin{align}
R&= 2 \left[(1-p)n_d+p\max\{(n_d-n_c)^+,n_c\}\right],\quad \text{VWI/WI}\label{Eq: LBF_corr1}
\end{align}
\begin{align}
R&=2(1-p)n_d+p\{(n_d-n_c)^++\max(n_d,n_c)\},\quad \text{MI/SI}.\label{Eq: LBF_corr2}
\end{align}

\end{theorem}

\begin{IEEEproof}
 The sum rates \eqref{Eq: LBF_corr1} and \eqref{Eq: LBF_corr2} are achieved by using the optimal scheme for the non-bursty IC when the two receivers are affected by interference~\cite{jafar10}, and by using uncoded transmission in absence of interference. 
\end{IEEEproof}
Table~\ref{Tab:global_corr1} summarizes the results of Theorems~\ref{Thm:converse-globalCSI_corr} and \ref{Thm:achiev-globalCSI_corr}. For global CSIRT and fully correlated $B_1^K$ and $B_2^K$, converse and achievability bounds coincide. Thus, \eqref{Eq: LBF_corr1} and \eqref{Eq: LBF_corr2} indicate the sum capacity.
\begin{table}[tbp]
\centering \scriptsize
\caption{Bounds on the sum capacity $C$ for global CSIRT when $B_1^K$ and $B_2^K$ are fully correlated.}
\label{Tab:global_corr1}
\begin{tabular}{cc}
\toprule
\textbf{Regions}&\textbf{Bounds}\\\midrule
\textbf{{VWI}}& {$2(n_d-pn_c)$}\\ \midrule
\textbf{WI}& {$2[(1-p)n_d+pn_c]$}\\ \midrule
\textbf{MI}& {$2(1-p)n_d+p(2n_d+n_c)$}\\ \midrule
\textbf{SI}& {$2(1-p)n_d+p(n_c)$}\\ \bottomrule
\end{tabular}
\end{table}

\subsection{Quasi-Static vs. Ergodic Setup}\label{CSIRT_QS_ER}
Similar to the average sum capacity for local CSIR defined in Section~\ref{CSIR_QS_ER}, we define the average sum capacity for global CSIRT when $B_1$ and $B_2$ are independent as
\begin{equation}\label{R_GlobalCSIRT_uncorr}
\begin{aligned}
\bar{C}=&p^2 \sup_{R}\{R\}+p(1-p)\sup_{(R,\Delta R(01))}\{R+{\Delta R}(01)\}+p(1-p)\sup_{(R,\Delta R(10))}\{R+{\Delta R}(10)\}\\
&{}+(1-p)^2\sup_{(R,\Delta R(00))}\{R+{\Delta R}(00)\}
\end{aligned}
\end{equation}
 where the suprema are over all rate tuples $(R,\Delta R(00), \Delta R(01),\Delta R(10))$ that satisfy Theorems \ref{Thm:conv_realiable_rates} and \ref{Thm:conv_BIC_global}. The intuition behind \eqref{R_GlobalCSIRT_uncorr} is the same as that behind \eqref{R_LocalCSIR} for local CSIR, but with global CSIRT the transmitters can adapt their rates $(R_i,\Delta R_i(V_i))$ to the interference state. For example, the first term on the right-hand side (RHS) of \eqref{R_GlobalCSIRT_uncorr} corresponds to the interference state $[1,1]$, in which case we transmit at total sum rate $R$; the second term corresponds to the interference state $[0,1]$, in which case we transmit at total sum rate $R+\Delta R(01)$; and so on.
  
Table~\ref{Tab:global_corr} summarizes the average sum capacity for the different interference regions. The average sum capacities for VWI and WI coincide with the sum capacities in the ergodic setup (see Table~\ref{Tab:global_uncorr}). In contrast, for MI and SI, the average sum capacities are smaller than the sum capacities in the ergodic~setup.

\begin{table}[tbp]
\centering \scriptsize
\caption{Average sum capacity when $B_1$ and $B_2$ are independent.}
\label{Tab:global_corr}
\begin{tabular}{cc}
\toprule
\textbf{Regions}&\textbf{Bounds}\\\midrule
\textbf{VWI}& {$2(n_d-pn_c)$}\\ \midrule
\textbf{WI}& {$2[(1-p^2)n_d+(1-2p)pn_c]$}\\ \midrule
\textbf{MI}& {$2n_d-pn_c(2-p)$}\\ \midrule
\textbf{SI}& {$2n_d(1-p)^2+pn_c(2-p)$}\\ \bottomrule
\end{tabular}
\end{table}

Similarly, in the fully correlated case, we define the average sum capacity as 
\begin{IEEEeqnarray}{lCl}\label{R_GlobalCSIRT_corr}
\bar{C} \eqdef p\sup_{R}\{R\}+(1-p)\sup_{(R,\Delta R(00))}\{(R+\Delta R(00))\}
\end{IEEEeqnarray}
 where  the suprema are over all rate pairs $(R,\Delta R(00))$ that satisfy Theorems \ref{Thm:conv_realiable_rates} and \ref{Thm:convBIC_global_corr}. The corresponding results are summarized in Table~\ref{Tab: QS_ER_Global_CSIRT_corr}.
\begin{table}[tbp]
\centering \scriptsize
\caption{Average sum capacity when $B_1$ and $B_2$ are fully correlated.}
\label{Tab: QS_ER_Global_CSIRT_corr}
\begin{tabular}{cc}
\toprule
\textbf{Regions}&\textbf{Bounds}\\  \midrule
\textbf{VWI}& {$2(n_d-pn_c)$}\\ \midrule
\textbf{WI}& {$2[(1-p)n_d+pn_c]$}\\ \midrule
\textbf{MI}& {$2(1-p)n_d+p(2n_d+n_c)$}\\ \midrule
\textbf{SI}& {$2(1-p)n_d+p(n_c)$}\\ \bottomrule
\end{tabular}
\end{table}

We observe that the average sum capacities coincide with the sum capacities of the ergodic setup. 

\section{Exploiting CSI}\label{Sec: Exploiting_CSI}
In this section, we study how the level of CSI affects the sum rate in the quasi-static and ergodic~setups.

For the quasi-static channel, Figures~\ref{Fig: Comparison_B=00_QS} and \ref{Fig: Comparison_B=01_QS} show the total sum capacity presented in Theorems~\ref{Thm:convBIC_local},~\ref{Thm:conv_BIC_global} and \ref{Thm:convBIC_global_corr}. Specifically, we plot the normalized total sum capacity $\tfrac{C+\Delta C}{n_d}$ versus $\alpha$, comparing scenarios of local CSIR/CSIRT and global CSIRT. We analyze separately the cases $\vect{B}=[0,0]$ and $\vect{B}=[0,1]$. For the case where $\vect{B}=[0,0]$ and global CSIRT, the total sum capacity is $2 n_d$ for all interference regions. For $\vect{B}=[0,0]$ and local CSIR/CSIRT, the total sum capacity is $2 n_d$ for VWI and VSI, but is strictly smaller in the remaining interference regions. Hence, in these regions global CSIRT outperforms local CSIR/CSIRT. For the case where $\vect{B}=[0,1]$, the total sum capacity is equal to $(n_d-n_c)^++\max(n_d,n_c)$ irrespective of the level of CSI.

\begin{figure}[tbp]
		\centering
		\includegraphics[width=0.47\linewidth]{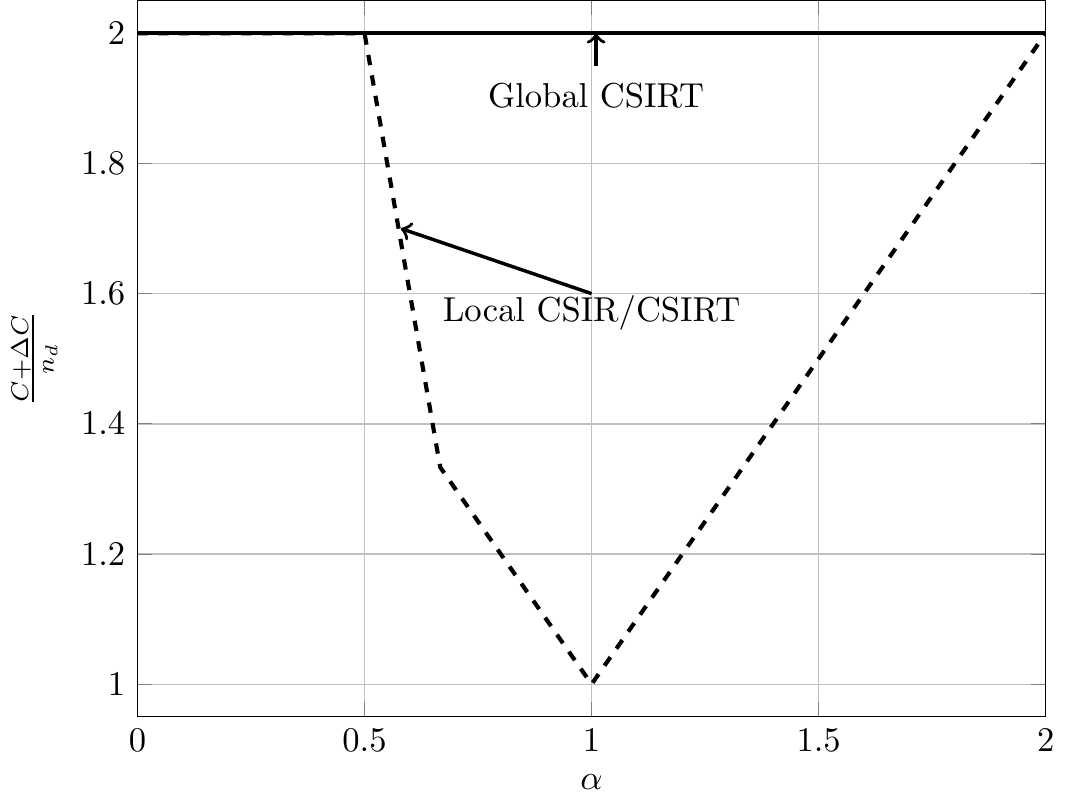}
		\caption{Total sum capacity for $\vect{B}=[0,0]$, for local CSIR/CSIRT and global CSIRT.}
		\label{Fig: Comparison_B=00_QS}
	   \end{figure}
	   \unskip
	   \begin{figure}[tbp]
    	\centering
			\includegraphics[width=0.47\linewidth]{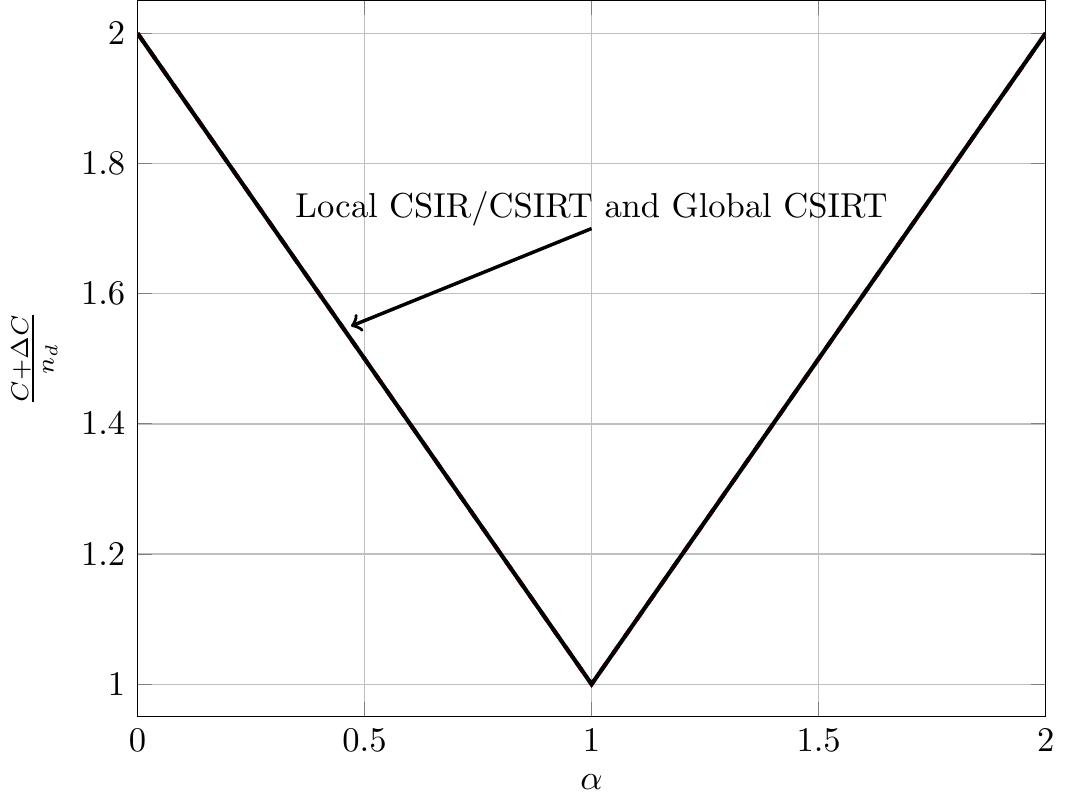}
		\caption{Total sum capacity for $\vect{B}=[0,1]$, for local CSIR/CSIRT and global CSIRT.}
		\label{Fig: Comparison_B=01_QS}
\end{figure}

We further observe that the opportunistic-capacity region for local CSIRT is equal to that for local CSIR. Thus, local CSI at the transmitter is not beneficial. As we shall see later, this is in stark contrast to the ergodic setup, where local CSI at the transmitter-side is beneficial. Intuitively, in the ergodic case the input distributions of $\mat {X}_1^K$ and $\mat {X}_2^K$ depend on the realizations of ${B}_1^K$ and ${B}_2^K$, respectively. Hence, adapting the input distributions to these realizations increases the sum capacity. In contrast, in the quasi-static case, the worst-case scenario (presence of interference) and the best-case scenario (absence of interference) are treated separately. Hence, there is no difference to the case of local CSIR.

For the ergodic setup, Figures~\ref{Fig: VWI}--\ref{Fig: SI} show the converse and achievability bounds presented in Theorems~\ref{Thm:converse-indCSI}, \ref{Thm:achiev-noCSI}, \ref{Thm:converse-fullCSI} and \ref{Thm:achiev-fullCSI}. We further include the results on local CSIRT presented in Section~\ref{Sec: local_CSIRT}.
Specifically, we plot the normalized sum capacity $\tfrac{C}{n_d}$ versus the probability of presence of interference $p$, comparing scenarios of local CSIR, local CSIRT and global CSIRT  when $B_1^K$ and $B_2^K$ are independent of each other. The shadowed areas correspond to the regions where achievability and converse bounds do not~coincide. 

\begin{figure}[tbp]
		\centering
        \includegraphics[width=0.45\linewidth]{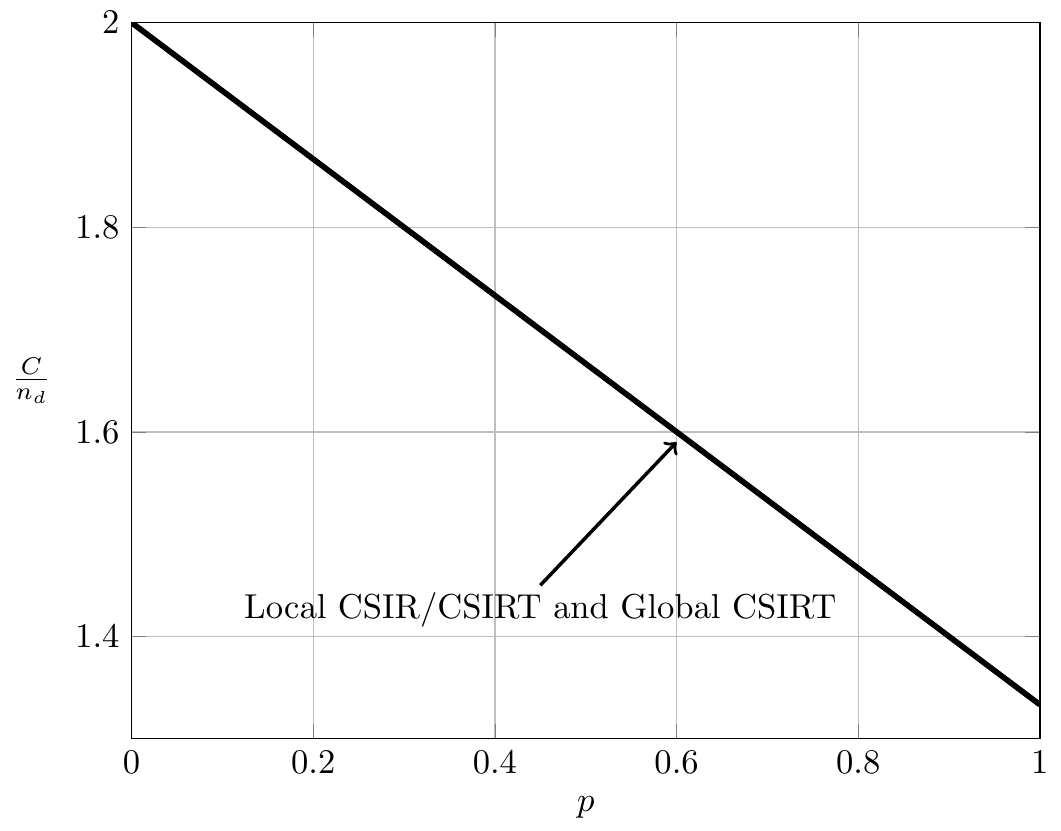}
		\caption{Sum capacity  for local CSIR/CSIRT and global CSIRT when $B_1^K$ and $B_2^K$ are independent and $\alpha=\frac{1}{3}$ (VWI).}
		\label{Fig: VWI}
		\end{figure}
		\unskip
	   \begin{figure}[tbp]
    	\centering
		\includegraphics[width=0.45\linewidth]{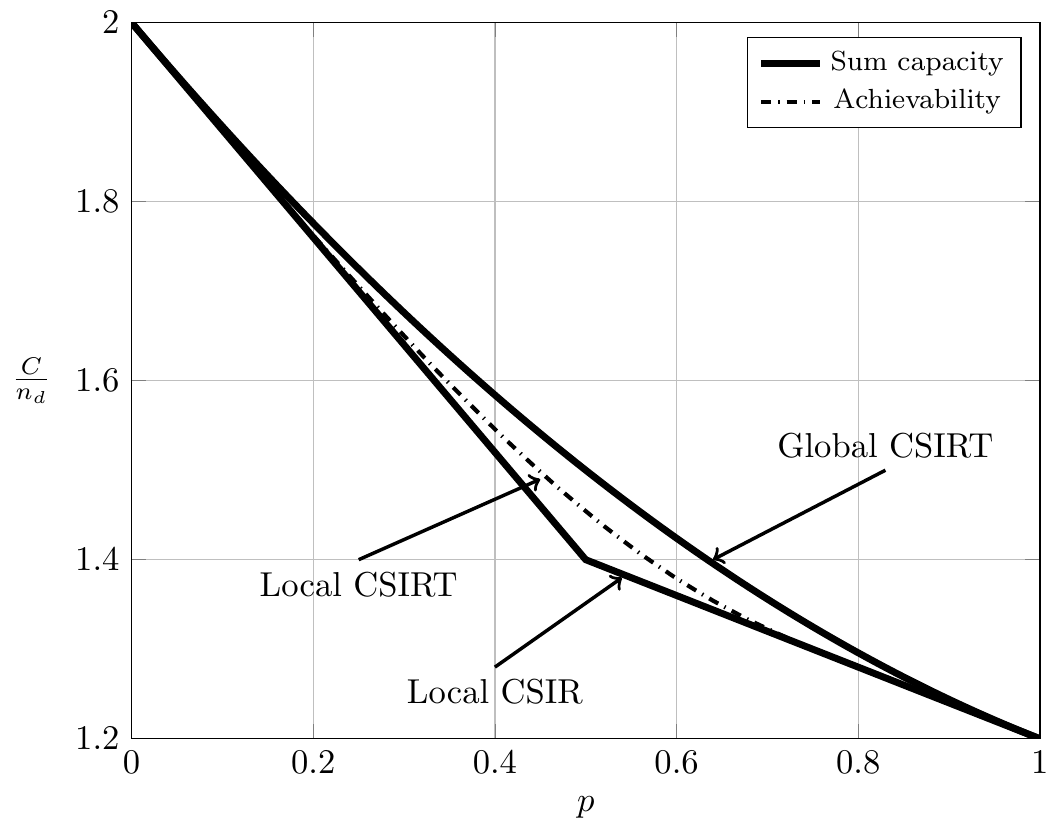}
		\caption{Sum capacity  for local CSIR/CSIRT and global CSIRT when $B_1^K$ and $B_2^K$ are independent and $\alpha=\frac{3}{5}$ (WI).}
		\label{Fig: WI}
\end{figure}
\unskip
\begin{figure}[tbp]
		\centering
		\includegraphics[width=0.45\linewidth]{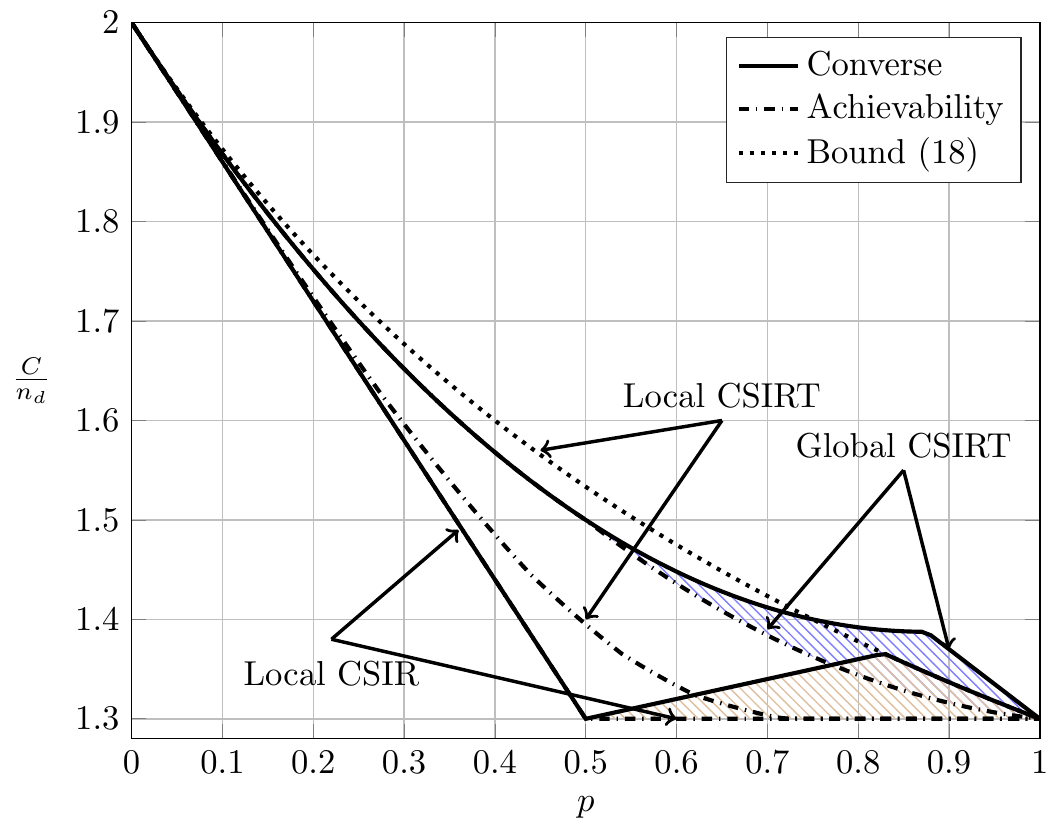}
		\caption{Sum capacity  for local CSIR/CSIRT and global CSIRT when $B_1^K$ and $B_2^K$ are independent and $\alpha=\frac{7}{10}$ (MI).}
		\label{Fig: MI}
		\end{figure}
		\unskip
		\begin{figure}[tbp]
	\centering
	       \includegraphics[width=0.45\linewidth]{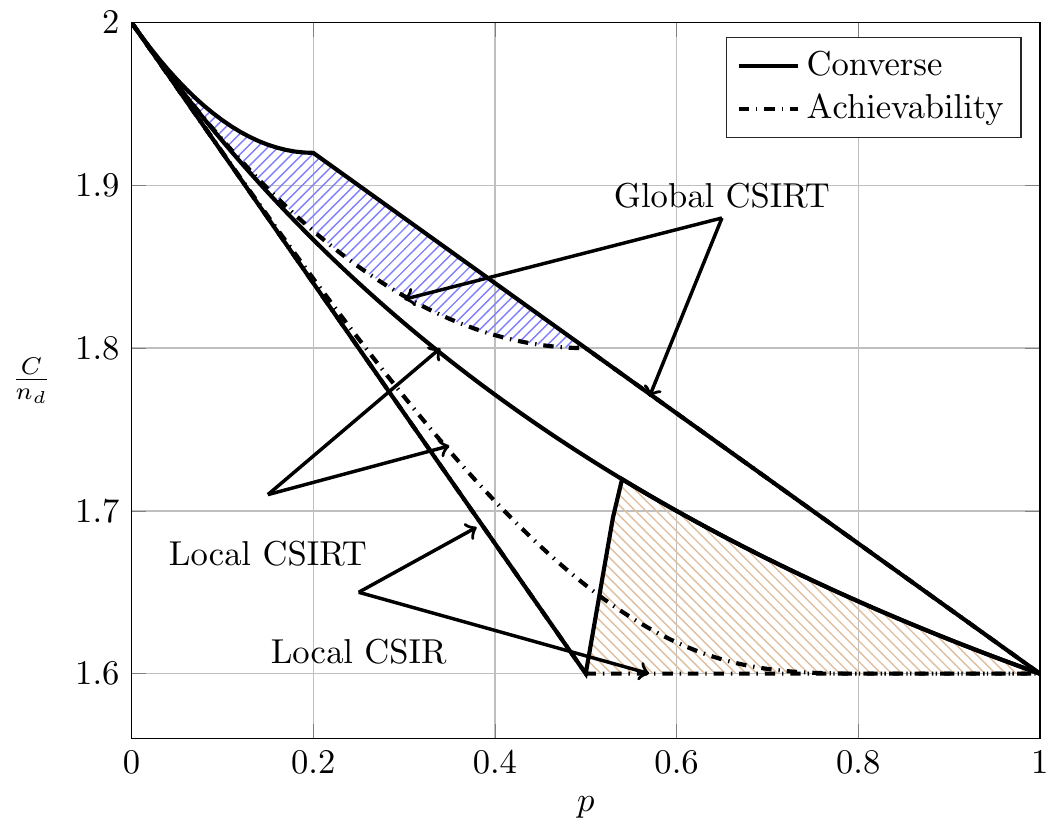}
			\caption{Sum capacity  for local CSIR/CSIRT and global CSIRT when $B_1^K$ and $B_2^K$ are independent and $\alpha=\frac{8}{5}$ (SI).}
			\label{Fig: SI}
\end{figure}

Figure~\ref{Fig: VWI} reveals that in the VWI region the sum capacity is equal to $2(n_d-p n_c)$, irrespective of the availability of CSI (see Figure \ref{Fig: VWI}). Thus, in this region access to global CSIRT is not beneficial compared to the local CSIR scenario. In the VSI region, the sum capacity of the non-bursty IC is equal to $2n_d$, which is that of two parallel channels without interference \cite[Sec.~II-A]{Avestimehr11}. Therefore, burstiness of the interference (and hence CSI) does not affect the sum capacity.
 
In the WI region, shown in Figure~\ref{Fig: WI}, the converse and achievability bounds for local CSIR and global CSIRT coincide and it is apparent that global CSIRT outperforms local CSIR.
In the MI and SI regions, the converse and achievability bounds only coincide for certain regions of $p$. Nevertheless, Figures~\ref{Fig: MI} and~\ref{Fig: SI} show that, in almost all cases, global CSIRT outperforms local CSIR. (For the case presented in Figure~\ref{Fig: MI} $\left(\alpha=\tfrac{7}{10}\right)$, we also present the local CSIRT converse bound \eqref{UB_TK}, although it is looser 
for some values of $p$, with respect to the one depicted for global CSIRT.) Local CSIRT outperforms local CSIR in all interference regions (except VWI). We stress again the fact that this was not the case in the quasi-static scenario, where both coincide.

We next consider the case where $B_1^K$ and $B_2^K$ are fully correlated. For this scenario, \cite{Wang13,Wang13_pre} studied the effect of perfect feedback on the bursty IC. For comparison, the non-bursty IC with feedback was studied by Suh~\etal in \cite{Suh11}, where it was demonstrated that the gain of feedback  becomes arbitrarily large for certain interference regions (VWI and WI) when the signal-to-noise-ratio increases. This gain corresponds to a better resource utilization and thereby a better resource sharing between users. Specifically, \cite{Wang13, Wang13_pre} (bursty IC) and \cite{Suh11} (non-bursty IC) assume that noiseless, delayed feedback is available from receiver $i$ to transmitter $i$ $(i=1,2)$. For the symmetric setup treated in this paper,~\cite[Th.~3.2]{Wang13} or \cite[Th.~3.2]{Wang13_pre} showed the following: 

\begin{theorem}[Channel capacity for the bursty IC with feedback \cite{Wang13, Wang13_pre}]\label{Thm:Feedback} 
 The sum capacity of the bursty IC with noiseless, delayed feedback is given by
 
\begin{equation}\label{Cap_Feedback}
C =
\begin{cases} 
      2n_d-2\frac{p}{1+p}n_c, & \alpha\leq 1, \\
      2\frac{1-p}{1+p}n_d+2\frac{p}{1+p}n_c& 1 < \alpha\leq 2, \\
      2(1-p)n_d+p n_c,& 2 < \alpha.\\
\end{cases}
\end{equation}

\end{theorem}
\begin{IEEEproof}
See \cite[Sec.~IV and V]{Wang13}, \cite[Sec.~IV and V, Appendices~A, C, D]{Wang13_pre}.
\end{IEEEproof}
Observe that \eqref{Cap_Feedback}  for $\alpha\leq 2$ coincides with \eqref{UB_TK}. This implies that local CSIRT can never outperform delayed feedback. Intuitively, feedback contains not only information about the channel state, but also about the previous symbols transmitted by the other transmitter, which can be exploited to establish a certain cooperation between the transmitters.
Figures~\ref{Fig: VWI_corr}--\ref{Fig: SI_corr} show the bounds on the normalized sum capacity, $\tfrac{C}{n_d}$, comparing the scenarios of local CSIR versus global CSIRT when the interference states are fully correlated, i.e., $B_1^K=B_2^K$. They further show the sum capacity for the case where the transmitters have noiseless  delayed feedback \cite{Wang13}. 
The shadowed areas correspond to the regions where achievability and converse bounds do not coincide.

\begin{figure}[tbp]
		\centering
		\includegraphics[width=0.5\linewidth]{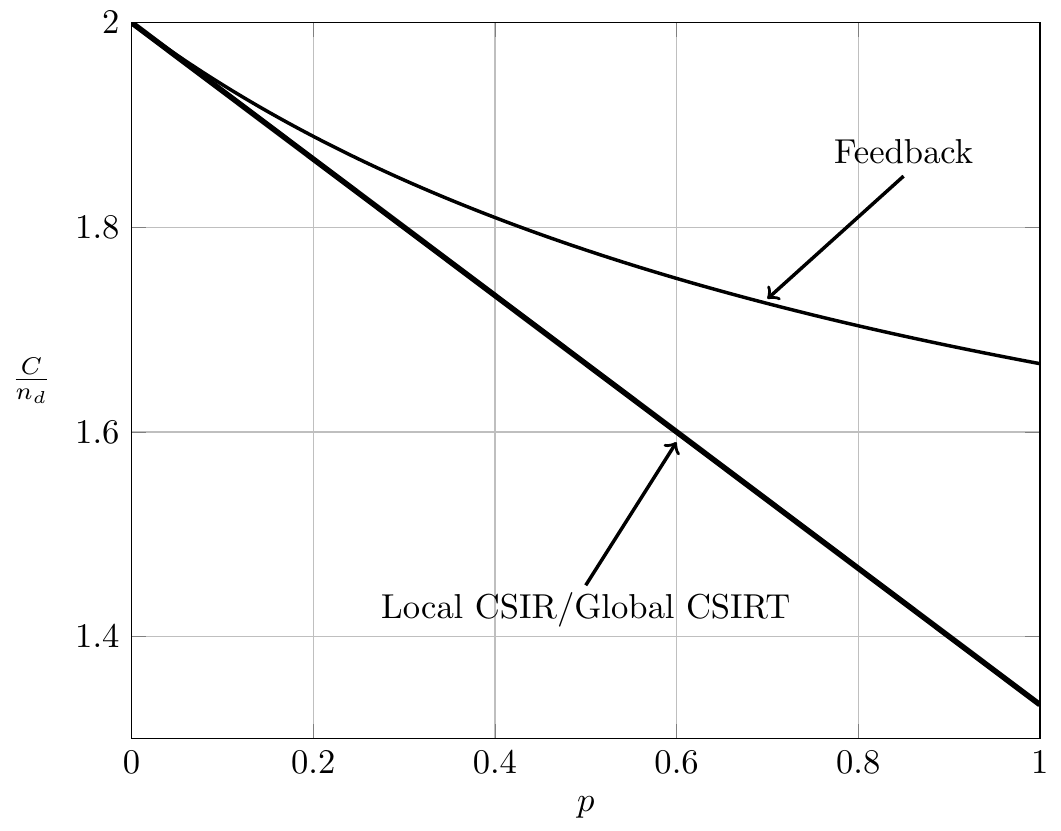}
		\caption{Sum capacity for local CSIR and  global CSIRT when $B_1^K=B_2^K$ and $\alpha=\frac{1}{3}$ (VWI).}
		\label{Fig: VWI_corr}
		\end{figure}
		\unskip
		\begin{figure}[tbp]
    	\centering
			\includegraphics[width=0.5\linewidth]{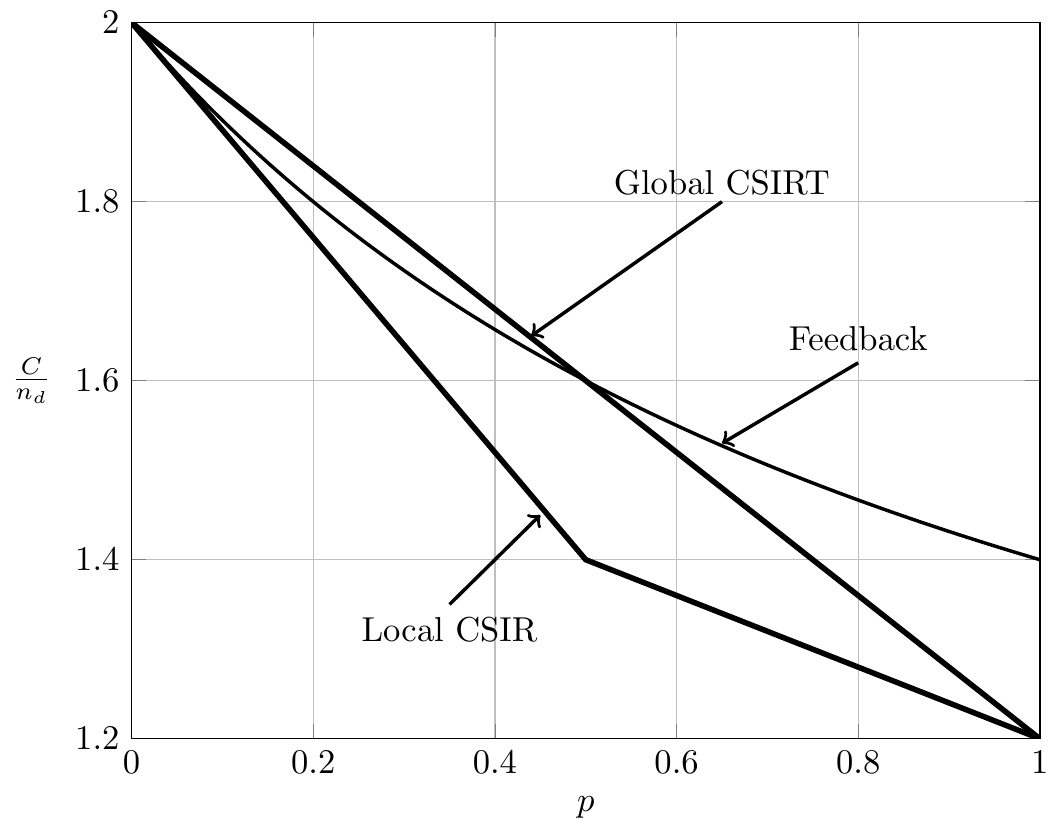}
		\caption{Sum capacity  for local CSIR and global CSIRT when $B_1^K=B_2^K$ and $\alpha=\frac{3}{5}$ (WI).}
		\label{Fig: WI_corr}
\end{figure}
\unskip
\begin{figure}[tbp]
		\centering
			\includegraphics[width=0.5\linewidth]{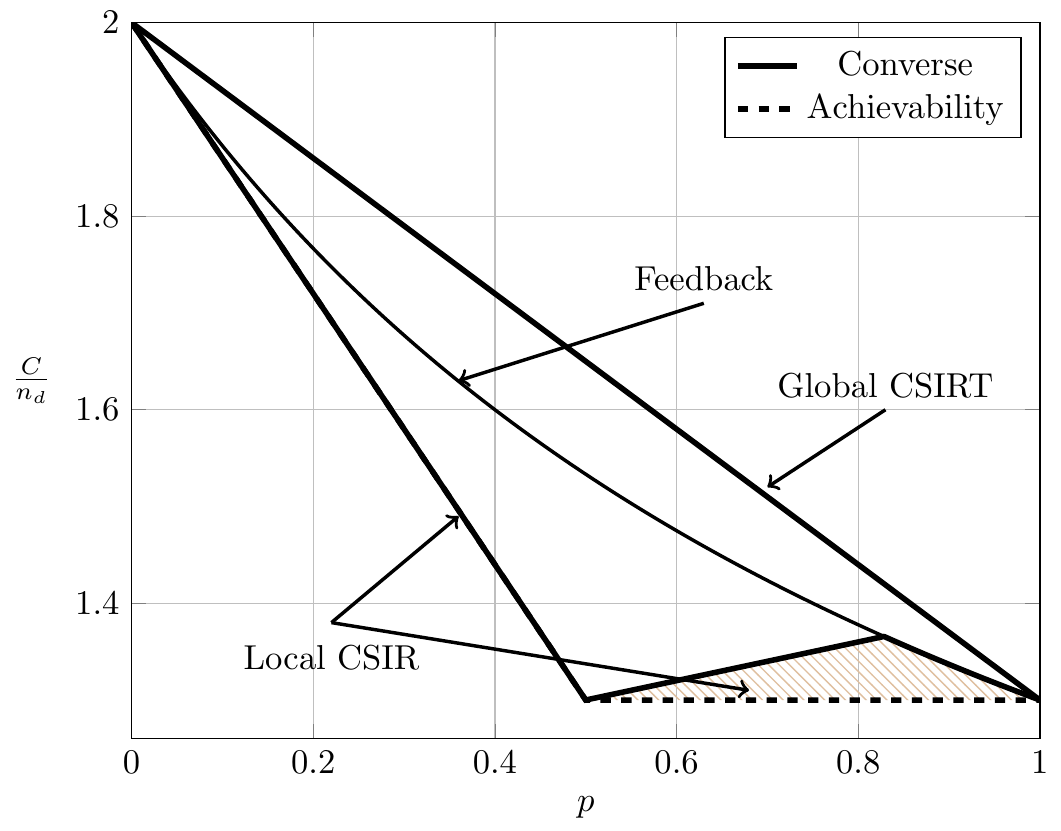}
		\caption{Sum capacity  for local CSIR and  global CSIRT when $B_1^K=B_2^K$ and $\alpha=\frac{7}{10}$ (MI).}
		\label{Fig: MI_corr}
		\end{figure}
		\unskip
		\begin{figure}[tbp]
	\centering
			\includegraphics[width=0.5\linewidth]{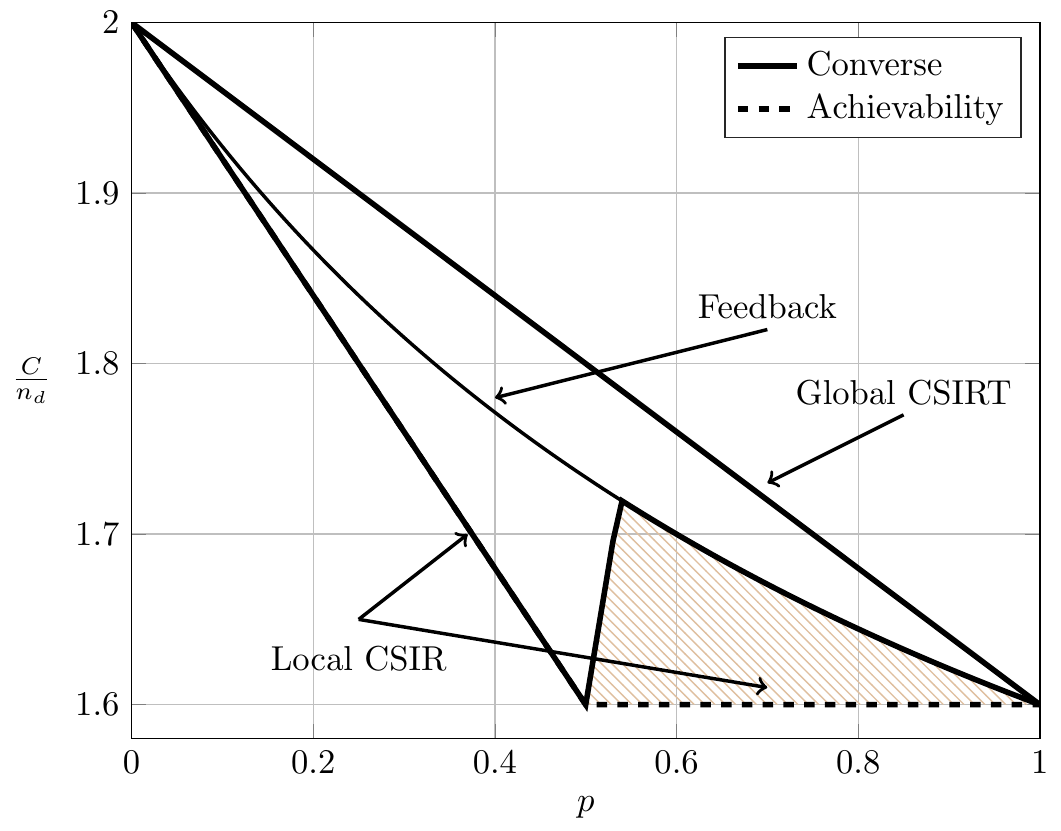}
			\caption{Sum capacity  for local CSIR and  global CSIRT when $B_1^K=B_2^K$ and $\alpha=\frac{8}{5}$ (SI).}
			\label{Fig: SI_corr}
\end{figure}

Figure~\ref{Fig: VWI_corr} reveals that feedback in the VWI region outperforms the non-feedback case, irrespective of the availability of CSI.  Wang \etal \cite{Wang13} have further shown that feedback also outperforms the non-feedback case in the VSI region. 
The order between global CSIRT and the feedback scheme is not obvious. There are regions where global CSIRT outperforms the feedback scheme and vice versa. Indeed, on the one hand, feedback contains information about the previous interference states and previous symbols transmitted by the other transmitter, permitting the resolution of collisions in previous transmissions. On the other hand, global CSIRT provides \emph{non-causal} information about the interference states, allowing a better adaptation of the transmission strategy to the interference burstiness.

\section{Exploiting Interference Burstiness} \label{Sec: Exploiting_Burstiness}
To better illustrate the benefits of interference burstiness, we show the normalized sum capacity as a function of $\alpha$, in order to appreciate all the interference regions. In the non-bursty IC ($p=1$), this curve corresponds to the well-known W-curve obtained by Etkin \etal in \cite{Etkin}. We next study how burstiness affects this curve in the different considered scenarios.

In the quasi-static setup, burstiness can be exploited by sending opportunistic messages. We consider the total sum capacity for the case where the worst-case rate R is maximized. For local CSIR/CSIRT, Theorem~\ref{Thm:convBIC_local} suggests that the use of an opportunistic code is only beneficial if the interference region is VWI or WI. For other interference regions there is no benefit. In contrast, for global CSIRT an opportunistic code is beneficial for all interference regions (except for VSI where the sum capacity corresponds to that of two parallel channels without interference). 

Figures~\ref{Fig:QS_uncorr_local} and \ref{Fig:QS_uncorr_global} illustrate these observations. Specifically, in Figures~\ref{Fig:QS_uncorr_local} and \ref{Fig:QS_uncorr_global} we show the normalized total sum capacity achieved under local CSIR/CSIRT and global CSIRT when the interference states are independent. We observe that, for local CSIR, the opportunistic rates $\Delta R_1(0)$ and $\Delta R_2(0)$,  are only positive in the VWI and WI regions.  In these regions, if only one of the receivers is affected by interference the sum capacity is given by the worst-case rate $R$ plus one opportunistic rate of the user which is not affected by interference. In absence of interference at both receivers, both receivers can decode opportunistic messages. Hence, the total sum capacity is equal to $C+\Delta C_1(0)+\Delta C_2(0)$. For global CSIRT we can observe that, when only one of the receivers is affected by interference, we achieve the same total sum capacity as in the local CSIR/CSIRT. However, in absence of interference at both receivers, we achieve the trivial upper bound corresponding to two parallel channels. The fully correlated scenario can be considered as a subset of the independent scenario. Indeed, for the case $\vect {B}=[0,0]$ and $\vect {B}=[1,1]$ we obtain the same total sum capacity as for the independent scenario. The main difference is that in the fully correlated scenario the interference states $\vect{B}=[0,1]$ and $\vect{B}=[1,0]$ are impossible.  

\begin{figure}[tbp]
		\centering
		\includegraphics[width=0.5\linewidth]{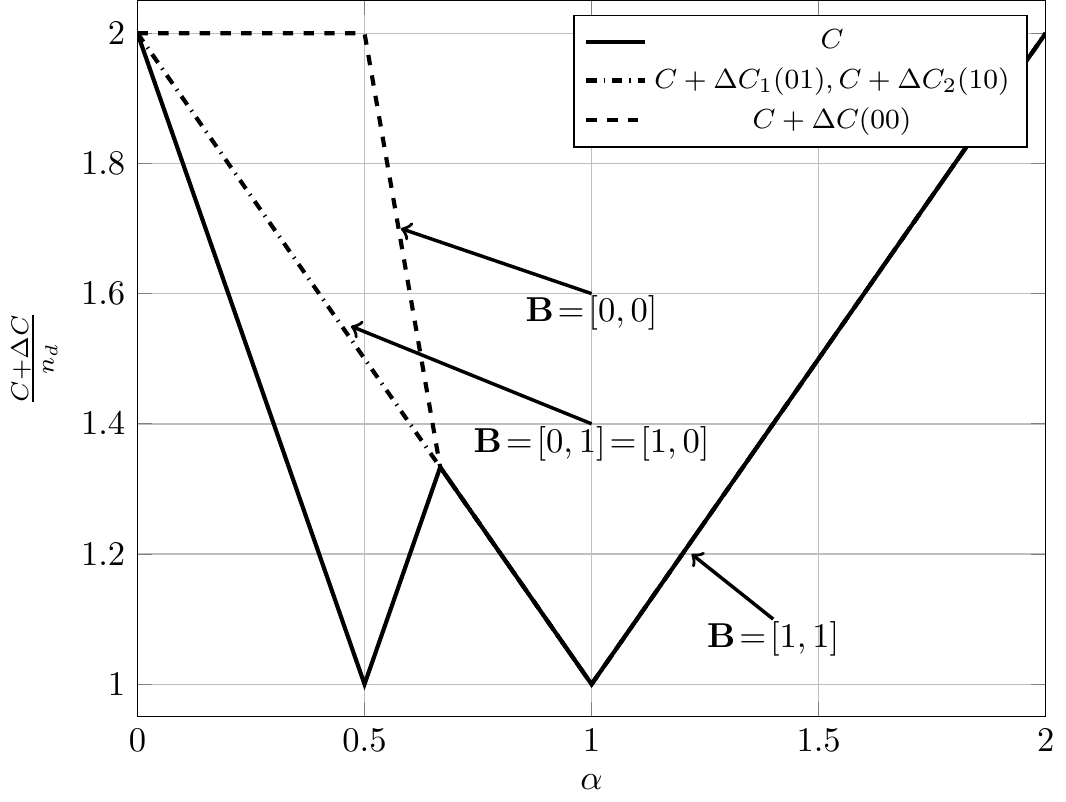}
		\caption{Normalized total sum capacity $\tfrac{C+\Delta C}{n_d}$ as a function of $\alpha$ for local CSIR/CSIRT when $B_1$ and $B_2$ are independent.}
		\label{Fig:QS_uncorr_local}
		\end{figure}
		\unskip
\begin{figure}[tbp]
		\centering
		\includegraphics[width=0.5\linewidth]{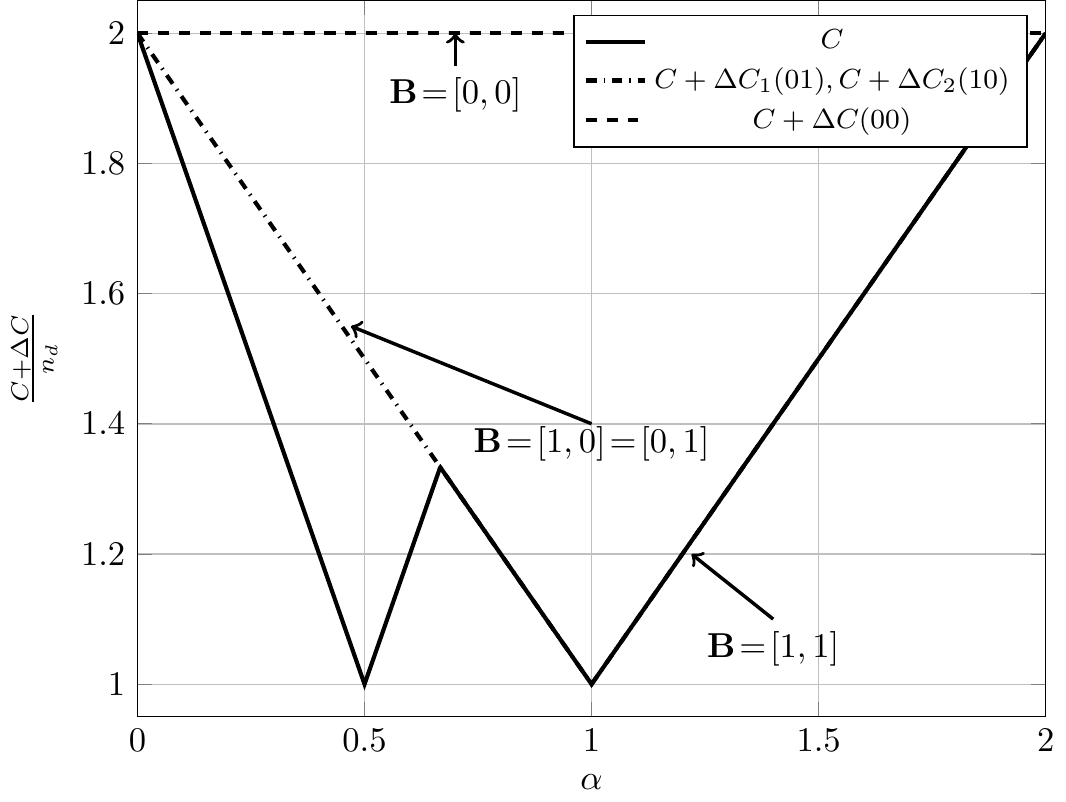}
		\caption{Normalized total sum capacity $\tfrac{C+\Delta C}{n_d}$ as a function of $\alpha$ for global CSIRT when $B_1$ and $B_2$ are independent.}
		\label{Fig:QS_uncorr_global}
\end{figure}

For the ergodic case, Figures~\ref{Fig: W_local_uncorr} and \ref{Fig: W_global_uncorr}  show the bounds on the normalized sum capacity, $\tfrac{C}{n_d}$, as a function of $\alpha$ when $B_1^K$ and $B_2^K$ are  independent.  The shadowed areas correspond to the regions where achievability and converse bounds do not coincide. We further show the W-curve. Observe that for $p\leq \tfrac{1}{2}$ the sum capacity as a function of $\alpha$ forms a V-curve instead of the W-curve. Further observe how the sum capacity approaches the W-curve as $p$ tends to one.

\begin{figure}[tbp]
		\centering
		\includegraphics[width=0.5\linewidth]{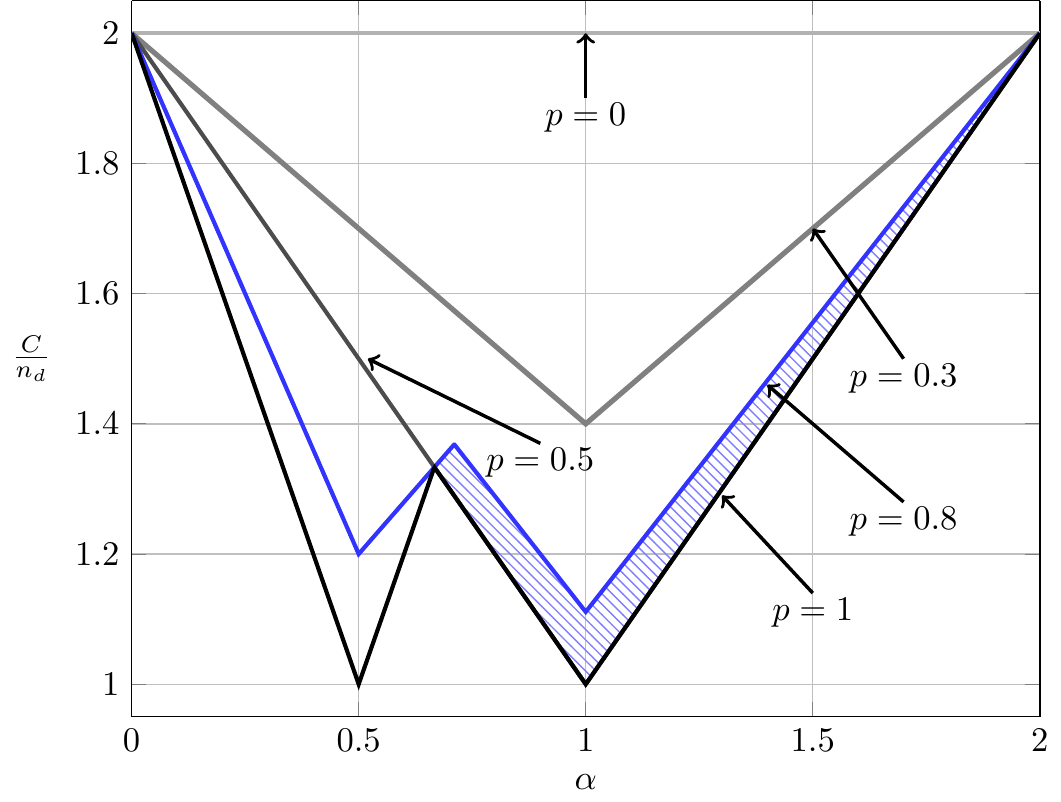}
		\caption{Normalized sum capacity $\tfrac{C}{n_d}$ as a function of $\alpha$ for local CSIR/CSIRT when $B_1^K$ and $B_2^K$ are~independent.}
		\label{Fig: W_local_uncorr}
\end{figure}
\unskip
\begin{figure}[tbp]
		\centering
		\includegraphics[width=0.5\linewidth]{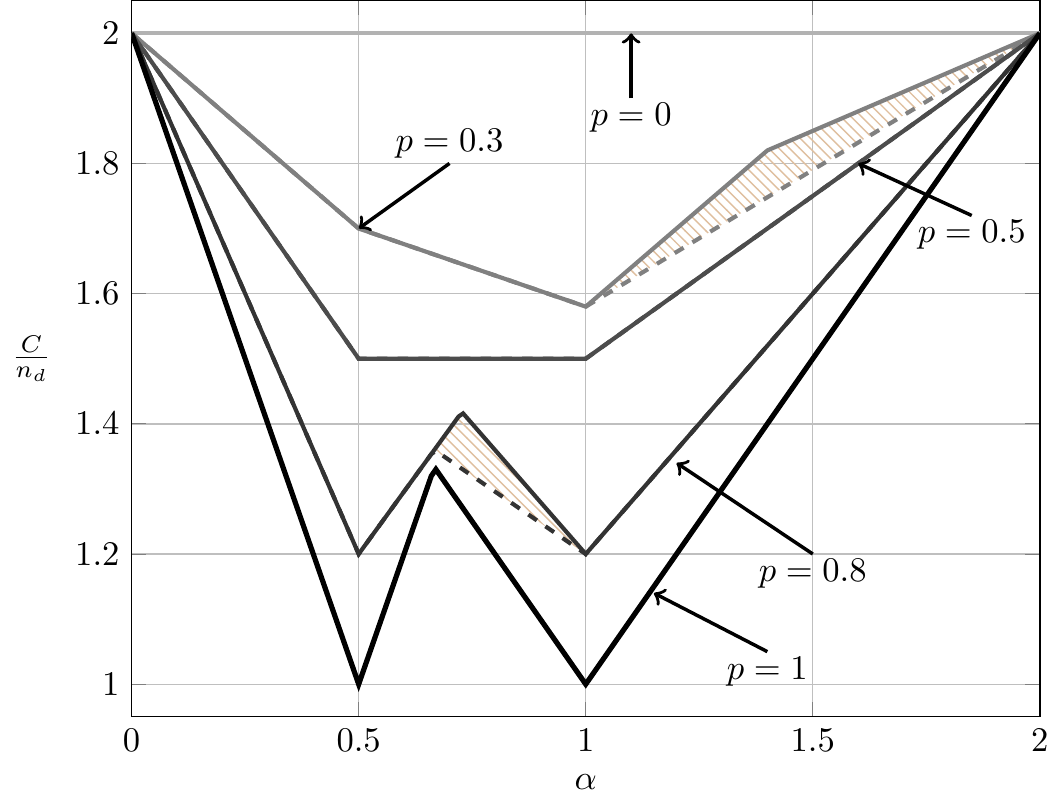}
		\caption{Normalized sum capacity $\tfrac{C}{n_d}$ as a function of $\alpha$ for global CSIRT when $B_1^K$ and $B_2^K$ are~independent.}
		\label{Fig: W_global_uncorr}
\end{figure}

In Figure~\ref{Fig: W_global_corr} we show the bounds on the normalized sum capacity, $\tfrac{C}{n_d}$, as a function of $\alpha$ for global CSIRT when $B_1^K$ and $B_2^K$ are fully correlated. (For local CSIR  the sum capacity is not affected by the correlation between $B_1^K$ and $B_2^K$, so the curve for $\tfrac{R}{n_d}$ as a function of $\alpha$  coincides with the one obtained in Figure~\ref{Fig: W_local_uncorr}.) We observe that, for all values of $p>0$, the sum capacity forms a W-curve similar to the W-curve for $p=1$. This is the case because, when both interference states are fully correlated, the bursty IC is a combination of an IC and two parallel channels.

We observe that for global CSIRT the burstiness of the interference is beneficial for all interference regions and all values of $p$. For local CSIR, burstiness is beneficial for all values of $p$ for VWI and WI. However, for MI and SI, burstiness is only of clear benefit for $p\leq\tfrac{1}{2}$. It is yet unclear whether burstiness is also beneficial in these interference regions when $p>\tfrac{1}{2}$. To shed some light on this question, note that evaluating the converse bound in \cite[Lemma~A.1]{Wang13_pre}, which yields \eqref{Eq: C_MI}, for inputs $\mat {X}_1^K$ and $\mat {X}_2^K$ that are temporally independent, we recover the achievability bound \eqref{Ach. CSIR2}. Since for MI/SI and $p\geq\tfrac{1}{2}$ this bound coincides with the rates achievable over the non-bursty IC, this implies that an achievability scheme can only exploit the burstiness of the interference in this regime if it introduces some temporal correlation (this observation is also revealed by considering the average sum capacity for the quasi-static case). In fact, for global CSIRT the achievability schemes proposed in Theorem~\ref{Thm:achiev-fullCSI} for MI and SI copy the same bits over several coherence blocks, i.e., they exhibit a temporal correlation, which cannot be achieved using temporally independent distributions. However, the temporal pattern of these bits requires knowledge of both interference states, so this approach cannot be adapted to the cases of local CSIR/CSIRT. 
In contrast, for global CSIRT in the fully correlated case where converse and achievability bounds coincide, it is not necessary to introduce temporal memory. This scenario is simpler, since in this case the channel exhibits only two channel states, a non-bursty IC and two parallel channels.
\begin{figure}[tbp]
\centering
\includegraphics[width=0.5\linewidth]{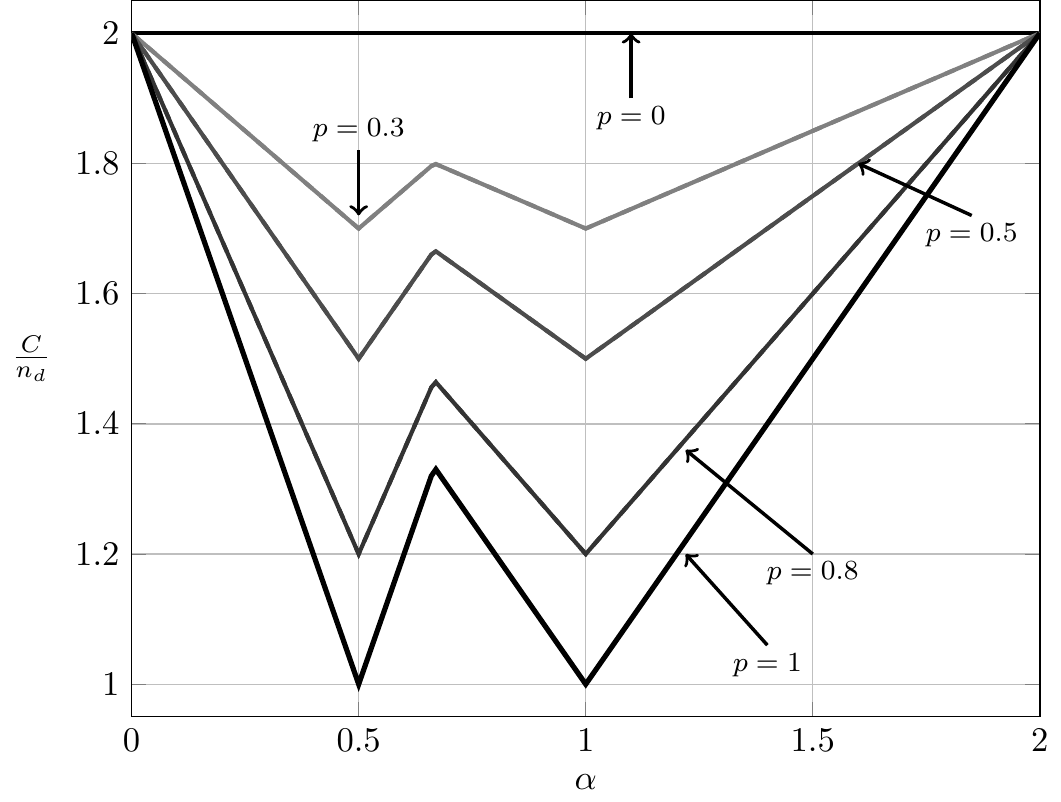}
\caption{Normalized sum capacity $\tfrac{C}{n_d}$  as a function of $\alpha$ for global CSIRT when $B_1^K=B_2^K$. }
\label{Fig: W_global_corr}
\end{figure}

\section{Summary and Conclusions}\label{Sec: Summary}

In this work, we considered a two-user bursty IC in which the presence/absence of interference is modeled by a block-i.i.d.\ Bernoulli process while the power of the direct and cross links remains constant during the whole transmission. This scenario corresponds, e.g., to a slow-fading scenario in which all the nodes can track the channel gains of the different links, but where the interfering links are affected by intermittent occlusions due to some physical process. While this model may appear over-simplified, it yields a unified treatment of several aspects previously studied in the literature and gives rise to several new results on the effect of the CSI in the achievable rates over the bursty IC. Our channel model encompasses both the quasi-static scenario studied in \cite{Diggavi06,Khude09} and the ergodic scenario (see, e.g.,  \cite{Wang13,Vahid14_dCSIT}).  While the model recovers several cases studied in the literature, it also presents scenarios which have not been previously analyzed. This is the case, for example, for the ergodic setup with local and global CSIRT. Our analysis in these scenarios does not yield matching upper and lower bounds for all interference and burstiness levels. Yet, examining the obtained results, we observe that the best strategies in these scenarios often require elaborated coding strategies for both users that feature memory across different interference. This fact probably explains why no previous results exist in these scenarios. Furthermore, several of our proposed achievability schemes require complex correlation among signal levels. Thus, while the LDM in general provides insights on the Gaussian IC, the proposed schemes may actually be difficult to convert to the Gaussian case.  

In the quasi-static scenario, the highest sum rate $R$ that can be achieved is limited by the worst realization of the channel and thus coincides with that of the (non-bursty) IC. We can however transmit at an increased (opportunistic) sum rate $R+\Delta R$ when there is no interference at any of the interfering links. For the ergodic setup, we showed that an increased rate can be obtained when local CSI is present at both transmitter and receiver, compared to that obtained when CSI is only available at the receiver side. This is in contrast to the quasi-static scenario, where the achievable rates for local CSIR and local CSIRT coincide. Featuring global CSIRT at all nodes yields an increased sum rate for both the quasi-static and the ergodic scenarios. In the quasi-static channel, global CSI yields increased opportunistic rates in all the regions except in the very strong interference region, which is equivalent to having two parallel channels with no interference.

Both in the quasi-static and ergodic scenarios, global CSI exploits interference burstiness for all interference regions (except for very strong interference), irrespective of the level of burstiness. When local CSI is available only at the receiver side, interference burstiness is of clear benefit if the interference is either weak or very weak, or if the channel is ergodic and interference is present at most half of the time. When local CSI is available at each transmitter and receiver and the channel is ergodic, interference burstiness is beneficial in all interference regions except in the very weak and very strong interference regions.

In order to compare the achievable rates of the quasi-static and ergodic setup, one can define the average sum rate of the quasi-static setup for local CSIR/CSIRT as $R+(1-p)(\Delta R_1(0) + \Delta R_2(0))$, with a similar definition for the average sum rate for global CSIRT. The average sum rate corresponds to a scenario where several codewords are transmitted over independent quasi-static bursty ICs. This, in turn, could be the case if a codeword spans several coherence blocks, but no coding is performed over these blocks. This is in contrast to the ergodic setup where coding is typically performed over different coherence blocks. By the law of large numbers, roughly a fraction of $p$ codewords experiences interference, the remaining codewords are transmitted free of interference. Consequently, an opportunistic transmission strategy achieves the rate $\displaystyle{p R + (1-p) (R+\Delta R_1(0) + \Delta R_2(0))}$, which corresponds to the average sum rate. Our results demonstrate that, for local CSIR, the average sum capacity, obtained by maximizing the average sum rate over all achievable rate pairs $\displaystyle{(R,\Delta R_1(0) + \Delta R_2(0))}$, coincides with the achievable rates in the ergodic setup for all interference regions. In contrast, for local CSIRT, the average sum capacity is strictly smaller than the sum capacity in the ergodic setup. For global CSIRT, average sum capacity and sum capacity coincide for all interference regions when the interference states are fully correlated, and they coincide for VWI and WI when the interference states are independent. For global CSIRT, MI/SI, and independent interference states, the average sum capacity is smaller than the sum capacity in the ergodic setup. In general, the average sum capacity defined for the quasi-static setup never exceeds the sum capacity in the ergodic setup. This is perhaps not surprising if we recall that the average sum capacity corresponds to the case where no coding is performed over coherence blocks. Interestingly, the average sum capacity is not always achieved by maximizing the worst-case rate. For small values of $p$, it is beneficial to reduce the worst-case rate in order to achieve a larger opportunistic rate.

In our work we considered both the case where the interference states of the two users are independent and the case where the interference states are fully correlated. In both ergodic and quasi-static setups, the results for local CSIR are independent of the correlation between interference states. For other CSI levels, dependence between the interference states helps in all interference regions except very weak and very strong interference regions.

\vspace{6pt} 



\section*{Acknowledgment}
Fruitful discussions with S.~Gherekhloo are gratefully acknowledged. We further thank the anonymous reviewers for their
insightful comments and suggestions.



\appendices
\section{Proofs for the Quasi-Static Case}\label{Sec: Proof QS}

\makeatletter 
\setcounter{figure}{0} 
\@addtoreset{figure}{section}
\renewcommand{\thefigure}{A\arabic{figure}}
\makeatletter

\makeatletter 
\setcounter{table}{0} 
\@addtoreset{table}{section}
\renewcommand{\thetable}{A\arabic{table}}
\makeatletter

\makeatletter 
\setcounter{equation}{0} 
\@addtoreset{equation}{section}
\renewcommand{\theequation}{A\arabic{equation}}
\makeatletter

We define $p_{\vect{b}}=\Pr\{\vect {B}=\vect{b}\}$. Clearly, when $B_1,B_2$ are independent, we have $p_{00}=(1-p)^2$, $p_{11}=p^2$ and $p_{01}=p_{10}=p(1-p)$, and when  $B_1,B_2$ are fully correlated $p_{00}=1-p$, $p_{11}=p$ and $p_{01}=p_{10}=0$.

The converse bounds in the quasi-static case are based on an information density approach \cite{verdu-han}. In particular, we define the information densities for the bursty IC
\begin{eqnarray}\label{i1(b1)}
\infr_1(\vect{x}_1^N,\vect{y}_1^N,\vect{b})\eqdef \infr_{\vect{X}_1^N\vect {Y}_1^N|\vect{B}}(\vect{x}_1^N;\vect {y}_1^N|\vect{b})=\log\frac{P_{\vect {Y}_1^N|\vect{X}_1^N,\vect{B}}(\vect {y}_1^N|\vect {x}_1^N,\vect{b})}{P_{\vect {Y}_1^N|\vect{B}}(\vect {y}_1^N|\vect{b})}
\end{eqnarray} 
\begin{eqnarray}\label{i2(b2)}
\infr_2(\vect{x}_2^N,\vect{y}_2^N,\vect{b})\eqdef \infr_{\vect{X}_2^N\vect {Y}_2^N|\vect{B}}(\vect{x}_2^N;\vect {y}_2^N|\vect{b})=\log\frac{P_{\vect {Y}_2^N|\vect{X}_2^N,\vect{B}}(\vect {y}_2^N|\vect {x}_2^N,\vect{b})}{P_{\vect {Y}_2^N|\vect{B}}(\vect {y}_2^N|\vect{b})}.
\end{eqnarray} 
Here and throughout the appendices, we use the notations $\vect{X}_i^N=\mat{X}_i$, $\vect{x}_i^N=\mat{x}_i$, $\vect{Y}_i^N=\mat{Y}_i$, and $\vect{y}_i^N=\mat{y}_i$ to highlight the fact that, in the quasi-static setting, we transmit $N$ symbols in one coherence~block.

We further consider the individual error events
\begin{eqnarray}\label{error_event}
\Ee_i(\Gamma_i)&\eqdef& \bigg\{\frac{1}{N}\infr_i(\vect{x}_i^N,\vect{y}_i^N,\vect{b}) \leq \Gamma_i\bigg\},\  i=1,2
\end{eqnarray}
and the joint error event
\begin{eqnarray}\label{joint_error_event}
\Ee_{12}(\Gamma)&\eqdef& \bigg\{\frac{1}{N}\left(\infr_1(\vect{x}_1^N,\vect{y}_1^N,\vect{b})+\infr_2(\vect{x}_2^N,\vect{y}_2^N,\vect{b})\right)\leq \Gamma\bigg\}.
\end{eqnarray}

The proofs of the converse results are based on the following lemmas.

\begin{lemma}[Verd\'u-Han lemma] \label{Verdu-Han}
Every $(N,R,P_e)$ code over a channel $P_{\vect{Y}^N|\vect{X}^N}$ satisfies

\begin{eqnarray}
P_e\geq \Pr\Bigl\{ \frac{1}{N}\infr_{\vect{X}^N\vect{Y}^N}(\vect{X}^N;\vect{Y}^N)\leq R-\gamma\Bigr\}-e^{-\gamma N}
\end{eqnarray}
for  every $\gamma>0$, where $X^N$  places probability mass $\frac{1}{2^{NR}}$ on each codeword and $\displaystyle{\infr_{\vect{X}^N\vect{Y}^N}(\vect{X}^N;\vect{Y}^N)\triangleq\log\tfrac{P_{\vect{Y}^N|\vect{X}^N}(\vect{y}^N|\vect{x}^N)}{P_{\vect{Y}^N}(\vect{y}^N)}}$.
\end{lemma}
\begin{IEEEproof}
See \cite[ (Th.~4)]{verdu-han}.
\end{IEEEproof}

\begin{lemma}\label{Lemma_e}
Suppose that $ \Pr\{\Ee_{12}(\Gamma) \big|\vect{B}=\vect{b}\}\to 0$ as $N\to\infty$. Then, for each pair $\vect{b}\in\{0,1\}^2$, the threshold $\Gamma$ must satisfy the following conditions:
\begin{itemize}
\item For $\vect{B}=[0,0]$, $\Gamma$ satisfies
\begin{eqnarray}
\Gamma &\leq& 2n_{d}. \label{P2P}
\end{eqnarray}

\item For $\vect{B}=[0,1]$ and $\vect{B}=[1,0]$, $\Gamma$ satisfies \eqref{P2P} and   
\begin{eqnarray}
\Gamma &\leq& (n_{d}-n_{c})^++\max(n_{d},n_{c}).\label{SZ_1}
\end{eqnarray}

\item For $\vect{B}=[1,1]$, $\Gamma$ satisfies \eqref{P2P} and  \eqref{SZ_1}, and   
\begin{eqnarray}
\Gamma &\leq& 2\max\{(n_{d}-n_{c})^+,n_{c}\}.\label{IC_1}
\end{eqnarray}

\end{itemize}
\end{lemma}
\begin{IEEEproof}
See Appendix \ref{App: Verdu-Han}.
\end{IEEEproof}

\subsubsection{Proof of Theorem~\ref{Thm:conv_realiable_rates}} \label{IC_Verdu-Han} In this section we prove the IC channel converse bounds for $p>0$. This proof assumes global CSIRT, hence the resulting bounds also apply to local CSIR and local CSIRT.
Let $\displaystyle{P_e^{(N)}=\Pr\{(\hat{W}_1\neq W_1 \cup \hat{W}_2 \neq W_2)\}}$, and let us denote by $P_{e_1}^{(N)}$ and $P_{e_2}^{(N)}$ the error probabilities at decoders one and two,~respectively:
\begin{eqnarray}
P_{e1}^{(N)}&\eqdef& \Pr\{\hat W_1\neq W_1\},\label{Pe1}\\
P_{e2}^{(N)}&\eqdef& \Pr\{\hat  W_2\neq W_2\}\label{Pe2}.
\end{eqnarray}
Clearly, the  error probabilities $P_e^{(N)}$, $P_{e1}^{(N)}$ and $P_{e2}^{(N)}$ are related by the following sets of inequalities
\begin{eqnarray}
\max\Bigl(P_{e1}^{(N)},P_{e2}^{(N)} \Bigr) \leq  P_e^{(N)} \leq P_{e1}^{(N)} + P_{e2}^{(N)} \leq 2\max\Bigl(P_{e1}^{(N)},P_{e2}^{(N)} \Bigr).
\label{Pe12bound}
\end{eqnarray}
Using these inequalities we conclude that
\begin{eqnarray}
P_e^{(N)} \geq\frac{1}{2}\Bigl( P_{e1}^{(N)} + P_{e2}^{(N)}\Bigr).
\label{Pebound}
\end{eqnarray}

We now rewrite  \eqref{Pe1} and \eqref{Pe2} as
\begin{eqnarray}
P_{e1}^{(N)}&=&\sum_{\vect{b}} p_{\vect{b}}\Pr\{\hat W_1\neq W_1|\vect{B}=\vect{b}\},\label{Pe1_1}\\
P_{e2}^{(N)}&=&\sum_{\vect{b}} p_{\vect{b}}\Pr\{\hat  W_2\neq W_2|\vect{B}=\vect{b}\}\label{Pe2_1}
\end{eqnarray}
and apply the Verd\'u-Han lemma (Lemma~\ref{Verdu-Han}) to each of the probability terms $\Pr\{\hat W_i\neq W_i|\vect{B}=\vect{b}\},\ i=1,2$, in \eqref{Pe1_1} and \eqref{Pe2_1}. This yields
\begin{eqnarray}
\Pr\{\hat W_1\neq W_1|\vect{B}=\vect{b}\}&\geq& \Pr \Bigl\{\frac{1}{N}\infr_1(\vect{x}_1^N,\vect{y}_1^N,\vect{b})\leq R_1-\gamma_{1}|\vect{B}=\vect{b}\Bigr\} -e^{-\gamma_1N},\label{verdu-han1}\\
\Pr\{\hat W_2\neq W_2|\vect{B}=\vect{b}\}&\geq& \Pr \Bigl\{\frac{1}{N}\infr_2(\vect{x}_2^N,\vect{y}_2^N,\vect{b})\leq R_2-\gamma_{2}|\vect{B}=\vect{b}\Bigr\} -e^{-\gamma_2N}.\label{verdu-han2}
\end{eqnarray}
We set $\Gamma_i=R_i-\gamma_i$ and $\Gamma=\Gamma_1+\Gamma_2=R-\gamma_1-\gamma_2$. Then, using the definition of $\Ee_{i}$ in \eqref{error_event}, we can write \eqref{verdu-han1} and \eqref{verdu-han2} as 
\begin{eqnarray}
\Pr\{\hat W_1\neq W_1|\vect{B}=\vect{b}\}&\geq& \Pr \{\Ee_1(\Gamma_1)|\vect{B}=\vect{b}\} -e^{-\gamma_1N},\label{verdu-han-1}\\
\Pr\{\hat W_2\neq W_2|\vect{B}=\vect{b}\}&\geq& \Pr \{\Ee_2(\Gamma_2)|\vect{B}=\vect{b}\} -e^{-\gamma_2N}.\label{verdu-han-2}
\end{eqnarray}
Comparing the joint error event $\Ee_{12}(\Gamma)$ in \eqref{joint_error_event} with $\Ee_1(\Gamma_1)$ and $\Ee_2(\Gamma_2)$ in \eqref{error_event}, it can be shown~that
\begin{eqnarray}
\Ee_1(\Gamma_1) \cap \Ee_2(\Gamma_2) &\subseteq& \Ee_{12}(\Gamma),\label{E12cap}\\
\Ee_1^c(\Gamma_1) \cap \Ee_2^c(\Gamma_2) &\subseteq& \Ee_{12}^c(\Gamma)\;\Rightarrow\;
 \Ee_{12}(\Gamma) \subseteq \Ee_1(\Gamma_1) \cup \Ee_2(\Gamma_2).\label{E12cup}
\end{eqnarray}
Using \eqref{E12cup} and the union bound, we thus obtain
\begin{equation}
\begin{aligned}
\Pr\{\Ee_{12}(\Gamma)|\vect{B}=\vect{b}\}&\leq \Pr\{\Ee_1(\Gamma_1)\cup \Ee_2(\Gamma_2)|\vect{B}=\vect{b}\}\\
&\leq \Pr\{\Ee_1(\Gamma_1)|\vect{B}=\vect{b}\}+\Pr\{\Ee_2(\Gamma_2)|\vect{B}=\vect{b}\}.
\end{aligned}
\label{union12bound}
\end{equation}
Combining this result with  \eqref{Pebound}, \eqref{verdu-han-1} and \eqref{verdu-han-2} gives
\begin{equation}
\begin{aligned}
P_e^{(N)} &\geq \frac{1}{2}\Bigl( P_{e1}^{(N)} + P_{e2}^{(N)}\Bigr)\\
               &\geq \frac{1}{2}\sum_{\vect{b}}p_{\vect{b}}\bigl( \Pr \{\Ee_1(\Gamma_1)|\vect{B}=\vect{b}\} +  \Pr \{\Ee_2(\Gamma_2)|\vect{B}=\vect{b}\}  - e^{-\gamma_1N}- e^{-\gamma_2N}\bigr)\\
               &\geq \frac{1}{2}\sum_{\vect{b}}p_{\vect{b}}\bigl( \Pr \{ \Ee_{12}(\Gamma)|\vect{B}=\vect{b}\} -e^{-\gamma_1N}-e^{-\gamma_2N}\bigr).
 \end{aligned}
\label{PeVHbound}
\end{equation}

The remainder of this section is devoted to an analysis of $\Pr \{ \Ee_{12}(\Gamma)|\vect{B}=\vect{b}\}$. Indeed, by \eqref{PeVHbound} we have for any $\gamma_{1},\gamma_{2}>0$ that
\begin{eqnarray}\label{limPEVHbound}
\lim\limits_{N\to\infty}P_e^{(N)}\geq \lim\limits_{N\to\infty}\frac{1}{2}\left[p_{11}\epsilon_{11}
+p_{00}\epsilon_{00}
+p_{10}\epsilon_{10}+p_{01}\epsilon_{01}\right],
\end{eqnarray}
where $\epsilon_{\vect{b}}\eqdef \Pr\{\Ee_{12}(\Gamma) \big|\vect{B}=\vect{b}\}$. When $p>0$, the probability $p_{11}$ is strictly positive both when $(B_1,B_2)$ are independent and when they are fully correlated. Since $p_{\vect{b}}$ does not depend on $N$, it follows that the only way that $\lim\limits_{N\to \infty} P_e^{(N)} =0$  is that $\epsilon_{11}\to 0$ as $N\to\infty$. The conditions on $R$ under which this happens are summarized in Lemma~\ref{Lemma_e}. Specifically, recalling that $\Gamma=R-(\gamma_1+\gamma_2)$, we obtain from Lemma~\ref{Lemma_e} that $P_e^{(N)}\to 0$ only if 
\begin{eqnarray}
R-(\gamma_1+\gamma_2)&\leq& 2n_d\label{R1}\\
R-(\gamma_1+\gamma_2)&\leq& (n_d-n_c)^++\max(n_d,n_c)\label{R2}\\
R-(\gamma_1+\gamma_2)&\leq& 2\max\{(n_d-n_c)^+,n_c\}\label{R3}.
\end{eqnarray}
Since ${\gamma_1,\gamma_2}> 0$  are arbitrary, we obtain the converse bounds \eqref{Eq: IC3} and \eqref{Eq: IC2} in  Theorem~\ref{Thm:conv_realiable_rates} from~\eqref{R1}--\eqref{R3} upon letting  $N\to\infty$ and then $\gamma_{1}\to 0$ and $\gamma_{2}\to 0$.

When $p=0$, the only positive probability is $p_{00}$. A necessary condition for $\lim\limits_{N\to \infty} P_e^{(N)} =0$ is that $\epsilon_{00}\to 0$ as $N\to\infty$. By following the same approach as for the case $p>0$, we obtain the converse bound \eqref{Eq: IC1} in Theorem~\ref{Thm:conv_realiable_rates}.


\subsubsection{Converse Proof of Theorem~\ref{Thm:convBIC_local}}\label{Proof:convBIC_local}
In this section, we analyze the opportunistic rate $\Delta R_1(b_1) + \Delta R_2(b_2),\ b_i\in\{0,1\}$ for local CSIRT and independent $B_1$ and $B_2$. Let us denote by $\Pehat{1(b_1)}$ and $\Pehat{2(b_2)}$ the error probabilities at decoders one and two, defined in \eqref{QS_Pr_1} and \eqref{QS_Pr_2}, i.e., 
\begin{eqnarray}
\Pehat{1(b_1)}&\eqdef&\Pr\{(\hat W_1,\Delta {\hat{W}_1}(B_1))\neq (W_1,\Delta {W_1}(B_1))|B_1=b_1\},\ b_1\in\{0,1\}\label{Eq: LocalPehat1},\\
\Pehat{2(b_2)}&\eqdef&\Pr\{(\hat  W_2, \Delta {\hat{W}_2}(B_2))\neq (W_2, \Delta {W_2}(B_2))|B_2=b_2\}, \ b_i\in\{0,1\}\label{Eq: LocalPehat2}.
\end{eqnarray}

Before we apply the Verd\'u-Han lemma, we have to deal with the fact that \eqref{Eq: LocalPehat1} and \eqref{Eq: LocalPehat2} are conditioned on two different variables but we need to analyze the probability of error jointly. To solve this problem, we expand the probability of error  \eqref{Eq: LocalPehat1} as
\begin{eqnarray}\label{QS_Local_Pr_3}
\Pehat{1(b_1)}=\sum\limits_{b_2=0,1}\Pr\{B_2=b_2\}\Pr\bigl\{(\hat{W}_1,\Delta {\hat{W}_1}(B_1))\neq(W_1,\Delta {W_1}(B_1))\big|\vect{B}=\vect{b}\bigr\}.
\end{eqnarray}
Since, by assumption,  $\Pr\{B_2=b_2\}\in (0,1)$, it follows that
\begin{eqnarray}
\Pr\bigl\{(\hat{W}_1,\Delta {\hat{W}_1}(B_1))\neq(W_1,\Delta {W_1}(B_1))|B_1=b_1\bigr\}\to  0\ \text{as}\ N\to \infty\nonumber
\end{eqnarray}
{if, and only if,}
\begin{eqnarray}
\Pr\bigl\{(\hat{W}_1,\Delta {\hat{W}_1}(B_1))\neq(W_1,\Delta {W_1}(B_1))|\vect{B}=\vect{b}\bigr\}\to  0,\ b_2\in\{0,1\}\ \text{as}\ N\to \infty. \label{QS_local_1}
\end{eqnarray}

We shall lower-bound \eqref{QS_Local_Pr_3} by considering only one of the two terms in the sum. Proceeding analogously for the second user and applying  the Verd\'u-Han lemma (Lemma~\ref{Lemma Verdu-Han}), we obtain
\begin{myequation}
\Pehat{1(b_1)} \geq \left(\Pr \Bigl\{\frac{1}{N}\infr_1(\vect{x}_1^N,\vect{y}_1^N,\vect{b})\leq R_1+\Delta R_1(B_1)-\gamma_{1}|\vect{B}=\vect{b}\Bigr\} -e^{-\gamma_1N}\right) \Pr\{B_2=b_2\},\quad b_2=0,1,\label{verdu-han3}
\end{myequation}
\begin{myequation}
\Pehat{2(b_2)} \geq \left(\Pr \Bigl\{\frac{1}{N}\infr_2(\vect{x}_2^N,\vect{y}_2^N,\vect{b})\leq R_2+\Delta R_2(B_2)-\gamma_{2}|\vect{B}=\vect{b}\Bigr\} -e^{-\gamma_2N}\right)\Pr\{B_1=b_1\},\quad b_1=0,1.\label{verdu-han4}
\end{myequation}
Let $\Gamma_i=R_i+\Delta R_i-\gamma_i$, $i=1,2$ and $\Gamma=R+\Delta {R_1}(B_1)+\Delta {R_2}(B_2)-(\gamma_1+\gamma_2)$. Then, \eqref{verdu-han3} and~\eqref{verdu-han4} can be written as
\begin{eqnarray}
\Pehat{1(b_1)} &\geq& \left(\Pr \{\Ee_1(\Gamma_1)|\vect{B}=\vect{b}\} -e^{-\gamma_{1}N}\right) \Pr\{B_2=b_2\},\quad b_2=0,1,\label{verdu-han-3}\\
\Pehat{2(b_2)} &\geq& \left(\Pr \{\Ee_2(\Gamma_2)|\vect{B}=\vect{b}\} -e^{-\gamma_{2}N}\right)\Pr\{B_1=b_1\},\quad b_1=0,1.\label{verdu-han-4}
\end{eqnarray}
Proceeding analogously as in \eqref{E12cap}--\eqref{PeVHbound}, and using that $\Pr\{B_i=b_i\}\geq \min\{p,1-p\}$, we~obtain

\begin{myequation}\label{local_PeVHbound_1}
\begin{aligned}
\Pehat{1(b_1)} + \Pehat{2(b_2)}
 &\geq  \Bigl(\Pr \{\Ee_1(\Gamma_1)|\vect{B}=\vect{b}\} + \Pr \{\Ee_2(\Gamma_2)|\vect{B}=\vect{b}\}  - e^{-\gamma_1N}-e^{-\gamma_{2}N}\Bigr) \min\{p,1-p\}\\
 &\geq  \Bigl(\Pr \{ \Ee_{12}(\Gamma)|\vect{B}=\vect{b}\} -e^{-\gamma_{1}N}-e^{\gamma_{2}N}\Bigr) \min\{p,1-p\}.   
 \end{aligned}     
\end{myequation}
 Since $\gamma_{1},\gamma_{2} >0$, the left-hand side (LHS) of \eqref{local_PeVHbound_1} only tends to zero as $N\to\infty$ if $\Pr(\Ee_{12}(\Gamma)|\vect{B}=\vect{b})\to 0$ as $N\to\infty$. It thus follows from Lemma~\ref{Lemma_e} that $\Pehat{1(b_1)} + \Pehat{2(b_2)}\to 0$ as $N\to\infty$ only if conditions \eqref{P2P}-\eqref{IC_1} are satisfied. Letting $\gamma_1\to 0$ and $\gamma_2 \to 0$ then gives the following constraints:
\begin{itemize}
\item For {$\vect {B}=[1,1]$}
\begin{eqnarray}
R_1+\Delta {R_1}{(1)}+R_2+\Delta {R_2}{(1)} &\leq& 2n_{d}\label{IC_local1}\\
R_1+\Delta {R_1}{(1)}+R_2+\Delta {R_2}{(1)} &\leq& (n_{d}-n_{c})^++\max(n_{d},n_{c})\label{IC_local3}\\
R_1+\Delta {R_1}{(1)}+R_2+\Delta {R_2}{(1)} &\leq& 2\max\{(n_{d}-n_{c})^+,n_{c}\}.\label{IC_local2}
\end{eqnarray} 

\item For {$\vect {B}=[0,0]$}, 
\begin{eqnarray}
R_1+\Delta {R_1}{(0)}+R_2+\Delta {R_2}{(0)} &\leq& 2n_{d}.\label{P2P_local}
\end{eqnarray}

\item For {$\vect {B}=[0,1]$}, using that ${\Delta R_2}{(1)}=0$,
\begin{eqnarray}
R_1+\Delta {R_1}{(0)}+R_2 &\leq& (n_{d}-n_{c})^++\max(n_{d},n_{c}).\label{S_local}
\end{eqnarray} 

\item For {$\vect {B}=[1,0]$}, using that ${\Delta R_1}{(1)}=0$, 
\begin{eqnarray}
R_1+R_2+\Delta {R_2}{(0)} &\leq& (n_{d}-n_{c})^++\max(n_{d},n_{c}).\label{Z_local}
\end{eqnarray}
 
\end{itemize}

The constraints \eqref{P2P_local}--\eqref{Z_local} yield \eqref{Eq: local_Delta0}--\eqref{Eq: local_Delta2}. This proves Theorem~\ref{Thm:convBIC_local}.

\subsubsection{Achievability Proof of Theorem~\ref{Thm:convBIC_local}}\label{Ach. apportunistic}
 
In this section, we present the achievability bounds in Theorem~\ref{Thm:convBIC_local} for the regions in which it is possible to transmit opportunistic messages, namely the VWI and WI regions. The presented bounds are valid for local CSIR and local CSIRT.
 
 \paragraph{Very Weak Interference}
 Transmitter 1 (Tx$_1$) and transmitter 2 (Tx$_2$) transmit in the most significant levels a block of $n_d(1-\alpha)$ bits, and they transmit in the least significant levels a block of $n_d\alpha$ bits. The same construction is used for both transmitters. Figure \ref{Fig: VWI1} depicts the signal levels of the transmitted signals (normalized by $n_d$) as observed at receiver 1 (Rx$_1$), when it is affected by interference.
 
 \begin{figure}[tbp]
 		\centering
 		\includegraphics[width=0.4\linewidth]{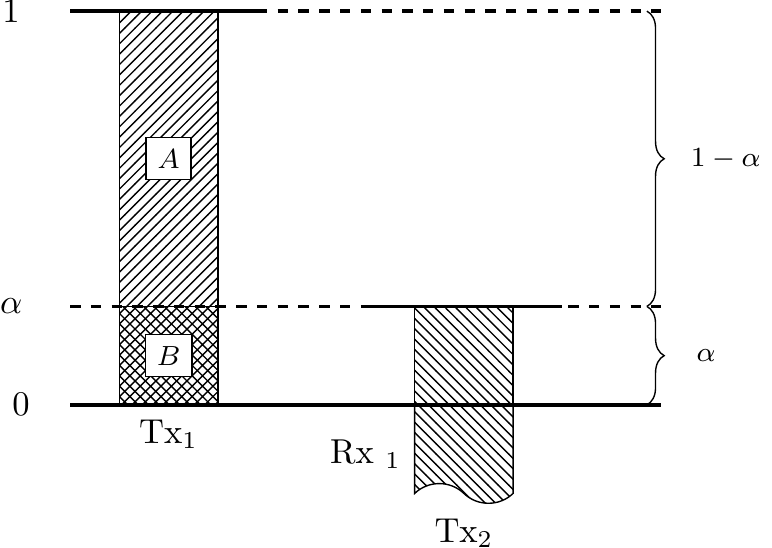}
 		\caption{Normalized signal levels at Rx$_1$ for $\alpha\leq\frac{1}{2}$.} 
 		\label{Fig: VWI1}
 \end{figure}
 At the receiver side, we have the following procedure:
 \begin{itemize}
 \item In presence of interference:
  decode block \framebox[0.3cm][c]{\footnotesize $A$} in the desired signal which is interference free, and treat the block \framebox[0.3cm][c]{\footnotesize $B$} as noise. We thus obtain the individual rate
\begin{eqnarray}\label{RVWI_IC}
 R_1=(n_{d}-n_{c})^+ \ \tfrac{\text{bits}}{\text{sub-channel use}}.
 \end{eqnarray}
 
 \item In absence of interference:
 decode blocks \framebox[0.3cm][c]{\footnotesize $A$} and \framebox[0.3cm][c]{\footnotesize $B$}. We thus obtain the individual rate
 \begin{eqnarray} \label{RVWI_BIC}
 R_1+\Delta R_1(0)=n_{d}\ \tfrac{\text{bits}}{\text{sub-channel use}}.
 \end{eqnarray}
where $\Delta R_1(0)=n_c \tfrac{\text{bits}}{\text{sub-channel use}}$ corresponds to the opportunistic rate.
\end{itemize}

The bounds \eqref{RVWI_IC} and \eqref{RVWI_BIC} coincide with the bounds for the bounds of user 2.
In order to obtain the possible sum rates according to the interference states, we combine \eqref{RVWI_IC} (which corresponds to $B_1=1$)  and  \eqref{RVWI_BIC} (which corresponds to $B_1=0$) to obtain the converse bounds \eqref{Eq: local_Delta0}--\eqref{Eq: local_Delta1}.

\paragraph{Weak Interference}
The symbol transmitted by Tx$_1$ (normalized by $n_d$) is depicted in Figure \ref{Fig: A2}a. Specifically, we transmit in the most significant levels a block of $n_d(1-\alpha)$ bits. In the subsequent levels we transmit a block of $n_d(2\alpha - 1)$ zeros, followed by $n_d(2-3\alpha)$ opportunistic bits. Finally, in the least significant levels, we transmit a block of $n_d(2\alpha-1)$ bits. The same construction is used for both transmitters.
  
 Figure \ref{Fig: A2}b depicts the normalized signal levels of the transmitted signals as observed by Rx$_1$.
At the receiver side, we have the following procedure:
 \begin{itemize}
 \item In presence of interference:
 The channel pushes the interference level by  $n_d-n_c$ bits. Thus, the least significant $2n_c-n_d$ bits of the desired signal (block \framebox[0.3cm][c]{\footnotesize $A$}) align with the zeros of the interference signal and can be decoded free from interference. Since $(n_d-n_c)\leq n_c$, the most significant $n_d-n_c$ bits (block \framebox[0.3cm][c]{\footnotesize $B$}) are also free from interference.
 Thus, we achieve the rate
 \begin{eqnarray}\label{RWI_IC}
 R_1&=& n_d-n_{c}+2n_c-n_d\nonumber\\
 &=& n_c \ \tfrac{\text{bits}}{\text{sub-channel use}}.
 \end{eqnarray}
 
 \item In absence of interference:
The bits in blocks \framebox[0.3cm][c]{\footnotesize $A$} , \framebox[0.3cm][c]{\footnotesize $B$} ,  and \framebox[0.3cm][c]{\footnotesize $D$} can be decoded free from interference. Thus, we achieve the rate
 \begin{eqnarray}\label{RWI_BIC}
 R_1+\Delta R_1(0)&=& n_d-n_{c}+2n_c-n_d+2n_d-3n_c\nonumber\\
 &=& 2(n_d-n_c) \ \tfrac{\text{bits}}{\text{sub-channel use}}
 \end{eqnarray}
 where $\Delta R_1(0)=2n_d-3n_c \tfrac{\text{bits}}{\text{sub-channel use}}$ corresponds to the opportunistic rate.
\end{itemize}

By symmetry, the bounds \eqref{RWI_IC} and \eqref{RWI_BIC} also apply for  the achievable rates of  user 2.
In order to obtain the possible sum rates according to the interference states, we combine \eqref{RWI_IC} (which corresponds to $B_1=1$)  and  \eqref{RWI_BIC} (which corresponds to $B_1=0$) to obtain the achievability  bounds in Theorem~\ref{Thm:convBIC_local}.

 \begin{figure}[tbp]
  	\centering
  	\subfigure[ \label{Fig: Coding_WI}]{\includegraphics[width=0.4\linewidth]{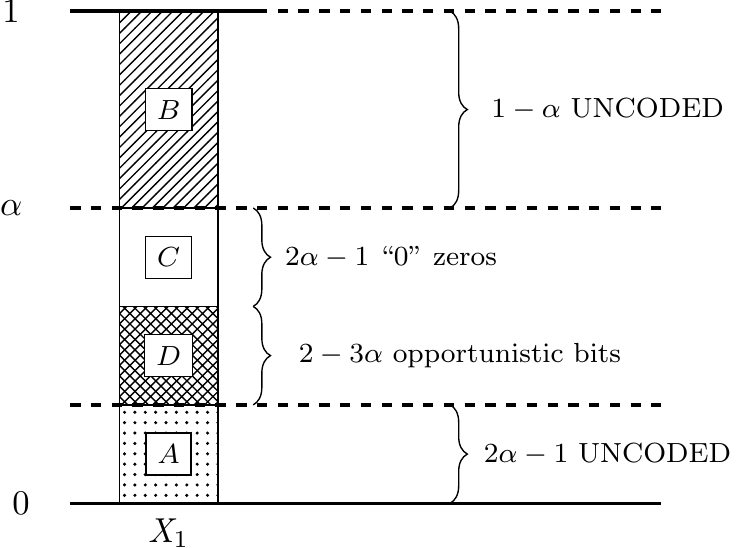}}\quad
    \subfigure[
     		\label{Fig: WI_BIC}]{\includegraphics[width=0.43\linewidth]{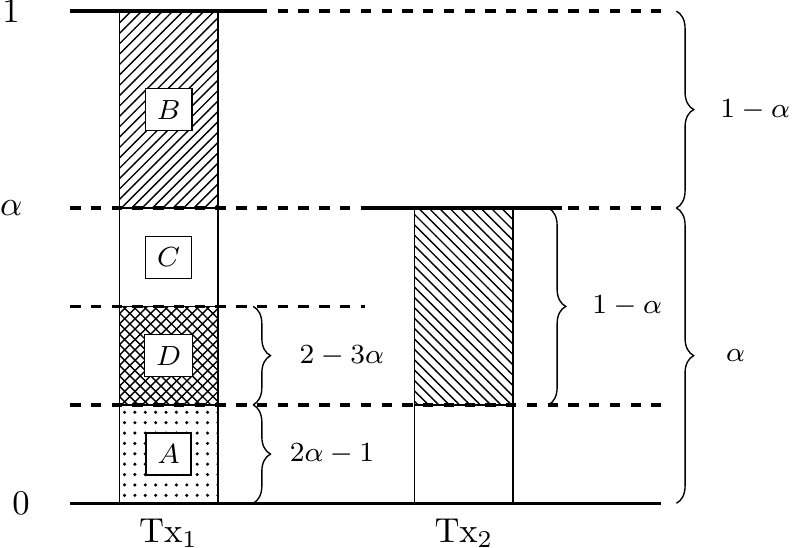}}\\	
  	\caption{(\textbf{a}) Normalized transmitted symbol at Tx$_1$; (\textbf{b}) Normalized signal levels at Rx$_1$.} \label{Fig: A2}
  \end{figure}

\subsubsection{Converse Proof of Theorem \ref{Thm:convBIC_local} when $B_1=B_2$} \label{Sec:Local_00_corr}

The proof of the converse bound \eqref{Eq: local_Delta0} for local CSIR when $B_1=B_2$ is similar to the proof when $B_1$ and $B_2$ are independent; see Appendix~\ref{Proof:convBIC_local}. However, to prove the converse bound \eqref{Eq: local_Delta1} for the case where $B_1=B_2$ we cannot simply reproduce the steps for the independent case. The reason is that, in the correlated case, we only have the interference states $[0,0]$ and $[1,1]$, but the derivation of \eqref{Eq: local_Delta1} for the independent case follows from the analysis of the states $\vect{B}=[0,1]$ and $\vect{B}=[1,0]$ (see~\eqref{S_local} and~\eqref{Z_local} in Appendix~\ref{Proof:convBIC_local}). 
To sidestep this problem, we follow a slightly different approach. Specifically, we combine the error probability of user $1$ when $\vect{B}=[0,0]$ with that of user $2$ when $\vect{B}=[1,1]$. This approach yields a tighter converse bound compared to the one obtained by simply considering $\vect{B}=[0,0]$ in both probabilities.

Consider $\Pehat{1(b_1)}$ and $\Pehat{2(b_2)}$ defined in \eqref{Eq: LocalPehat1} and \eqref{Eq: LocalPehat2}. Applying the Verd\'u-Han lemma (Lemma~\ref{Verdu-Han}) with $\Gamma_1=R_1+\Delta {R_1}{(0)}-\gamma_1$ and $\Gamma_2=R_2-\gamma_2$, and using \eqref{QS_Local_Pr_3}, we obtain the lower~bounds
\begin{eqnarray}
\Pehat{1(0)} &\geq& \left(\Pr \Bigl\{\Ee_1(\Gamma_1)|\vect{B}=[0,0]\Bigr\} -e^{-\gamma_1N}\right) \Pr\{B_2=0\}\label{verdu-han30}\\
\Pehat{2(1)} &\geq& \left(\Pr \Bigl\{\Ee_2(\Gamma_2)|\vect{B}=[1,1]\Bigr\} -e^{-\gamma_2N}\right)\Pr\{B_1=1\}.\label{verdu-han40}
\end{eqnarray}
Note that compared to the derivation in Section~\ref{Proof:convBIC_local},  the two error events $\Ee_{1}(\Gamma_1)$ and $\Ee_{2}(\Gamma_2)$ are conditioned on different interference states.
In order to derive a joint error event for $\Ee_{1}(\Gamma_1)$ and $\Ee_{2}(\Gamma_2)$, we use the next lemma.

\begin{lemma}\label{Lemma_LocalCSIR}
For local CSIR,  the information density $\infr_i$, $i=1,2$  depends only on $(\vect{x}_i^N,\vect{y}_i^N)$ and the corresponding state $b_i$, i.e., 
\begin{eqnarray}
\infr_1(\vect{x}_1^N;\vect {y}_1^N,[b_1,0])=\infr_1(\vect{x}_1^N;\vect {y}_1^N,[b_1,1])&\eqdef& \infr_1(\vect{x}_1^N,\vect{y}_1^N,b_1)\label{i1local}\\
\infr_2(\vect{x}_2^N;\vect {y}_2^N,[0,b_2])=\infr_2(\vect{x}_2^N;\vect {y}_2^N,[1,b_2])&\eqdef& \infr_2(\vect{x}_2^N,\vect{y}_2^N,b_2).\label{i2local}
\end{eqnarray}

\end{lemma}
\begin{IEEEproof}
We prove \eqref{i1local} for user 1. By the definition of the information density \eqref{i1(b1)}, it follows that
\begin{eqnarray}\label{i1localCSIR1}
\infr_1(\vect{x}_1^N,\vect{y}_1^N,[b_1,b_2])=\log\frac{P_{\vect {Y}_1^N|\vect{X}_1^N,\vect{B}}(\vect {y}_1^N|\vect {x}_1^N,[b_1,b_2])}{P_{\vect {Y}_1^N|\vect{B}}(\vect {y}_1^N| [b_1,b_2])}
\end{eqnarray}
Evaluating $\infr_1$ for $\vect{B}=[0,b_2]$, $b_2=0,1$ and $\vect{B}=[1,b_2]$, $b_2=0,1$ we obtain that both cases are independent of $b_2$. The identity \eqref{i1local} can be proven in the same way.
\end{IEEEproof}

We next analyze the probability terms in \eqref{verdu-han30} and \eqref{verdu-han40}.  It follows from \eqref{i1local} in Lemma~\ref{Lemma_LocalCSIR}  that $\infr_1(\vect{x}_1^N,\vect{y}_1^N,[0,b_2])$ is independent of $b_2$. Consequently,
\begin{equation}
\begin{aligned}
\Pr\{
\Ee_{1}(\Gamma_1)\big|\vect{B}=[0,0]\} & =\E{\mathds{1}\left\{\frac{1}{N}\infr_1(\vect{X}_1^N,\vect{Y}_1^N,[0,0]) \leq \Gamma_1\right\}}\\
&=\E{\mathds{1}\left\{\frac{1}{N}\infr_1(\vect{X}_1^N,\vect{Y}_1^N,[0,1]) \leq \Gamma_1\right\}}\\
&=\Pr\{\Ee_{1}(\Gamma_1)\big|\vect{B}=[0,1]\}.
\end{aligned}
\label{PGamma1B00}
\end{equation}
Analogously, using \eqref{i2local} in \eqref{verdu-han40}, we obtain
\begin{eqnarray}
\Pr\{\Ee_{2}(\Gamma_2)\big|\vect{B}=[1,1]\}
=\Pr\{\Ee_{2}(\Gamma_2)\big|\vect{B}=[0,1]\}.
\label{PGamma2B11}
\end{eqnarray}
Adding \eqref{verdu-han30} and \eqref{verdu-han40}, using \eqref{PGamma1B00} and \eqref{PGamma2B11}, and lower-bounding $\Pr\{B_1=1\}$ and $\Pr\{B_2=0\}$  by $\min\{p,1-p\}$, we obtain
\begin{myequation}
\begin{aligned}
\Pehat{1(0)} + \Pehat{2(1)}
&\geq    \bigl(\Pr \{\Ee_1(\Gamma_1)|\vect{B}=[0,1]\} +  \Pr \{\Ee_2(\Gamma_2)|\vect{B}=[0,1]\}  - e^{-\gamma_{1}N}- e^{-\gamma_{2}N}\bigr)\min\{p,1-p\}\\
 &\geq  \Bigl(\Pr \{ \Ee_{12}(\Gamma)|\vect{B}=[0,1]\} -e^{-\gamma_{1}N}-e^{\gamma_{2}N}\Bigr)\min\{p,1-p\}  
 \end{aligned}
\label{local_PeVHbound_corr}
\end{myequation}
where $\Gamma=\Gamma_1+\Gamma_2$. We next apply Lemma~\ref{Lemma_e} with $\Gamma=R+\Delta R_1(0)+\Delta R_2(0)-(\gamma_{1}+\gamma_{2})$.  Since $\min\{p,1-p\}$ is strictly positive for $0<p<1$, and since $-e^{-\gamma_{1}N}-e^{-\gamma_{2}N}\to 0$ as $N\to\infty$ for any fixed $\gamma_{1},\gamma_{2}>0$,  a necessary condition for \eqref{local_PeVHbound_corr} going to zero is that $\Pr \{ \Ee_{12}(\Gamma)|\vect{B}=[0,1]\}\to 0$ as $N\to\infty$. This is the case if, and only if, \eqref{SZ_1} in Lemma~\ref{Lemma_e} is fulfilled. Since $\gamma_1,\gamma_2>0$ are arbitrary, we conclude the proof by letting $\gamma_1\to 0$ and $\gamma_2 \to 0$ and using that ${\Delta R_2}{(1)}=0$ to obtain
\begin{eqnarray}
R_1+\Delta {R_1}{(0)}+R_2 &\leq& (n_{d}-n_{c})^++\max(n_{d},n_{c}).\label{S_local_corr}
\end{eqnarray} 
Given the symmetry of the problem, a bound on $\Delta {R_2}{(0)}$ follows by swapping the roles of users $1$ and $2$, yielding in this case
\begin{eqnarray}
R_1+R_2+\Delta {R_2}{(0)} &\leq& (n_{d}-n_{c})^++\max(n_{d},n_{c}).\label{Z_local_corr}
\end{eqnarray} 
Finally, combining \eqref{S_local_corr} and \eqref{Z_local_corr}, we obtain the bound \eqref{Eq: local_Delta1} in Theorem \ref{Thm:convBIC_local} for the fully correlated~scenario.
 
\subsubsection{Converse Proof of Theorem~\ref{Thm:conv_BIC_global}}\label{Proof:convBIC_global}
In this section, we analyze the opportunistic rates $\{\Delta R(\mathbf{b}), \mathbf{b}\in\{0,1\}^2\}$ for global CSIRT and independent $B_1$ and $B_2$. Let us denote by $\Pehat{1(\vect{b})}$ and $\Pehat{2(\vect{b})}$ the error probabilities at decoders 1 and 2, defined in \eqref{QS_Pr_1} and~\eqref{QS_Pr_2},~namely,
\begin{eqnarray}
\Pehat{1(\vect{b})}&\eqdef&\Pr\{(\hat W_1,\{\Delta {\hat{W}_1}{(\vect{B}})\})\neq (W_1,\{\Delta {W_1}{(\vect{B}})\})|\vect{B}=\vect{b}\},\quad \vect{b}\in \{0,1\}^2\label{Eq: global_Pehat1},\\
\Pehat{2(\vect{b})}&\eqdef&\Pr\{(\hat  W_2, \{\Delta {\hat{W}_2}{(\vect{B}})\})\neq (W_2, \{\Delta {W_2}{(\vect{B}})\})|\vect{B}=\vect{b}\},\quad \vect{b}\in \{0,1\}^2\label{Eq: global_Pehat2}.
\end{eqnarray}
We shall follow analogous steps as in Section~\ref{Proof:convBIC_local} and set $\Gamma_i=R_i+\Delta {R_i}{(\vect{B})}-\gamma_i$, $i=1,2$, and $\Gamma=R+\Delta {R}{(\vect{B})}-(\gamma_1+\gamma_2)$.
Proceeding analogously as in \eqref{E12cap}--\eqref{union12bound}, we obtain
\begin{eqnarray}\label{global_PeVHbound}
\Pehat{1(\vect{b})} + \Pehat{2(\vect{b})} 
&\geq&  \Pr \{ \Ee_{12}(\Gamma)|\vect{B}=\vect{b}\} -e^{-\gamma_{1}N}-e^{-\gamma_{2}N}.
\end{eqnarray}
By invoking Lemma~\ref{Lemma_e} for fixed (but arbitrary) $\gamma_1,\gamma_2> 0$, and letting then $\gamma_{1}\to 0$ and $\gamma_{2}\to 0$, we obtain that the RHS of  \eqref{global_PeVHbound} vanishes as $N\to\infty$ only if the following constraints are satisfied:
 \begin{itemize}
 \item For  $\vect{B}=[1,1]$, 
 \begin{eqnarray}
 R_1+\Delta {R_1}{(11)}+R_2+ \Delta {R_2}{(11)}&\leq& 2n_d\label{IC_Global1}\\
  R_1+\Delta {R_1}{(11)}+R_2+ \Delta {R_2}{(11)}&\leq& (n_{d}-n_{c})^++\max(n_{d},n_{c}) \label{IC_Global3}\\
  R_1+\Delta {R_1}{(11)}+R_2+ \Delta {R_2}{(11)} &\leq& 2\max\{(n_{d}-n_{c})^+,n_{c}\}\label{IC_Global2}.
 \end{eqnarray}
 
 \item For $\vect{B}=[0,0]$, 
  \begin{eqnarray}
  R_1+\Delta {R_1}{(00)}+ R_2+\Delta {R_2}{(00)} &\leq& 2n_{d}.\label{P2P_global}
  \end{eqnarray}
  
   \item For $\vect{B}=[0,1]$, 
 \begin{eqnarray}
  R_1+\Delta {R_1}{(01)}+ R_2+\Delta {R_2}{(01)} &\leq& (n_{d}-n_{c})^++\max(n_{d},n_{c}).\label{S_global}
 \end{eqnarray} 
 
 \item For $\vect{B}=[1,0]$,  
 \begin{eqnarray}
 R_1+\Delta {R_1}{(10)}+R_2+\Delta {R_2}{(10)} &=& (n_{d}-n_{c})^++\max(n_{d},n_{c}).\label{Z_global}
 \end{eqnarray} 
 
 \end{itemize}
This proves the converse bounds in Theorem~\ref{Thm:conv_BIC_global}.

\subsubsection{Achievability Proof of Theorem~\ref{Thm:conv_BIC_global}}\label{Ach. apportunistic_global_uncorr}
 In this section, we present the achievability schemes for global CSIRT when $B_1$ and $B_2$ are independent. In contrast to the local CSIR/CSIRT case, we can adapt our transmission strategy to the interference states.
  
 When $\vect{B}=[0,0]$, the capacity-achieving scheme consists of sending uncoded bits in all $n_d$ level. We thus achieve the sum rate $R+\Delta {R}{(00)} =2n_d  \ \tfrac{\text{bits}}{\text{sub-channel use}}$.

 When $\vect{B}=[0,1]$ or $\vect{B}=[1,0]$, the achievability schemes coincide with the schemes described in Section~\ref{Ach. apportunistic}. In this case, we can only send opportunistic messages when we have VWI or WI.

\paragraph{Very Weak Interference}
Consider the achievability scheme depicted in Figure~\ref{Fig: VWI1}. By \eqref{RVWI_IC} and \eqref{RVWI_BIC},
 \begin{eqnarray}
 R_1+\Delta {R_1}{(01)}&=& R_2+\Delta R_2(10)=n_d \ \tfrac{\text{bits}}{\text{sub-ch.use}}\label{RVWI_BIC_guncorr1}\\
 R_1+\Delta {R_1}{(10)}&=& R_2+\Delta R_2(01)=n_d-n_c \tfrac{\text{bits}}{\text{sub-ch.use}}.\label{RVWI_BIC_guncorr11}
 \end{eqnarray}
 This proves the achievability bounds in Theorem~\ref{Thm:conv_BIC_global} for VWI.

\paragraph{Weak Interference}
 Consider the achievability scheme depicted in Figure~\ref{Fig: A2}a. By \eqref{RWI_IC} and \eqref{RWI_BIC},
 \begin{eqnarray}
 R_1+\Delta {R_1}{(01)}=R_2+\Delta {R_2}{(10)}
 &=& 2(n_d-n_c) \ \tfrac{\text{bits}}{\text{sub-ch.use}}\label{RWI_BIC_guncorr1}\\
 R_1+\Delta {R_1}{(10)}=R_2+\Delta {R_1}{(01)}
 &=& n_c\ \tfrac{\text{bits}}{\text{sub-ch.use}}.\label{RWI_BIC_guncorr11}
 \end{eqnarray}
 Combining \eqref{RWI_BIC_guncorr1} and \eqref{RWI_BIC_guncorr11}, we obtain the achievability  bounds in Theorem~\ref{Thm:conv_BIC_global} for WI.

\section{Proofs for the Ergodic Case}\label{Sec: Proof ER}
\makeatletter 
\setcounter{figure}{2} 
\@addtoreset{figure}{section}
\renewcommand{\thefigure}{B\arabic{figure}}
\makeatletter

\makeatletter 
\setcounter{table}{0} 
\@addtoreset{table}{section}
\renewcommand{\thetable}{B\arabic{table}}
\makeatletter

\makeatletter 
\setcounter{equation}{0} 
\@addtoreset{equation}{section}
\renewcommand{\theequation}{B\arabic{equation}}
\makeatletter

\subsubsection{Proof of \eqref{UB_TK}  in Theorem~\ref{Thm:converse-indCSI}}\label{Ap: UB_local_CSI}
The bound \eqref{UB_TK}  coincides with \cite[Th.~3.1]{Wang13}. However, \cite[Th.~3.1]{Wang13} derives \eqref{UB_TK} for the considered channel model with $T=1$ and feedback. In this section we show that \eqref{UB_TK} also holds  for general $T$ in the no-feedback case.
We follow along the lines of the proof of \cite[Thm~3.1]{Wang13}. We begin by applying Fano's inequality to obtain
\begin{myequation}
\label{UB_R}
\begin{aligned}
N(R_1-\epsilon_{1K})\leq&
 I(W_1;\mat{Y}_1^K\big|B_1^K)\\
 =& \sum\limits_{k=1}^{K}\left[H(\mat{Y}_{1,k}\big|\mat{Y}_{1}^{k-1},B_1^K)-H(\mat{Y}_{1,k}\big|W_1,\mat{Y}_{1}^{k-1},B_1^K)\right]\\
\stackrel{(a)}{=}&\sum_{k=1}^{K}\left[H(\mat{Y}_{1,k}\big|\mat{Y}_{1}^{k-1},B_{1,k},B_1^{k-1},B_{1,k+1}^K)-H(B_{1,k}\dmat{S}_{n_c}\mat{X}_{2,k}\big|\{B_{1,\ell}\dmat{S}_{n_c}\mat{X}_{2,\ell}\}_{\ell=1}^{k-1},W_1,B_1^K)\right]\\
=&\sum_{k=1}^{K}\Bigl[(1-p)H(\mat{Y}_{1,k}\big|\mat{Y}_{1}^{k-1},B_{1,k}=0,B_1^{k-1},B_{1,k+1}^K)+pH(\mat{Y}_{1,k}\big|\mat{Y}_{1}^{k-1},B_{1,k}=1,B_1^{k-1},B_{1,k+1}^K)\Bigr.\\
&\qquad\Bigl.{}-pH(\dmat{S}_{n_c}\mat{X}_{2,k}\big|\{B_{1,\ell}\dmat{S}_{n_c}\mat{X}_{2,\ell}\}_{\ell=1}^{k-1}, W_1,B_{1,k}=1,B_{1,k+1}^K,B_1^{k-1})\Bigr]\\
\stackrel{(b)}{\leq}&\sum_{k=1}^{K}\Bigl[(1-p)H(\dmat{S}_{n_d}\mat{X}_{1,k}|B_{1,k}=0)+pH(\mat{Y}_{1,k}|B_{1,k}=1)\\
&\qquad\Bigl.{}-pH(\dmat{S}_{n_c}\mat{X}_{2,k}\big|\{B_{1,\ell}\dmat{S}_{n_c}\mat{X}_{2,\ell}\}_{\ell=1}^{k-1},B_1^{k-1})\Bigr]
\end{aligned}
\end{myequation}
where $\epsilon_{1K} \to 0$ as $K\to\infty$. Here, $(a)$ follows  because ($W_1,B_1^K$) determine $\mat{X}_1^K$, so we can subtract the contribution of $\mat{X}_1^K$ in the second entropy and by evaluating the entropy for different interference states.  Step $(b)$ follows because $(B_1^{k-1},\vect{X}_2^k)$ are independent of $(B_{1,k}^K,W_1)$ (which in turn follows because $\mat {X}_2^K$ only depends on $(B_2^K,W_2)$, which is independent of $(B_1^K,W_1)$) 
and because conditioning reduces entropy.

Likewise, we have
\begin{equation}
\label{UB_pR}
\begin{aligned}
 N(R_2-\epsilon_{2K}) 
&\leq I(W_2;\mat{Y}_2^K\big|B_2^K)\\
&\stackrel{(a)}{\leq} I(W_2;\mat{Y}_1^K,\mat{Y}_2^K\big|W_1,B_1^K,B_2^K)\\
&=H(\mat{Y}_{1}^K,\mat{Y}_{2}^K\big|W_1,B_1^{K},B_2^{K})\\
&= \sum_{k=1}^{K}H(\mat{Y}_{1,k},\mat{Y}_{2,k}\big|W_1,B_1^{K},B_2^{K},\mat{Y}_{1}^{k-1},\mat{Y}_{2}^{k-1})\\
&\stackrel{(b)}{\leq} \sum_{k=1}^{K}H(\dmat{S}_{n_c}\mat{X}_{2,k},\dmat{S}_{n_d}\mat{X}_{2,k}\big|W_1,B_1^{K},\{B_{1,\ell}\dmat{S}_{n_c}\mat{X}_{2,\ell}\}_{\ell=1}^{k-1})\\
&\stackrel{(c)}{\leq} \sum_{k=1}^{K}\!\Bigl[H(\dmat{S}_{n_c}\mat{X}_{2,k}\big|\{B_{1,\ell}\dmat{S}_{n_c}\mat{X}_{2,\ell}\}_{\ell=1}^{k-1},B_1^{k-1}) +H(\dmat{S}_{n_d}\mat{X}_{2,k}\big|\dmat{S}_{n_c}\mat{X}_{2,k})\Bigr]
\end{aligned}
\end{equation}
where $\epsilon_{2K}\to 0$ as $K\to\infty$. Here, $(a)$ follows because $W_2$, $W_1$ and $B_1^K$ are independent. Step (b) follows because $(W_1,B_1^K)$ determines $\mat {X}_1^K$, so we can subtract its contribution from $(\mat{Y}_{1,k},\mat{Y}_{2,k})$, because $\mat{Y}_{1,k}\oplus \dmat{S}_{n_d}\mat {X}_{1,k}=B_{1,k}\dmat{S}_{n_c}\vect{X}_{2,k}$ has a lower entropy than $\dmat{S}_{n_c}\vect{X}_{2,k}$, and because conditioning reduces entropy. Step $(c)$ follows by the chain rule, and because conditioning reduces entropy.

Combining \eqref{UB_R} and \eqref{UB_pR} yields
\begin{myequation}\label{Eq: UB_(1+p)R}
\begin{aligned}
N( R_1+pR_2)-N(\epsilon_{1K}+p\epsilon_{2K}) \leq& \sum_{k=1}^{K}\!\Bigl[(1-p)H(\dmat{S}_{n_d}\mat{X}_{1,k}|B_{1,k}=0)\\
&\qquad{}+pH(\mat{Y}_{1,k}|B_{1,k}=1)+pH(\dmat{S}_{n_d}\mat{X}_{2,k}|\dmat{S}_{n_c}\mat{X}_{2,k})\Bigr].
\end{aligned}
\end{myequation}
By maximizing the individual entropies in \eqref{Eq: UB_(1+p)R} over all input distributions, dividing both sides of \eqref{Eq: UB_(1+p)R} by $N=KT$, and by letting then $K$ tend to infinity, we obtain that
\begin{align}\label{UB_TK1}
 &R_1 + pR_2 \leq (1-p) n_d + p [(n_d-n_c)^+ +\max(n_d,n_c)].\!
\end{align}

By symmetry, the same bound also holds for $R_2+pR_1$. Thus, by averaging over the two cases, it follows that \eqref{UB_TK1} is also an upper bound on $(R_1+R_2)(1+p)/2$. The final result \eqref{UB_TK} follows by dividing \eqref{UB_TK1} by $\tfrac{1+p}{2}$.

\subsubsection{Achievability Proof of Theorem~\ref{Thm:achiev-noCSI}}\label{Ap: Ach_Partial_CSI}
In this section, we describe the achievability schemes that yield the rates presented in Theorem~\ref{Thm:achiev-noCSI} for local CSIR. The bursty IC described in Section~\ref{Sec: ChannelModel} is treated here as a set of $n_d$ parallel sub-channels.

\paragraph{Scheme 1 (VWI; WI, MI for $0\leq p \leq \frac{1}{2}$)}\label{Ap:Ach-NoCSI-1}
The achievability scheme is illustrated in Figure~\ref{Fig: A3}a. In the figure, we present the normalized received signal at Rx$_1$, i.e., we represent graphically the time-$k$ channel output $\mat{Y}_{1,k}$ given by \eqref{Eq: Y1_det_int}, where the signal level from Tx$_1$ corresponds to $\dmat{S}_{n_d}\mat{X}_{1,k}$ and the signal level from Tx$_2$ corresponds to $\dmat{S}_{n_c}\mat{X}_{2,k}$, both normalized by $n_d$. In our scheme, the upper $n_d-n_c$ sub-channels (block \framebox[0.3cm][c]{\footnotesize$A$} in the figure) carry uncoded data (rate $1$ bits/sub-channel use), while in the lower $n_c$ channels (block \framebox[0.3cm][c]{\footnotesize$B$} in the figure) a capacity-achieving code of blocklength $N=KT$ for a \emph{binary erasure channel (BEC)} with erasure probability $p$ is used (with asymptotic rate $1-p$ bits/sub-channel use) \cite[Sec.~7.1.5]{Cover}. Block \framebox[0.3cm][c]{\footnotesize$A$} is received free of interference and can be directly decoded at the receiver. Block \framebox[0.3cm][c]{\footnotesize$B$} is affected by interference with probability (w.p.) $p$. Since the fading state $B_{i,k}$ is known to the $i$-th receiver, interfered slots are treated as erasures. Consequently, when $K$ tends to infinity, user $i$ achieves the rate $R_i = (n_d-n_c) + (1-p) n_c$. The sum rate $R$ is thus given by

\begin{align}
  R = 2 (n_d - p n_c), \quad n_d\geq n_c.
\end{align}
This scheme is tight for VWI and for WI and MI when $p\leq \tfrac{1}{2}$.

\begin{figure}[tbp]
	\centering
	\subfigure[ \label{Fig: VWeak}]{\includegraphics[width=0.23\linewidth]{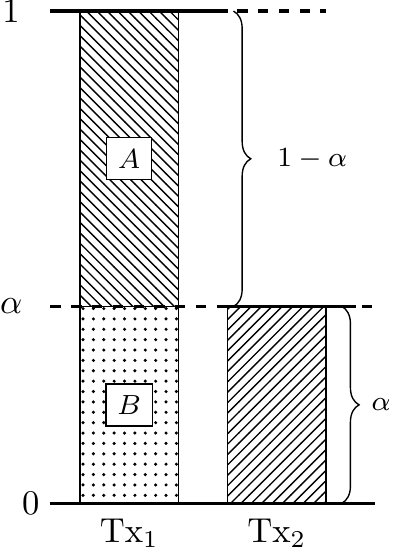}}\quad
  \subfigure[ \label{Fig: Weak2}]{\includegraphics[width=0.3\linewidth]{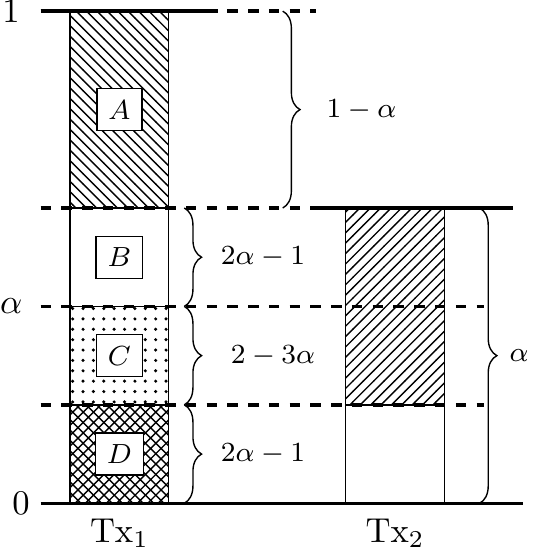}}\\	
	\caption{Normalized signal levels at Rx$_1$. (\textbf{a}) VWI; WI; MI, $p\leq\tfrac{1}{2}$; (\textbf{b}) WI, $p>\tfrac{1}{2}$.} \label{Fig: A3}
\end{figure}

\paragraph{Scheme 2 (WI, $\frac{1}{2} < p \leq 1$)}\label{Ap:Ach-NoCSI-2}
We next consider the achievability scheme illustrated in Figure~\ref{Fig: A3}b. In blocks \framebox[0.3cm][c]{\footnotesize$A$} and \framebox[0.3cm][c]{\footnotesize$B$} uncoded data is transmitted (rate $1$ bits/sub-channel use), block \framebox[0.3cm][c]{\footnotesize$C$} carries the deterministic all-zeros sequence (rate $0$ bit/sub-channel use) and in block \framebox[0.3cm][c]{\footnotesize$D$} a capacity-achieving code for the BEC (with asymptotic rate $1-p$ bits/sub-channel use) is used. As in Scheme~1, blocks \framebox[0.3cm][c]{\footnotesize$A$} and \framebox[0.3cm][c]{\footnotesize$B$} can be decoded without interference, and  block \framebox[0.3cm][c]{\footnotesize$D$} is decoded by treating interfered symbols as erasures. The rate achieved by this scheme at user $i$ is $R_i=(n_d-n_c) + (2n_c-n_d) + (1-p)(2n_d-3n_c)$, so
\begin{align}
  R = 4(n_d-n_c) + p (6n_c-4n_d), \quad \tfrac{2n_d}{3} \geq n_c\geq \tfrac{n_d}{2}.
\end{align}

\paragraph{Scheme 3 (SI, $0 \leq p \leq \frac{1}{2}$)}\label{Ap:Ach-NoCSI-3}
We use an achievability scheme similar to Scheme~1. Now, the upper $2n_d-n_c$ sub-channels carry a capacity-achieving code for a BEC with erasure probability $p$, and the lower $n_c-n_d$ sub-channels carry uncoded data. Consequently, when $K$ tends to infinity, user $i$ achieves the rate $R_i = (n_c-n_d) + (1-p) (2n_d-n_c)$. The sum rate $R=R_1+R_2$ is thus given by
\begin{align}
  R = 2(1-2p)n_d + 2p n_c, \quad 2n_d \geq n_c \geq n_d.
\end{align}
This proves Theorem~\ref{Thm:achiev-noCSI}.

\subsubsection{Proof of Theorem~\ref{Thm:converse-fullCSI}}\label{Proof_UBF}
In this section, we prove the converse bounds for global CSIRT and independent $B_1^K$ and $B_2^K$.

\paragraph{Converse Bound \eqref{UBF1} for Global CSIRT}\label{Proof_UBF1}
By Fano's inequality, we have

\begin{equation}\label{UB_R1F}
\begin{aligned}
 N(R_1-\epsilon_{1K})&  
  \leq I(W_1;\mat{Y}_1^K|\vect{B}^K)\\
&\stackrel{(a)}{=}\sum_{k=1}^{K} \left[H(\mat{Y}_{1,k}|\mat{Y}_{1}^{k-1},\vect{B}^K)-H(B_{1,k}\dmat{S}_{n_c}\mat {X}_{2,k}|W_1,\mat{Y}_{1}^{k-1},\vect{B}^K)\right]\\
&=\sum_{k=1}^{K} \Bigl[(1-p)H(\mat {Y}_{1,k}|\mat{Y}_{1}^{k-1},B_{1,k}=0,B_{1}^{k-1},B_{1,k+1}^K,B_2^K)\Bigr.\\
&\qquad\quad{}+pH(\mat {Y}_{1,k}|\mat{Y}_{1}^{k-1},B_{1,k}=1,B_{1}^{k-1},B_{1,k+1}^K,B_2^K)\\
&\qquad\quad\Bigl.{}-pH(\dmat{S}_{n_c}\mat {X}_{2,k}|W_1,\mat{Y}_{1}^{k-1},B_{1,k}=1,B_{1}^{k-1},B_{1,k+1}^K,B_2^K)\Bigr]\\
&\leq\sum_{k=1}^{K}\Bigl[(1-p)H(\dmat{S}_{n_d}\mat{X}_{1,k}|B_{1,k}=0)+pH(\mat{Y}_{1,k}|B_{1,k}=1)\\
&\qquad\quad\Bigl.{}-pH(\dmat{S}_{n_c}\mat{X}_{2,k}|W_1,\mat{Y}_{1}^{k-1},B_{1,k}=1,B_1^{k-1},B_{1,k+1}^{K},B_2^{K})\Bigr]
\end{aligned}
\end{equation}
where $\epsilon_{1K} \to 0$ as $K\to\infty$. Here, $(a)$ follows because $(W_1,\vect{B}^K)$ determines $\mat {X}_{1,k}$, so we can subtract its contribution from the second entropy.  Likewise,
\begin{myequation}\label{UB_R2F}
\begin{aligned}
N(R_2-\epsilon_{2K}) 
\leq & I(W_2;\mat{Y}_2^K|\vect{B}^K)\\
\stackrel{(a)}{\leq} &I(W_2;\mat{Y}_1^K,\mat{Y}_2^K|W_1,\vect{B}^K)\\
{=}& \sum_{k=1}^{K} H(\mat{Y}_{1,k},\mat{Y}_{2,k}|W_1,\mat{Y}_{1}^{k-1},\mat{Y}_{2}^{k-1},\vect{B}^K)\\
\stackrel{(b)}{\leq} &\sum_{k=1}^{K}H(B_{1,k}\dmat{S}_{n_c}\mat{X}_{2,k},\dmat{S}_{n_d}\mat{X}_{2,k}|W_1,\mat{Y}_{1}^{k-1},\vect{B}^{K})\\
 \stackrel{(c)}{\leq}&
\sum_{k=1}^{K}\!\Bigl[(1-p)H(\dmat{S}_{n_d}\mat{X}_{2,k}|,B_{1,k}=0) +pH(\dmat{S}_{n_c}\mat{X}_{2,k}|W_1,\mat{Y}_{1}^{k-1},B_{1,k}=1,B_1^{k-1},B_{1,k+1}^{K},B_{2}^{K})\Bigr.\\
& \qquad\,\Bigl. 
{}+pH(\dmat{S}_{n_d}\mat{X}_{2,k}|\dmat{S}_{n_c}\mat{X}_{2,k},B_{1,k}=1)\Bigr]
\end{aligned}
\end{myequation}
where $\epsilon_{2K}\to 0$ as $K\to\infty$. Here, step $(a)$ follows because $W_2$ and $(W_1,B_1^K)$ are independent. Step $(b)$ follows because $(W_1,\vect{B}^K)$ determines $\mat {X}_{1,k}$, so we can subtract its contribution from $\mat{Y}_{1,k}$ and $\mat{Y}_{2,k}$, and because conditioning reduces entropy. Step $(c)$ follows by evaluating the entropies for different interference states and because conditioning reduces entropy.
Combining \eqref{UB_R1F} and \eqref{UB_R2F} yields
\begin{equation}
\label{Eq: UB_(1+p)RF}
\begin{aligned}
 N(R_1+R_2)-N(\epsilon_{1K}+\epsilon_{2K})
&\leq \sum_{k=1}^{K}\!\left[(1-p)\left(H(\dmat{S}_{n_d}\mat{X}_{1,k}|B_{1,k}=0)+H(\dmat{S}_{n_d}\mat{X}_{2,k}|B_{1,k}=0)\right)\right.\\
&\qquad\quad\left.{}+pH(\mat{Y}_{1,k}|B_{1,k}=1)+pH(\dmat{S}_{n_d}\mat{X}_{2,k}|\dmat{S}_{n_c}\mat{X}_{2,k},B_{1,k}=1)\right].
\end{aligned}
\end{equation}
By maximizing the entropies in \eqref{Eq: UB_(1+p)RF} over all input distributions, dividing by $N=KT$, and  letting $K$ tend to infinity, we obtain that
\begin{equation}\label{UB_TKF1}
R\leq 2(1-p) n_d +  p \max(n_d,n_c) + p (n_d-n_c)^+
\end{equation}
which is \eqref{UBF1}.

\paragraph{Converse Bound \eqref{UBF2} for Global CSIRT}\label{Proof_UBF2}

Let $\vect{b}^K$ denote the realizations of the interference states $\vect{B}^K$. We label the set of time indices where the pair $({b}_{1,k},{b}_{2,k})$ takes the value (0,1) by $\dmat {A}$; (1,1) by $\dmat {B}$;  (1,0) by $\dmat {C}$; and  (0,0) by $\dmat {D}$. We denote the length of each of these states by $j_{\dmat{A}}, j_{\dmat{B}}, j_{\dmat{C}}$ and $j_{\dmat{D}}$, respectively. For example, 
\begin{equation*}
\dmat{A}\eqdef\{i=1,\ldots,K:\vect{b}_k=[1,1]\}
\end{equation*}
and 
\begin{equation*}
\displaystyle{j_{\dmat{A}}=\sum_{k^=1}^K\mathds{1}\{\vect{B}=[1,1]\}}.
\end{equation*} 
These states are schematically  shown in Figure~\ref{Fig: Full CSI}, where shaded areas correspond to $b_i=1$.

\begin{figure}[tbp]
		\centering
		\includegraphics[width=0.5\linewidth]{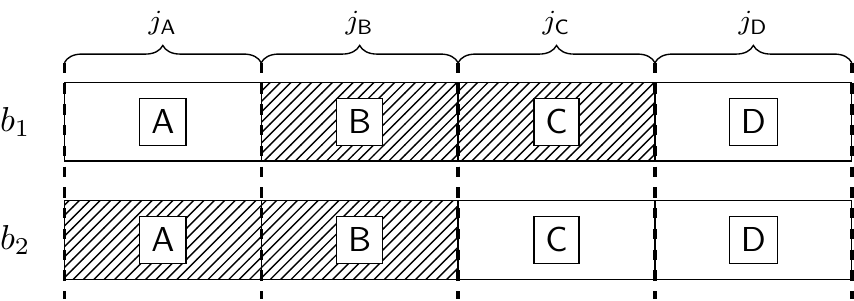}
		\caption{Possible interference states.} 
		\label{Fig: Full CSI}
\end{figure}

For global CSIRT, $(\mat {X}_1^K, \mat {X}_2^K)$ may depend on $\vect{B}^K=\vect{b}^K$. We shall denote by $\mat{X}_i^{\dmat{A}}, \mat{X}_i^{\dmat{B}}, \mat{X}_i^{\dmat{C}}$ and $\mat{X}_i^{\dmat{D}}$ the $\mat{X}_{1,k}$'s with indices in $\dmat{A},\dmat{B}, \dmat{C}$ and $\dmat{D}$. For example, $\mat {X}_i^{\dmat{A}}=\{\mat {X}_{i,k}: k\in\dmat{A}\}$. At time $k$, the interference states $\vect{B}_{k}=\vect {b}_k$ can be in one of the 4 possible cases, as depicted in Figure~\ref{Fig: Full CSI}. The converse bound \eqref{UBF2} is proved as follows. We begin by applying Fano's inequality to obtain

\begin{equation}
\label{Eq:UB_FullCSI_VWI_WI}
\begin{aligned}
&N(R_1+R_2)-N(\epsilon_{1K}+\epsilon_{2K})\\
&\qquad\qquad\qquad \leq I(W_1;\mat{Y}_1^K|\vect{B}^K)+I(W_2;\mat{Y}_2^K|\vect{B}^K)\\
&\qquad\qquad\qquad=\sum\limits_{\vect {b}\in\{{0},{1}\}^K}\mathcal{P}(\vect{B}=\vect {b}^K)\left [ I(W_1;\mat{Y}_1^K|\vect{B}^K=\vect {b}^K)+I(W_2;\mat{Y}_2^K|\vect{B}^K=\vect {b}^K)\right] 
\end{aligned}
\end{equation}
where $\epsilon_{1K}\to 0$ and $\epsilon_{2K}\to 0$ as $N\to \infty$. For every $\vect{b}^K$, we have
\begin{equation}
\label{Eq:Full_I1}
\begin{aligned}
& I(W_1;\mat{Y}_1^K|\vect{B}^K\!=\!\vect {b}^K)+\!I(W_2;\mat{Y}_2^K|\vect{B}^K\!=\!\vect {b}^K)\\
&=H(\mat{Y}_1^K|\vect{B}^K\!=\vect {b}^K)\!-\!H(\mat{Y}_1^K|W_1,\vect{B}^K\!=\!\vect {b}^K)\!+\!H(\mat{Y}_2^K|\vect{B}^K\!=\!\vect {b}^K)\!-\!H(\mat{Y}_2^K|W_2,\vect{B}^K\!=\!\vect {b}^K)\\
& \stackrel{(a)}{=}H(\mat{Y}_1^{\dmat{C}}|\vect{B}^K\!=\!\vect {b}^K)\!+\!H(\mat{Y}_1^{\dmat{A}}\!,\!\mat{Y}_1^{\dmat{B}}|\mat{Y}_1^{\dmat{C}},\vect{B}^K\!=\!\vect {b}^K)+H(\mat{Y}_1^{\dmat{D}}|\mat{Y}_1^{\dmat{A}},\mat{Y}_1^{\dmat{B}},\mat{Y}_1^{\dmat{C}},\vect{B}^K\!=\!\vect {b}^K)\\
&\quad\:\, {}-H(\dmat{S}_{n_c}\mat{X}_2^{\dmat{B}},\dmat{S}_{n_c}\mat{X}_2^{\dmat{C}}|\vect{B}^K\!=\!\vect {b}^K)\\
&\quad\:\,{}+H(\mat{Y}_2^{\dmat{A}}|\vect{B}^K\!=\!\vect {b}^K)+H(\mat{Y}_2^{\dmat{B}}\!,\!\mat{Y}_2^{\dmat{C}}|\mat{Y}_2^{\dmat{A}},\vect{B}^K\!=\!\vect {b}^K)+H(\mat{Y}_2^{\dmat{D}}|\mat{Y}_2^{\dmat{A}}, \mat{Y}_2^{\dmat{B}},\mat{Y}_2^{\dmat{C}},\vect{B}^K\!=\!\vect {b}^K)\\
&\quad\:\,{}-H(\dmat{S}_{n_c}\mat{X}_1^{\dmat{A}},\dmat{S}_{n_c}\mat{X}_1^{\dmat{B}}|\vect{B}^K\!=\!\vect {b}^K)\\
&\stackrel{(b)}{\leq}H(\mat{Y}_1^{\dmat{C}}|\vect{B}^K\!=\!\vect {b}^K)\!+\!H(\mat{Y}_1^{\dmat{A}}\!,\!\mat{Y}_1^{\dmat{B}}|\vect{B}^K\!=\!\vect {b}^K)+H(\mat{Y}_1^{\dmat{D}}|\vect{B}^K\!=\!\vect {b}^K)-H(\dmat{S}_{n_c}\mat{X}_2^{\dmat{B}},\dmat{S}_{n_c}\mat{X}_2^{\dmat{C}}|\vect{B}^K\!=\!\vect {b}^K)\\
&\quad\:\,{}+H(\mat{Y}_2^{\dmat{A}}|\vect{B}^K\!=\!\vect {b}^K)+H(\mat{Y}_2^{\dmat{B}}\!,\!\mat{Y}_2^{\dmat{C}}|\vect{B}^K\!=\!\vect {b}^K)+H(\mat{Y}_2^{\dmat{D}}|\vect{B}^K\!=\!\vect {b}^K)-H(\dmat{S}_{n_c}\mat{X}_1^{\dmat{A}},\dmat{S}_{n_c}\mat{X}_1^{\dmat{B}}|\vect{B}^K\!=\!\vect {b}^K)
\end{aligned}
\end{equation}
where step $(a)$ follows by the chain rule for entropy and because $(W_1, \vect{B}^K)$ determines $\mat {X}_1^K$, so we can subtract its contribution from the second and fourth entropy. Step $(b)$ follows because conditioning reduces entropy. We next upper-bound \eqref{Eq:Full_I1} by combining the positive and negative entropies in areas $\dmat{B}$ and $\dmat{C}$ for user 1 and user 2; and areas $\dmat{A}$ and $\dmat{B}$ for user 2 and user 1:
\begin{equation}
\label{Eq:Full_I2}
\begin{aligned}
 & I(W_1;\mat{Y}_1^K|\vect{B}^K\!=\!\vect {b}^K)\!+\!I(W_2;\mat{Y}_2^K|\vect{B}^K\!=\!\vect {b}^K)\\
&\qquad\stackrel{(a)}{\leq} H(\mat{Y}_1^{\dmat{C}}|\vect{B}^K\!=\!\vect {b}^K)+H(\mat{Y}_1^{\dmat{A}},\mat{Y}_1^{\dmat{B}}|\dmat{S}_{n_c}\mat{X}_1^{\dmat{A}},\dmat{S}_{n_c}\mat{X}_1^{\dmat{B}},\vect{B}^K\!=\!\vect {b}^K)+H(\mat{Y}_1^{\dmat{D}}|\vect{B}^K\!=\!\vect {b}^K)\\
&\qquad\quad\:\, {}+H(\mat{Y}_2^{\dmat{A}}|\vect{B}^K\!=\!\vect {b}^K)+H(\mat{Y}_2^{\dmat{B}}, \mat{Y}_2^{\dmat{C}}|\dmat{S}_{n_c}\mat{X}_2^{\dmat{B}},\dmat{S}_{n_c}\mat{X}_2^{\dmat{C}},\vect{B}^K\!=\!\vect {b}^K)+H(\mat{Y}_2^{\dmat{D}}|\vect{B}^K\!=\!\vect {b}^K)\\
&\qquad\;{\leq}\; H(\mat{Y}_1^{\dmat{C}}|\vect{B}^K\!=\!\vect {b}^K)+H(\mat{Y}_1^{\dmat{A}}|\dmat{S}_{n_c}\mat{X}_1^{\dmat{A}},\vect{B}^K\!=\!\vect {b}^K)+H(\mat{Y}_1^{\dmat{B}}|\dmat{S}_{n_c}\mat{X}_1^{\dmat{B}},\vect{B}^K\!=\!\vect {b}^K)\\
&\qquad\;\quad\,{}+H(\mat{Y}_1^{\dmat{D}}|\vect{B}^K\!=\!\vect {b}^K)+H(\mat{Y}_2^{\dmat{A}}|\vect{B}^K\!=\!\vect {b}^K)+H(\mat{Y}_2^{\dmat{B}}|\dmat{S}_{n_c}\mat{X}_2^{\dmat{B}},\vect{B}^K\!=\!\vect {b}^K)\\
&\qquad\;\quad\,{}+H(\mat{Y}_2^{\dmat{C}}|\dmat{S}_{n_c}\mat{X}_2^{\dmat{C}},\vect{B}^K\!=\!\vect {b}^K)+H(\mat{Y}_2^{\dmat{D}}|\vect{B}^K\!=\!\vect {b}^K)
\end{aligned}
\end{equation}
where step $(a)$ follows because $H(F)-H(G)\leq H(F|G)$ for any random variables $F$ and $G$. 
By maximizing the entropies in \eqref{Eq:Full_I2} over all input distributions, we obtain 
\begin{equation}
\label{Eq:UB_FullCSI_VWI_WI_1}
\begin{aligned}
 I(W_1;\mat{Y}_1^K|\vect{B}^K\!=\!\vect {b}^K)\!+\!I(W_2;\mat{Y}_2^K\vect{B}^K\!=\!\vect {b}^K)
&\leq j_{\dmat{A}}T[(n_{d}-n_{c})^++\max(n_{d},n_{c})]\\
&\quad{}+2j_{\dmat{B}}T\max\{(n_{d}-n_{c})^+,n_{d}\}\\
&\quad{}+j_{\dmat{C}}T[(n_{d}-n_{c})^++\max(n_{d},n_{c})]+2j_{\dmat{D}}T(n_{d}).
\end{aligned}
\end{equation}
By dividing \eqref{Eq:UB_FullCSI_VWI_WI_1} by $N=KT$, and taking the limit as $K\to\infty$, we obtain
\begin{equation}
\label{Eq:UB_full_CSI_a<2/3}
\begin{aligned}
R_1+R_2
& \stackrel{(a)}{\leq} \lim\limits_{K\to\infty}\frac{1}{KT}\sum_{\vect{b}^K}\mathcal{P}\{\vect{B}^K\!=\!\vect {b}^K\}\Bigl[I(W_1;\mat{Y}_1^K|\vect{B}^K\!=\!\vect {b}^K)+I(W_2;\mat{Y}_2^K|\vect{B}^K\!=\!\vect {b}^K)\Bigr]\\
&=\lim\limits_{K\to\infty}\frac{1}{K}\Bigl[\E{j_{\mathsf{A}}[(n_{c}-n_{d})^++\max(n_{d},n_{c})]+2j_{\mathsf{B}}[\max\{(n_{d}-n_{c})^+,n_{c}\}]}\\
&\qquad\qquad\;\,\,\,\Bigl.{}+\E{j_{\mathsf{C}}[(n_{d}-n_{c})^++\max(n_{d},n_{c})]+2j_{\mathsf{D}}n_{d}}\Bigr]
\end{aligned}
\end{equation}
where $(a)$ follows because $(\epsilon_{1K}+\epsilon_{2K})\to 0$ as $K\to\infty$. Next, we apply the \emph{dominated convergence theorem (DCT)} \cite[Sec.~1.34]{Rudin_1987} to interchange limit and expectation. By the law of large numbers, we have that $\frac{j_{\mathsf{A}}}{K}\to p(1-p)$, $\frac{j_{\mathsf{B}}}{K}\to p^2$,
$\frac{j_{\mathsf{C}}}{K}\to p(1-p)$, and
$\frac{j_{\mathsf{B}}}{K}\to (1-p)^2$ almost surely as $K\to\infty$. By replacing these probabilities in \eqref{Eq:UB_full_CSI_a<2/3}, we thus obtain
\begin{equation}\label{Eq:UB_full_CSI_a<2/3_1}
R
\leq 2p(1-p)[(n_{d}-n_{c})^++\max(n_{d},n_{c})]+2p^2\max\{(n_{d}-n_{c})^+,n_{c}\}+2(1-p)^2n_d.
\end{equation}
This yields \eqref{UBF2}.

\subsubsection{Proof of Theorem~\ref{Thm:achiev-fullCSI}}
\label{Ap: Ach_Full_CSI}
In this section, we present the achievability schemes for global CSIRT and independent $B_1^K$ and $B_2^K$. Let $\vect{b}^K$ denote the realizations of the interference states $\vect{B}^K$, and define $j_{\min} \triangleq \min(j_{\mathsf{A}}, j_{\mathsf{B}}, j_{\mathsf{C}})$. Consider the following achievable schemes. 

\paragraph{Scheme 1 (MI, $ 0 \leq p \leq 1$)}
\label{Ap:Ach-FullCSI-1}
Both transmitters employ uncoded transmission in the first $j_{\min}$ indices of regions $\mathsf{A}$ and $\mathsf{C}$, respectively, and in the whole region $\mathsf{D}$. Tx$_1$ copies the first $j_{\min}$ indices of region $\mathsf{A}$ in region $\mathsf{B}$, while Tx$_2$ copies the first $j_{\min}$ indices of region $\mathsf{C}$ in $\mathsf{B}$, aligned with those of user 1. The remaining indices are treated as a non-bursty IC attaining rate $r_{\text{ic}}=n_d-\frac{n_c}{2}$~\cite{jafar10}.

To illustrate the decoding process, Figure~\ref{Fig: WI_memory} shows the different normalized signals at the Rx$_1$ when $j_{\dmat{A}}=j_{\dmat{B}}=j_{\dmat{C}}=j_{\dmat{D}}=1$. 
Tx$_1$ transmits the signals \framebox[0.3cm][c]{\footnotesize $1$} , \framebox[0.3cm][c]{\footnotesize $3$} , and \framebox[0.3cm][c]{\footnotesize $4$}, in channel state $\dmat{A}$ and $\dmat{B}$, $\dmat{C}$, and $\dmat{D}$, respectively. Similarly, Tx$_2$ transmits the signal \framebox[0.3cm][c]{\footnotesize $2$} in states $\dmat{B}$ and $\dmat{C}$. Rx$_1$ has access to a clean copy of signal \framebox[0.3cm][c]{\footnotesize $1$} in region $\mathsf{A}$, which can then be subtracted in state $\dmat{B}$ to recover the interfering signal \framebox[0.3cm][c]{\footnotesize $2$}. Since Tx$_2$ transmits the same signal in state $\dmat{C}$, the interference can then be canceled. Hence, signals \framebox[0.3cm][c]{\footnotesize $3$} and \framebox[0.3cm][c]{\footnotesize $4$} are recovered. 
For a given interference state  and general $\dmat{A}$ and $\dmat{B}$, $\dmat{C}$, and $\dmat{D}$,  the rate attained by user $i$ with this scheme is
\begin{align}
  R_i(\vect{b}^K) = n_d \tfrac{2j_{\min}}{K}  + n_d \tfrac{j_{\mathsf{D}}}{K} + r_{\text{ic}} \tfrac{j_{\mathsf{A}}+j_{\mathsf{B}}+j_{\mathsf{C}}-3 j_{\min}}{K}. \label{eqn:WI-memory-1}
\end{align}

\begin{figure}[tbp]
\centering
\includegraphics[width=0.6\linewidth]{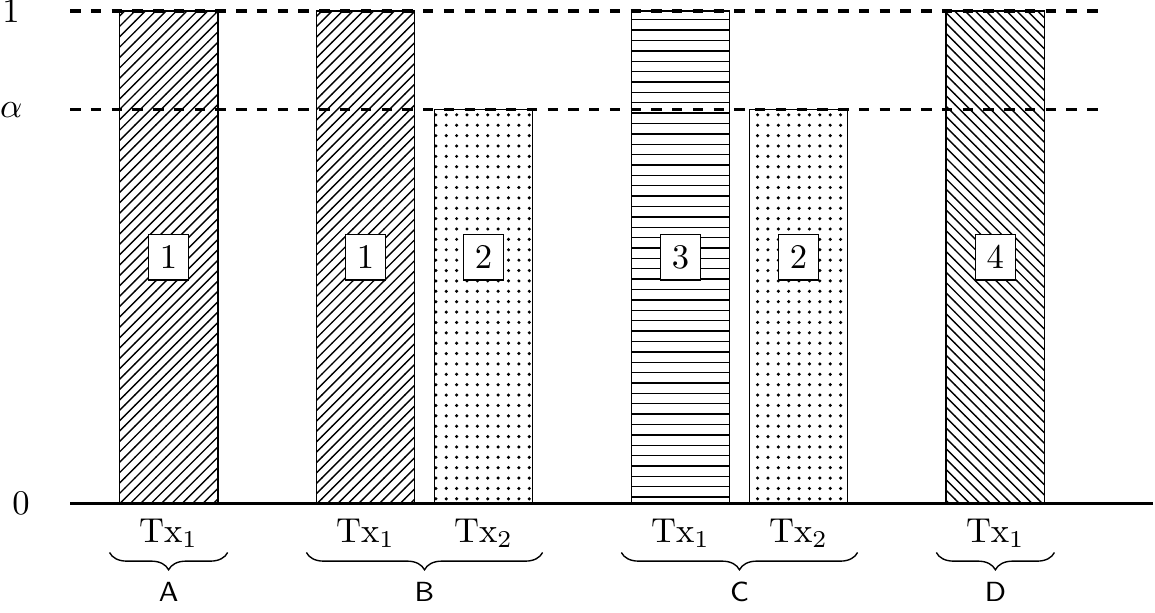}
\caption{Normalized by $n_d$ signal levels at Rx$_1$ for MI and $j_{\mathsf{A}}=j_{\mathsf{B}}=j_{\mathsf{C}}=j_{\mathsf{D}}$.}
\label{Fig: WI_memory}
\end{figure}
Averaging \eqref{eqn:WI-memory-1} over $\vect{B}^K$, and letting $K\to\infty$, we obtain for the sum rate
\begin{equation}
\begin{aligned}
  R 
  &= \lim_{K\to\infty} 2\E{n_d \tfrac{2j_{\min}}{K}  + n_d \tfrac{j_{\mathsf{D}}}{K} + r_{\text{ic}} \tfrac{j_{\mathsf{A}}+j_{\mathsf{B}}+j_{\mathsf{C}}-3 j_{\min}}{K}}\\
  &= 4 n_d p_{\min}  +  2n_d (1-p)^2  +  \bigl(2n_d-{n_c}\bigr)\bigl(2p-p^2-3  p_{\min})\label{eqn:WI-memory-3}
\end{aligned}
\end{equation}
where we changed the order of limit and expectation by appealing to the DCT, and used that, by the law of large numbers, $\frac{J_{\mathsf{A}}}{K}\rightarrow p(1-p)$, $\frac{J_{\mathsf{B}}}{K}\rightarrow p^2$, $\frac{J_{\mathsf{C}}}{K}\rightarrow p(1-p)$ and $\frac{J_{\mathsf{D}}}{K}\rightarrow (1-p)^2$ almost surely as~$K\rightarrow\infty$.

\paragraph{Scheme 2 (SI, $0 \leq p \leq 1$)}
\label{Ap:Ach-FullCSI-2}
Both transmitters employ uncoded transmission in the first $j_{\min}$ indices of states $\mathsf{A}$ and $\mathsf{C}$. Tx$_1$ copies the lowest $2n_d-n_c$ sub-channels of the first $j_{\min}$ indices of region $\mathsf{A}$  into the highest $2n_d-n_c$ sub-channels and uses uncoded transmission in the lowest $n_c-n_d$ sub-channels of the corresponding sub-region in $\mathsf{B}$. Tx$_2$ proceeds analogously but from  region $\mathsf{C}$ to $\mathsf{B}$. Both transmitters employ uncoded transmission in region $\mathsf{D}$ and treat the remaining indices as a non-bursty IC~\cite{jafar10} with rate $\frac{n_c}{2}$. 

To illustrate the decoding process, Figure~\ref{Fig: SI_memory} shows the different normalized signals at the Rx$_1$ when $j_{\dmat{A}}=j_{\dmat{B}}=j_{\dmat{C}}=j_{\dmat{D}}=1$.
Tx$_1$ transmits the signals (\framebox[0.3cm][c]{\footnotesize $1$}~, \framebox[0.3cm][c]{\footnotesize $2$})~, (\framebox[0.3cm][c]{\footnotesize $1$}~, \framebox[0.3cm][c]{\footnotesize $3$})~, \framebox[0.3cm][c]{\footnotesize $5$} and \framebox[0.3cm][c]{\footnotesize $6$} in channel state $\dmat{A}$ and $\dmat{B}$, $\dmat{C}$, and $\dmat{D}$, respectively. Similarly, Tx$_2$ transmits the signal (\framebox[0.3cm][c]{\footnotesize $4$}~, \framebox[0.3cm][c]{\footnotesize $7$}) and (\framebox[0.3cm][c]{\footnotesize $4$}~, \framebox[0.3cm][c]{\footnotesize $8$}) in states $\dmat{B}$ and $\dmat{C}$, respectively. Rx$_1$ has access to a clean copy of signals \framebox[0.3cm][c]{\footnotesize $1$} and \framebox[0.3cm][c]{\footnotesize $2$} in region $\mathsf{A}$, signal \framebox[0.3cm][c]{\footnotesize $1$} can then be subtracted in state $\dmat{B}$ to recover the interfering signals \framebox[0.3cm][c]{\footnotesize $4$} and \framebox[0.3cm][c]{\footnotesize $7$}. In state $\dmat{B}$, Rx$_1$ has access to signal \framebox[0.3cm][c]{\footnotesize $3$}. Since Tx$_2$ transmits signal \framebox[0.3cm][c]{\footnotesize $4$} in state $\dmat{C}$, the interference can then be canceled. Hence, signal \framebox[0.3cm][c]{\footnotesize $5$} can be recovered. Finally, signal \framebox[0.3cm][c]{\footnotesize $6$} is recovered without interference.
For a given interference state, and general $j_{\mathsf{A}}, j_{\mathsf{B}},j_{\mathsf{C}},j_{\mathsf{D}}$, the rate attained by user $i$ with this scheme is
\begin{align}
  R_i(\vect{b}^K) = (n_d+n_c) \tfrac{2j_{\min}}{K}  + n_d \tfrac{j_{\mathsf{D}}}{K} + r_{\text{ic}} \tfrac{j_{\mathsf{A}}+j_{\mathsf{B}}+j_{\mathsf{C}}-3 j_{\min}}{K}. \label{eqn:SI-memory-1}
\end{align}
Averaging \eqref{eqn:SI-memory-1} over $\vect{B}^K$, and letting $K\to\infty$, we obtain for the sum rate
\begin{align}
  R
  &= \lim_{K\to\infty} 2\E{(n_d+n_c) \tfrac{2j_{\min}}{K}  + n_d \tfrac{j_{\mathsf{D}}}{K} + r_{\text{ic}} \tfrac{j_{\mathsf{A}}+j_{\mathsf{B}}+j_{\mathsf{C}}-3 j_{\min}}{K}}\notag\\
  &= 2(n_d+n_c) p_{\min}  + 2n_d (1-p)^2 + {n_c} \bigl(2p - p^2 - 3  p_{\min}\bigr).\label{eqn:SI-memory-2}
\end{align}
where we changed the order of limit and expectation by appealing to the DCT, and used that, by the law of large numbers, $\frac{J_{\mathsf{A}}}{K}\rightarrow p(1-p)$, $\frac{J_{\mathsf{B}}}{K}\rightarrow p^2$, $\frac{J_{\mathsf{C}}}{K}\rightarrow p(1-p)$ and $\frac{J_{\mathsf{D}}}{K}\rightarrow (1-p)^2$ almost surely as $K\rightarrow\infty$.

\begin{figure}[tbp]
\centering
\includegraphics[width=0.6\linewidth]{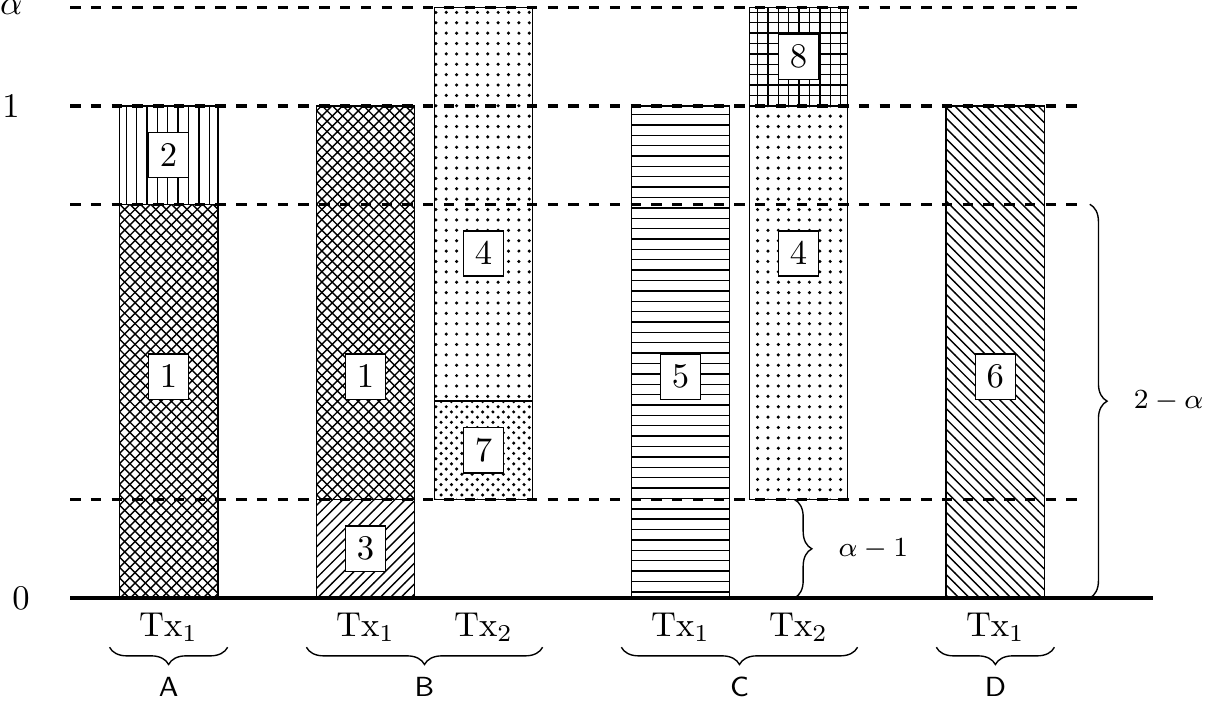}
\caption{Normalized by $n_d$ signal levels at Rx$_1$ for SI.}
\label{Fig: SI_memory}
\end{figure}

\subsubsection{Proof of Theorem~\ref{Thm:converse-globalCSI_corr}}\label{Proof_Global_corr}
The converse bound \eqref{UBF_corr1} for global CSIRT follows similar steps as in Appendix~\ref{Proof_UBF1}  but considering $\displaystyle{B_1^K=B_2^K=B^K}$.
We next present the converse bound \eqref{UBF_corr2} for global CSIRT when $B_1^K=B_2^K$. This bound follows by giving the  extra information $(B^K\dmat{S}_{n_c}\mat {X}_1^K)$ to Rx$_1$. By Fano's inequality, we~have
\begin{equation}
\label{UB_R1F2}
\begin{aligned}
N(R_1-\epsilon_{1K}) 
& \leq I(W_1;\mat{Y}_1^K|{B}^K)\\
& \leq I(W_1;\mat{Y}_1^K,B^K\dmat{S}_{n_c}\mat{X}_1^K|{B}^K)\\
&  = I(W_1;B^K\dmat{S}_{n_c}\mat{X}_1^K|{B}^K)+I(W_1;\mat{Y}_1^K|B^K\dmat{S}_{n_c}\mat{X}_1^K,{B}^K)\\
 & =  H(B^K\dmat{S}_{n_c}\mat{X}_1^K|{B}^K)+H(\mat{Y}_1^K|B^K\dmat{S}_{n_c}\mat{X}_1^K,{B}^K)-H(\mat{Y}_1^K|W_1,B^K\dmat{S}_{n_c}\mat{X}_1^K,{B}^K)\\
& =  H(B^K\dmat{S}_{n_c}\mat{X}_1^K|{B}^K)+H(\mat{Y}_1^K|B^K\dmat{S}_{n_c}\mat{X}_1^K,{B}^K)-H(B^K\dmat{S}_{n_c}\mat{X}_2^K|{B}^K)
\end{aligned}
\end{equation}
where $\epsilon_{1K}\to 0$ as $K\to \infty$.
 Analogously, by giving the extra information $(B^K\dmat{S}_{n_c}\mat {X}_2^K)$ to Rx$_2$, we obtain
\begin{align}
\label{UB_R2F2}
N(R_2-\epsilon_{2K}) 
{\leq}  H(B^K\dmat{S}_{n_c}\mat{X}_2^K|{B}^K)+H(\mat{Y}_2^K|B^K\dmat{S}_{n_c}\mat{X}_2^K,{B}^K)-H(B^K\dmat{S}_{n_c}\mat{X}_1^K|{B}^K)
\end{align}
where $\epsilon_{2K}\to 0$ as $K\to \infty$. Thus, \eqref{UB_R1F2} and \eqref{UB_R2F2} yield
\begin{equation}
\label{Eq: UB_RF2}
\begin{aligned}
&N(R_1+R_2)-N(\epsilon_{1K}+\epsilon_{2K})\\
&\qquad\qquad\leq  H(\mat{Y}_1^K|B^K\dmat{S}_{n_c}\mat{X}_1^K,{B}^K)+H(\mat{Y}_2^K|B^K\dmat{S}_{n_c}\mat{X}_2^K,{B}^K)\\
&\qquad\qquad{=} \sum_{k=1}^{K}\left[H(\mat{Y}_{1,k}|\mat{Y}_1^{k-1},B^K\dmat{S}_{n_c}\mat{X}_1^K,{B}^K)+H(\mat{Y}_{2,k}|\mat{Y}_2^{k-1},B^K\dmat{S}_{n_c}\mat{X}_2^K,{B}^K)\right]\\
&\qquad\qquad{\leq} \sum_{k=1}^{K}\left[H(\mat{Y}_{1,k}|B_k\dmat{S}_{n_c}\mat{X}_{1,k},{B}_k)+H(\mat{Y}_{2,k}|B_k\dmat{S}_{n_c}\mat{X}_{2,k},{B}_k)\right]\\
&\qquad\qquad{\leq} \sum_{k=1}^{K}\!\Bigl[(1-p)\left(H(\dmat{S}_{n_d}\mat{X}_{1,k}|B_k=0)+H(\dmat{S}_{n_d}\mat{X}_{2,k}|B_k=0)\right)\\
&\qquad\qquad\qquad\;\:{}+p(H(\mat{Y}_{1,k}|\dmat{S}_{n_c}\mat{X}_{1,k},{B}_k=1)+H(\mat{Y}_{2,k}|\dmat{S}_{n_c}\mat{X}_{2,k},{B}_k=1))\Bigr] 
\end{aligned}
\end{equation}
where we have used that conditioning reduces entropy.
By maximizing the entropies in \eqref{Eq: UB_RF2} over all input distributions, dividing by $N=KT$, and letting $K$ tend to infinity, we obtain that
\begin{equation}\label{UB_TKF3}
R\leq 2(1-p) n_d +  2p\max\{(n_d-n_c)^+,n_c\}.
\end{equation}
This proves \eqref{UBF_corr2}.

\section{Proof of Lemma~\ref{Lemma_e}}\label{App: Verdu-Han}

\makeatletter 
\setcounter{equation}{0} 
\@addtoreset{equation}{section}
\renewcommand{\theequation}{C\arabic{equation}}
\makeatletter

In this section, we prove the Lemma~\ref{Lemma_e}. To this end, we first introduce definitions and properties that will be used in the proof of the lemma.

\begin{definition}[Sup-entropy rate] \label{def div-entropy}
The sup-entropy  rate $\overline{H}(Y)$ is defined as the limsup in probability of $\frac{1}{N}\log\frac{1}{P_{Y^N}(Y^N)}$.
Analogously,  the conditional sup-entropy rate $\overline{H}(Y|X)$ is the limsup in probability (according to $\{P_{X^NY^N}\}$) of $\frac{1}{N}\log\frac{1}{P_{Y^N|X^N}(Y^N|X^N)}$. \label{def entropy}
\end{definition}

\begin{lemma}[Sup-entropy rate properties]\label{Lemma Verdu-Han}
Suppose (X,Y) takes values in $(\set {X}, \set {Y})$. The sup-entropy rate has the following properties:
\begin{eqnarray}
\overline{H}(Y|X)<\overline{H}(Y) \label{lemma 10}\\
0\leq\overline{H}(Y)<\log|\set {Y}| \label{lemma 5}
\end{eqnarray}
where $|\set {Y}|$ denotes the cardinality of $Y$.
\end{lemma}
\begin{IEEEproof}
Property \eqref{lemma 10} follows directly from properties (c) and (d) of \cite[Th.~8]{verdu-han}. Property \eqref{lemma 5} is equal to property (e) in \cite[Th.~8]{verdu-han}.
\end{IEEEproof}

We recall the information densities $\infr_1(\vect{x}_1^N,\vect{y}_1^N,\vect{b})$ and $\infr_2(\vect{x}_2^N,\vect{y}_2^N,\vect{b})$ defined in \eqref{i1(b1)} and \eqref{i2(b2)}, respectively. By decomposing the logarithms and applying the Bayes rule to both probability terms, we obtain
 \begin{equation}
 \begin{aligned}
\infr_i (\vect{x}_i^N,\vect{y}_i^N,\vect{b})&=\log{P_{\vect {Y}_i^N|\vect{X}_i^N,\vect{B}}(\vect {y}_i^N|\vect {x}_i^N,\vect{b})}-\log{P_{\vect {Y}_i^N|\vect{B}}(\vect {y}_i^N|\vect{b})}\\
&=\log{P_{\vect {X}_i^N|\vect{Y}_i^N,\vect{B}}(\vect {x}_i^N|\vect {y}_i^N,\vect{b})}-\log{P_{\vect {X}_i^N|\vect{B}}(\vect {x}_i^N|\vect{b})}.
 \end{aligned}
 \label{ii(b)-2} 
 \end{equation} 
To shorten notation, we shall omit the arguments and write $\infr_i \triangleq \infr_i(\vect{x}_i^N,\vect{y}_i^N,\vect{b})$, $i=1,2$ wherever the arguments are clear from the context.

Recall the error events $\Ee_{i}(\Gamma_i) \eqdef \left\{\tfrac{1}{n}\infr_i \leq \Gamma_i \right\}$, $i=1,2$, and $\Ee_{12} (\Gamma) \eqdef \left\{\tfrac{1}{n}\infr_1 + \tfrac{1}{n}\infr_2\leq \Gamma \right\}$, with $\Gamma=\Gamma_1+\Gamma_2$, as defined in \eqref{error_event} and \eqref{joint_error_event}, respectively.  We first note that
\begin{eqnarray} \label{Proof_Pe_LB}
\Ee_1 \cap \Ee_2 &\subseteq& \Ee_{12}\label{LB1}\\
\Ee_1 \cap \Ee_2 &=& \Ee_{1} \setminus\{\Ee_{1}\cap\Ee_{2}^c\} \supseteq \Ee_{1} \setminus\{\Ee_{2}^c\}\label{LB2}
\end{eqnarray}
where \eqref{LB1} follows because the conditions $\frac{1}{N}\infr_1\leq \Gamma_1$ and $\frac{1}{N}\infr_2\leq \Gamma_2$ imply that $\frac{1}{N}(\infr_1+\infr_{2})\leq \Gamma_1+\Gamma_2$. Then, \eqref{LB2} follows by applying basic set operations.
Using \eqref{LB1} and \eqref{LB2}, and computing the probability of the corresponding events, we obtain 
\begin{eqnarray}\label{Pe_LB}
\Pr\{\Ee_{12}\}\geq \Pr\{\Ee_1\}-\Pr\{ \Ee_2^c\}.
\end{eqnarray}

For clarity of exposition, we define
\begin{eqnarray}
\epsilon_{\vect{b}} \triangleq
\Pr\bigg\{\frac{1}{N}\left(\infr_1+\infr_2\right)\leq \Gamma\;\Big|\;  \vect{B}=\vect{b}\bigg\}\label{Pi12}
\end{eqnarray}
and analyze the necessary conditions on $\Gamma$ such that $\epsilon_{\vect{b}}\to 0$ as $N\to \infty$. We next consider separately the four possible realizations of $\vect{B}=\vect{b}$.

\subsubsection{Case $\vect{B}=\left[0,0\right]$} \label{Case: 00}
When $\vect{B}=[0,0]$, the channel corresponds to two parallel channels with no interference links. Then, the underlying distribution of the probability \eqref{Pi12} is
\begin{align}
&P_{\vect {X}_1^N,\vect {X}_2^N,\vect {Y}_1^N,\vect {Y}_2^N | \vect{B}}(\vect {x}_1^N,\vect {x}_2^N,\vect {y}_1^N,\vect {y}_2^N | \vect{b}) 
\nonumber\\\qquad\qquad\qquad 
&\qquad\qquad\qquad= P_{\vect {X}_1^N|\vect{B}}(\vect {x}_1^N|\vect{b})P_{\vect {X}_2^N|\vect{B}}(\vect {x}_2^N|\vect{b})\mathds{1}\{\vect {y}_1^N=\mathsf{S}_{n_{d}}\vect {x}_1^N\}\mathds{1}\{\vect {y}_2^N=\mathsf{S}_{n_{d}}\vect {x}_2^N\}
\label{case00-underlyingP}
\end{align}
as the outputs $\vect {y}_1^N$ and $\vect {y}_2^N$ must coincide with the corresponding inputs according to the deterministic model.
To prove the constraint \eqref{P2P}, we use \eqref{ii(b)-2} in \eqref{Pi12} to obtain 
\begin{eqnarray}
\epsilon_{00}
&=& {\Pr}\biggl\{
-\frac{1}{N}\log P_{\vect {X}_1^N|\vect{B}}(\vect {X}_1^N|\vect{B})-\frac{1}{N}\log P_{\vect {X}_2^N|\vect{B}}(\vect {X}_2^N|\vect{B})
\leq \Gamma \Big|\;  \vect{B}=\left[0,0\right] \biggr\}\label{case00-eps00-1}
\end{eqnarray}
where we used that, according to \eqref{case00-underlyingP}, $\log P_{\vect {X}_i^N|\vect {Y}_i^N,\vect{B}}(\vect{X}_i^N|\vect{Y}_i^N,\vect{B})=0$ w.p. 1, for $i=1,2$.

We consider now the conditional sup-entropy rates $\overline{H}(\vect{X}_i^N|\vect{B})$, $i=1,2$. According to \eqref{lemma 5} in Lemma~\ref{Lemma Verdu-Han}, we have that $\overline{H}(\vect {X}_i^N|\vect{B}) < n_d$, $i=1,2$. With these considerations, if we set $\Gamma=2n_d+2\delta$ for some arbitrary $\delta>0$ in \eqref{case00-eps00-1}, we obtain
\begin{myequation}
\begin{aligned}
\epsilon_{00} 
\geq &   {\Pr}\left\{- \frac{1}{N}
\log P_{\vect {X}_1^N|\vect{B}}(\vect {X}_1^N|\vect{B}) - \frac{1}{N}\log P_{\vect {X}_2^N|\vect{B}}(\vect {X}_2^N|\vect{B}\leq 2n_d+2\delta\;\Big|\;  \vect{B}=\left[0,0\right] \right\}\\
\geq &  {\Pr}\left\{- \frac{1}{N}
\log P_{\vect {X}_1^N|\vect{B}}(\vect {X}_1^N|\vect{B}) - \frac{1}{N}\log P_{\vect {X}_2^N|\vect{B}}(\vect {X}_2^N|\vect{B})< \overline{H}(\vect{X}_1^N|\vect{B})+\overline{H}(\vect{X}_2^N|\vect{B})+2\delta\;\Big|\;  \vect{B}=\left[0,0\right] \right\}\\
\geq &  {\Pr}\left\{ -\frac{1}{N}
\log P_{\vect {X}_1^N|\vect{B}}(\vect {X}_1^N|\vect{B})< \overline{H}(\vect{X}_1^N|\vect{B})+\delta\;\Big|\;  \vect{B}=\left[0,0\right]\right\}\\
& {}-  {\Pr}\left\{ -\frac{1}{N}
\log P_{\vect {X}_2^N|\vect{B}}(\vect {X}_2^N|\vect{B})\geq \overline{H}(\vect{X}_2^N|\vect{B})+ \delta\;\Big|\;  \vect{B}=\left[0,0\right]\right\}
\label{case00-eps00-2}
\end{aligned}
\end{myequation}
where the last step follows from \eqref{Pe_LB}.

Recalling the definitions of the conditional sup-entropy rates $\overline{H}(\vect{X}_i^N|\vect{B})$ we have that, for any $\delta>0$,
\begin{align}
\lim_{N\to\infty} & {\Pr}\Bigl\{- \frac{1}{N}\log P_{\vect {X}_i^N|\vect{B}}(\vect {X}_i^N|\vect{B})\geq \overline{H}(\vect {X}_i^N|\vect{B}) + \delta\;\Big|\;  \vect{B}=\left[0,0\right]\Bigr\} = 0\label{case00-Hbar}, \quad i=1,2.
\end{align}
This implies that the first probability on the RHS of \eqref{case00-eps00-2} tends to 1 as $N\to\infty$, and the second probability on the RHS of \eqref{case00-eps00-2} tends to 0 as $N\to\infty$. We conclude that for any $\Gamma > 2n_d$ the lower bound in \eqref{case00-eps00-2} tends to 1  as  $N\to\infty$. Thus, $\epsilon_{00}\to 0$ as $N\to\infty$ only if $\Gamma \leq 2n_d$.

\subsubsection{Case $\vect{B}=\left[0,1\right]$} \label{Case: 01}
When $\vect{B}=\left[0,1\right]$, the channel corresponds to a two-user IC where only one of the transmitters interferes its non-intended receiver. In this case, the underlying distribution in \eqref{Pi12} is given by 
\begin{myequation}
\begin{aligned}
&P_{\vect {X}_1^N,\vect {X}_2^N,\vect {Y}_1^N,\vect {Y}_2^N|\vect{B}}(\vect {x}_1^N,\vect {x}_2^N,\vect {y}_1^N,\vect {y}_2^N|\vect{b}) \\
&\qquad\qquad\qquad= P_{\vect {X}_1^N|\vect{B}}(\vect {x}_1^N|\vect{b})P_{\vect {X}_2^N|\vect{B}}(\vect {x}_2^N|\vect{b})\mathds{1}\{\vect {y}_1^N=\mathsf{S}_{n_{d}}\vect {x}_1^N\}\mathds{1}\{\vect {y}_2^N=\mathsf{S}_{n_{d}}\vect {x}_2^N\oplus \mathsf{S}_{n_{c}}\vect {x}_1^N\}
\label{case01-underlyingP}
\end{aligned}
\end{myequation}
We next prove the constraints \eqref{P2P} and \eqref{SZ_1} in Lemma~\ref{Lemma_e}.

\paragraph{Proof of Constraint \eqref{P2P}} \label{case01-P2P}
We lower-bound the probability $\epsilon_{01}$ by that of 2 parallel channels and follow the steps in Appendix~\ref{Case: 00}. Indeed, by using \eqref{ii(b)-2} in \eqref{Pi12} and lower-bounding $\log P_{\vect {X}_i^N|\vect {Y}_i^N,\vect{B}}(\vect{X}_i^N|\vect{Y}_i^N,\vect{B})\leq0$, $i=1,2$, we obtain that
\begin{eqnarray}
\epsilon_{01}
&\geq& {\Pr}\biggl\{
-\frac{1}{N}\log P_{\vect {X}_1^N|\vect{B}}(\vect {X}_1^N|\vect{B})-\frac{1}{N}\log P_{\vect {X}_2^N|\vect{B}}(\vect {X}_2^N|\vect{B})
\leq \Gamma \Big|\;  \vect{B}=\left[0,1\right] \biggr\}.\label{case01-eps-1}
\end{eqnarray}
The RHS of \eqref{case01-eps-1} coincides with \eqref{case00-eps00-1} conditioned in $\vect{B}=[0,1]$. The proof then follows the one in Appendix~\ref{Case: 00}, with the probabilities and sup-entropy rates conditioned on $\vect{B}=[0,1]$ instead of $\vect{B}=[0,0]$.

\paragraph{Proof of Constraint \eqref{SZ_1}} \label{case01-SZ}
According to \eqref{case01-underlyingP}, the following identities hold w.p. 1:
\begin{itemize}[leftmargin=21pt,labelsep=7pt]
\item [(i1)]  $\vect {Y}_2^N\oplus \mathsf{S}_{n_{d}}\vect {X}_2^{N} = \mathsf{S}_{n_{c}}\vect {X}_1^N$
\item [(i2)]  $P_{\vect {Y}_2^N|\vect {X}_2^N,\vect{B}}(\vect {Y}_2^N|\vect {X}_2^N,[0,1]) = P_{\mathsf{S}_{n_{c}}\vect {X}_1^N|\vect{B}}(\vect {Y}_2^N\oplus \mathsf{S}_{n_{d}}\vect {X}_2^{N}|\vect{B}=[0,1])$
\item [(i3)]  $P_{\vect {Y}_1^N|\vect {X}_1^N,\vect{B}}(\vect {Y}_1^N|\vect {X}_1^N, [0,1])=1$
\end{itemize}
Using \eqref{ii(b)-2} in \eqref{Pi12} and the identities (i1)--(i3), we obtain
\begin{equation}
\begin{aligned}
\epsilon_{01} &= {\Pr}\biggl\{\frac{1}{N}
\log{P_{\vect {Y}_1^N|\vect{X}_1^N,\vect{B}}(\vect{Y}_1^N|\vect{X}_1^N,\vect{B})}-\frac{1}{N}\log{P_{\vect {Y}_1^N|\vect{B}}(\vect{Y}_1^N|\vect{B})}\\
&\qquad\quad{}+ \frac{1}{N}
\log{P_{\vect {Y}_2^N|\vect{X}_2^N,\vect{B}}(\vect{Y}_2^N|\vect{X}_2^N,\vect{B})}-\frac{1}{N}\log{P_{\vect {Y}_2^N|\vect{B}}(\vect{Y}_2^N|\vect{B})} \leq \Gamma\Big|\;  \vect{B}=\left[0,1\right]\biggr\}\\
 &= {\Pr}\biggl\{
-\frac{1}{N}\log P_{\mathsf{S}_{n_{d}}\vect {X}_1^{N}|\vect{B}}(\mathsf{S}_{n_{d}}\vect {X}_1^{N}|\vect{B})
\\
&\qquad\quad
 {}+ \frac{1}{N}\log P_{\mathsf{S}_{n_{c}}\vect {X}_1^N|\vect{B}}(\mathsf{S}_{n_{c}}\vect {X}_1^N|\vect{B}) -\frac{1}{N}\log P_{\vect {Y}_2^N|\vect{B}}(\vect {Y}_2^N|\vect{B})
\leq \Gamma \Big|\;  \vect{B}=\left[0,1\right]\biggr\}\label{case01-eps-2}
\end{aligned}
\end{equation}
We next define $\mathsf{\tilde{L}}_{d}\eqdef \dmat{L}_{d}\mathsf{S}_{n_d}$ and apply the chain rule of probability to obtain
\begin{equation}
\begin{aligned}
&\log P_{\mathsf{S}_{n_{d}}\vect {X}_1^{N}|\vect{B}}(\mathsf{S}_{n_{d}}\vect {x}_1^{N}|\vect{b}) \\
&\qquad\quad= 
\log P_{\mathsf{S}_{n_{c}}\vect {X}_1^N|\vect{B}}(\mathsf{S}_{n_{c}}\vect {x}_1^N|\vect{b}) + \log P_{\mathsf{\tilde{L}}_{(n_{d}-n_{c})^+}\vect {X}_1^N | \mathsf{S}_{n_{c}}\vect {X}_1^N,\vect{B}}(\mathsf{\tilde{L}}_{(n_{d}-n_{c})^+}\vect {x}_1^N | \mathsf{S}_{n_{c}}\vect {x}_1^N,\vect{b}).
\label{case01-chainrule}
\end{aligned}
\end{equation}
Using \eqref{case01-chainrule} in \eqref{case01-eps-2} and canceling the term $\log P_{\mathsf{S}_{n_{c}}\vect {X}_1^N|\vect{B}}(\mathsf{S}_{n_{c}}\vect {X}_1^N|\vect{B})$, we obtain

\begin{equation}
\begin{aligned}
\epsilon_{01} 
=  & {\Pr}\left\{- \frac{1}{N} 
 \log P_{\mathsf{\tilde{L}}_{(n_{d}-n_{c})^+}\vect {X}_1^N | \mathsf{S}_{n_{c}}\vect {X}_1^N, \vect{B}}(\mathsf{\tilde{L}}_{(n_{d}-n_{c})^+}\vect {X}_1^N | \mathsf{S}_{n_{c}}\vect {X}_1^N, \vect{B}) \right.\\
 & \quad\;\,\left.{} - \frac{1}{N}\log P_{\vect {Y}_2^N|\vect{B}}(\vect {Y}_2^N|\vect{B}) \leq \Gamma\;\Big|\;  \vect{B}=\left[0,1\right] \right\}.
\label{case01-eps-3}
\end{aligned}
\end{equation}

Consider the sup-entropy rates $\overline{H}\bigl( \mathsf{\tilde{L}}_{(n_{d}-n_{c})^+}\vect {X}_1^N | \mathsf{S}_{n_{c}}\vect {X}_1^N,\vect{B}\bigr)$ and $\overline{H}\bigl( \vect {Y}_2^N | \vect{B} \bigr)$.
By \eqref{lemma 10} and \eqref{lemma 5} in Lemma~\ref{Lemma Verdu-Han}, we have that
\begin{eqnarray}
\overline{H}\bigl( \mathsf{\tilde{L}}_{(n_{d}-n_{c})^+}\vect {X}_1^N | \mathsf{S}_{n_{c}}\vect {X}_1^N,\vect{B}\bigr) \leq \overline{H}\bigl( \mathsf{\tilde{L}}_{(n_{d}-n_{c})^+}\vect {X}_1^N| \vect{B} ) & < & (n_d-n_c)^+ \\
\overline{H}\bigl( \vect {Y}_2^N | \vect{B} \bigr) & < & \max(n_d,n_c). 
\end{eqnarray}
Let $\displaystyle{\Gamma=(n_d-n_c)^++\max(n_d,n_c)+2\delta}$ for some arbitrary $\delta>0$. It follows that $\displaystyle{\Gamma \geq \overline{H}\bigl( \mathsf{\tilde{L}}_{(n_{d}-n_{c})^+}\vect {X}_1^N| \mathsf{S}_{n_{c}}\vect {X}_1^N,\vect{B})+\overline{H}\bigl( \vect {Y}_2^N | \vect{B} \bigr)+2\delta}$, so \eqref{case01-eps-3} can be lower-bounded as
\begin{myequation}
\begin{aligned}
\epsilon_{01} 
 &\geq   {\Pr}\Biggl\{- \frac{1}{N} 
  \log P_{\mathsf{\tilde{L}}_{(n_{d}-n_{c})^+}\vect {X}_1^N | \mathsf{S}_{n_{c}}\vect {X}_1^N, \vect{B}}(\mathsf{\tilde{L}}_{(n_{d}-n_{c})^+}\vect {X}_1^N | \mathsf{S}_{n_{c}}\vect {X}_1^N, \vect{B}) - \frac{1}{N}\log P_{\vect {Y}_2^N|\vect{B}}(\vect {Y}_2^N|\vect{B})\Biggr. \\
  &\quad\qquad\,\Biggl. {}<\overline{H}\bigl( \mathsf{\tilde{L}}_{(n_{d}-n_{c})^+}\vect {X}_1^N| \mathsf{S}_{n_{c}}\vect {X}_1^N,\vect{B})+\overline{H}\bigl( \vect {Y}_2^N | \vect{B} \bigr)+2\delta\;\Big|\;  \vect{B}=\left[0,1\right] \Biggr\}\\
&\geq   {\Pr}\Biggl\{ -\frac{1}{N} 
 \log P_{\mathsf{\tilde{L}}_{(n_{d}-n_{c})^+}\vect {X}_1^N | \mathsf{S}_{n_{c}}\vect {X}_1^N, \vect{B}}(\mathsf{\tilde{L}}_{(n_{d}-n_{c})^+}\vect {X}_1^N | \mathsf{S}_{n_{c}}\vect {X}_1^N, \vect{B}) \\
 & \quad\qquad\,< \overline{H}\bigl( \mathsf{\tilde{L}}_{(n_{d}-n_{c})^+}\vect {X}_1^N| \mathsf{S}_{n_{c}}\vect {X}_1^N,\vect{B} ) +\delta\;\Big|\;  \vect{B}=\left[0,1\right]\Biggr\}\\
&\quad\;{}-{\Pr}\left\{ -\frac{1}{N}
\log P_{\vect {Y}_2^N|\vect{B}}(\vect {Y}_2^N|\vect{B})\geq \overline{H}\bigl( \vect {Y}_2^N | \vect{B} \bigr)+ \delta\;\Big|\;  \vect{B}=\left[0,1\right]\right\}
\label{case01-eps-4}
\end{aligned}
\end{myequation}
where the second step follows from \eqref{Pe_LB}. By the definition of the conditional sup-entropy rate, it follows that the first probability on the RHS of \eqref{case01-eps-4} tends to 1 as $N\to\infty$, and the second probability on the RHS of \eqref{case01-eps-4} tends to 0 as $N\to\infty$. This implies that $\epsilon_{01} \to 0$ as $N\to\infty$ only if $\Gamma \leq (n_d-n_c)^++\max(n_d,n_c)$ and proves conditions \eqref{P2P} and \eqref{SZ_1} in Lemma~\ref{Lemma_e}.

\begin{remark} \label{Ch_ZS} 
Given the symmetry of the problem, the constraints \eqref{P2P} and \eqref{SZ_1} for $\vect{B}=[1,0]$ are proven by swapping the roles of users 1 and 2, and following the same steps as for $\vect{B}=[0,1]$.
\end{remark}

\subsubsection{Case $\vect{B}=\left[1,1\right]$}\label{Case: 11}
This scenario corresponds to a non-bursty IC. The underlying distribution in \eqref{Pi12} is given by 
\begin{equation}
\begin{aligned}
&P_{\vect {X}_1^N,\vect {X}_2^N,\vect {Y}_1^N,\vect {Y}_2^N|\vect{B}}(\vect {x}_1^N,\vect {x}_2^N,\vect {y}_1^N,\vect {y}_2^N|\vect{b})\\
&\qquad = P_{\vect {X}_1^N|\vect{B}}(\vect {x}_1^N|\vect{b})P_{\vect {X}_2^N|\vect{B}}(\vect {x}_2^N|\vect{b})\mathds{1}\{\vect {y}_1^N=\mathsf{S}_{n_{d}}\vect {x}_1^N\oplus \mathsf{S}_{n_{c}}\vect {x}_2^N\}\mathds{1}\{\vect {y}_2^N=\mathsf{S}_{n_{d}}\vect {x}_2^N\oplus \mathsf{S}_{n_{c}}\vect {x}_1^N\} \label{underlyingP}
\end{aligned}
\end{equation}
where the last step follows from the deterministic model since, for given $\vect {x}_1^N$ and $\vect {x}_2^N$, the outputs $\vect {y}_1^N$ and $\vect {y}_2^N$ are given by the equations appearing in the corresponding indicator functions. We next obtain the constraints \eqref{P2P}--\eqref{IC_1} in Lemma~\ref{Lemma_e}.

\paragraph{Proof of Constraint \eqref{P2P}} \label{Ch_P2P}
To prove this constraint, we lower-bound the probability $\epsilon_{11}$ by that of 2 parallel channels. Indeed, using \eqref{ii(b)-2} in \eqref{Pi12}, we obtain that

\begin{equation}
\begin{aligned}
\epsilon_{11}&= {\Pr}\Bigg\{
\frac{1}{N}\Bigg(\log P_{\vect {X}_1^N|\vect {Y}_1^N,\vect{B}}(\vect {X}_1^N|\vect {Y}_1^N,\vect{B})-\log P_{\vect {X}_1^N|\vect{B}}(\vect {X}_1^N|\vect{B})\Bigg.\\
&\qquad\quad\, \Bigg.{}+\log P_{\vect {X}_2^N|\vect {Y}_2^N,\vect{B}}(\vect {X}_2^N|\vect {Y}_2^N,\vect{B})-\log P_{\vect {X}_2^N|\vect{B}}(\vect {X}_2^N|\vect{B})\Bigg)
\leq\Gamma \Big|\;  \vect{B}=\left[1,1\right]\Bigg\}\\
&\geq {\Pr}\biggl\{
-\frac{1}{N}\log P_{\vect {X}_1^N|\vect{B}}(\vect {X}_1^N|\vect{B})-\frac{1}{N}\log P_{\vect {X}_2^N|\vect{B}}(\vect {X}_2^N|\vect{B})
\leq \Gamma \Big|\;  \vect{B}=\left[1,1\right] \biggr\}\label{case11-eps11-1}
\end{aligned}
\end{equation}
where the inequality follows because $\log P_{\vect {X}_i^N|\vect {Y}_i^N,\vect{B}}(\vect{X}_i^N|\vect{Y}_i^N,\vect{B})\leq0, \; i=1,2$. As this expression coincides with \eqref{case00-eps00-1} conditioned on $\vect{B}=[1,1]$, the proof then follows the one in Appendix~\ref{Case: 00}, with the probabilities and sup-entropy rates conditioned on $\vect{B}=[1,1]$ instead of $\vect{B}=[0,0]$.

\paragraph{Proof of Constraint \eqref{SZ_1}}\label{Ch_S}
We next lower-bound the probability $\epsilon_{11} $ by that of an interference channel, in which only one of the transmitters interferes its non-intended receiver. Using the information densities $i_1$ and $i_2$ in \eqref{ii(b)-2}, we have that 
\begin{myequation}
\begin{aligned}
\infr_1+\infr_2 & = 
\log{P_{\vect {Y}_1^N|\vect{X}_1^N,\vect{B}}(\vect {y}_1^N|\vect {x}_1^N,\vect{b})}-\log{P_{\vect {Y}_1^N|\vect{B}}(\vect {y}_1^N|\vect{b})} +\log{P_{\vect {Y}_2^N|\vect{X}_2^N,\vect{B}}(\vect {y}_2^N|\vect {x}_2^N,\vect{b})}-\log{P_{\vect {Y}_2^N|\vect{B}}(\vect {y}_2^N|\vect{b})}\\
&=\log P_{\vect {Y}_1^N|\vect {X}_1^N,\vect {X}_2^N,\vect{B}}(\vect {y}_1^N|\vect {x}_1^N,\vect {x}_2^N,\vect{b})-\log P_{\vect {Y}_1^N|\vect {X}_2^N,\vect{B}}(\vect {y}_1^N|\vect {x}_2^N,\vect{b})\\
& \quad {}+\log P_{\vect {Y}_2^N|\vect {X}_2^N,\vect{B}}(\vect {y}_2^N|\vect {x}_2^N,\vect{b})-\log P_{\vect {Y}_2^N,\vect{B}}(\vect {y}_2^N|\vect{b})-\log\frac{P_{\vect {X}_1^N|\vect {Y}_1^N,\vect {X}_2^N,\vect{B}}(\vect {x}_1^N|\vect {y}_1^N,\vect {x}_2^N,\vect{b})}{P_{\vect {X}_1^N|\vect {Y}_1^N,\vect{B}}(\vect {x}_1^N|\vect {y}_1^N,\vect{b})}\label{case11-eps11-2}
\end{aligned}
\end{myequation}
where the second step follows from adding and subtracting

\begin{equation*}
\frac{1}{N}\log \frac{P_{\vect {Y}_1^N|\vect {X}_1^N,\vect {X}_2^N,\vect{B}}(\vect {y}_1^N|\vect {x}_1^N,\vect {x}_2^N,\vect{b})}{P_{\vect {Y}_1^N|\vect {X}_2^N,\vect{B}}(\vect {y}_1^N|\vect {x}_2^N,\vect{b})}
\end{equation*}
 and simplifying the resulting terms via the Bayes rule and using that $P_{\vect {X}_1^N|\vect {X}_2^N,\vect{B}}(\vect {x}_1^N|\vect {x}_2^N,\vect{b}) = P_{\vect {X}_1^N|\vect{B}}(\vect {x}_1^N|\vect{b})$ since $\vect {X}_1^N$ and $\vect {X}_2^N$ are independent conditioned on $\vect{B}$.

According to the underlying distribution \eqref{underlyingP}, the following identities hold w.p. 1:
\begin{itemize}[leftmargin=21pt,labelsep=7pt]
\item [(i1)] $\vect {Y}_1^N\oplus\mathsf{S}_{n_{c}}\vect {X}_2^N = \mathsf{S}_{n_{d}}\vect {X}_1^{N}$
\item [(i2)]  $\vect {Y}_2^N\oplus \mathsf{S}_{n_{d}}\vect {X}_2^{N} = \mathsf{S}_{n_{c}}\vect {X}_1^N$
\item [(i3)]  $P_{\vect {Y}_1^N|\vect {X}_2^N,\vect{B}}(\vect {Y}_1^N|\vect {X}_2^N,\vect{B}=[1,1])=P_{\dmat{S}_{n_{d}}\vect {X}_1^{N}|\vect{B}}(\vect {Y}_1^N\oplus\dmat{S}_{n_{c}}\vect {X}_2^N|\vect{B}=[1,1])$
\item [(i4)]  $P_{\vect {Y}_2^N|\vect {X}_2^N,\vect{B}}(\vect {Y}_2^N|\vect {X}_2^N,\vect{B}=[1,1]) = P_{\mathsf{S}_{n_{c}}\vect {X}_1^N|\vect{B}}(\vect {Y}_2^N\oplus \mathsf{S}_{n_{d}}\vect {X}_2^{N}|\vect{B}=[1,1])$
\item [(i5)]  $P_{\vect {Y}_1^N|\vect {X}_1^N,\vect {X}_2^N,\vect{B}}(\vect {Y}_1^N|\vect {X}_1^N, \vect {X}_2^N,\vect{B}=[1,1])=1$
\item [(i6)]  $P_{\vect {X}_1^N|\vect {Y}_1^N,\vect {X}_2^N,\vect{B}}(\vect {X}_1^N|\vect {Y}_1^N,\vect {X}_2^N,\vect{B}=[1,1])=1$.
\end{itemize}
Using \eqref{case11-eps11-2} and the identities (i1)--(i6), we obtain for \eqref{Pi12}
\begin{myequation}
\begin{aligned}
\epsilon_{11} 
&= {\Pr}\biggl\{
-\frac{1}{N}\log P_{\mathsf{S}_{n_{d}}\vect {X}_1^{N}|\vect{B}}(\mathsf{S}_{n_{d}}\vect {X}_1^{N}|\vect{B}) + \frac{1}{N}\log P_{\mathsf{S}_{n_{c}}\vect {X}_1^N|\vect{B}}(\mathsf{S}_{n_{c}}\vect {X}_1^N|\vect{B})\\
&\qquad\quad{}-\frac{1}{N}\log P_{\vect {Y}_2^N|\vect{B}}(\vect {Y}_2^N|\vect{B})+\frac{1}{N}\log{P_{\vect {X}_1^N|\vect {Y}_1^N,\vect{B}}(\vect {X}_1^N|\vect {Y}_1^N,\vect{B}})
\leq \Gamma \;\Big|\;  \vect{B}=\left[1,1\right] \biggr\}.
\label{case11-eps11-3}
\end{aligned}
\end{myequation}
Using \eqref{case01-chainrule} in \eqref{case11-eps11-3}, canceling the term $\log P_{\mathsf{S}_{n_{c}}\vect {X}_1^N|\vect{B}}(\mathsf{S}_{n_{c}}\vect {X}_1^N|\vect{B})$, and using that $\log{P_{\vect {X}_1^N|\vect {Y}_1^N,\vect{B}}(\vect {X}_1^N|\vect {Y}_1^N,\vect{B})} \leq 0$, we obtain the lower bound
\begin{myequation}
\epsilon_{11}  \geq  {\Pr}\biggl\{- \frac{1}{N}
\log P_{\mathsf{\tilde{L}}_{(n_{d}-n_{c})^+}\vect {X}_1^N | \mathsf{S}_{n_{c}}\vect {X}_1^N, \vect{B}}(\mathsf{\tilde{L}}_{(n_{d}-n_{c})^+}\vect {X}_1^N | \mathsf{S}_{n_{c}}\vect {X}_1^N, \vect{B}) - \frac{1}{N} \log P_{\vect {Y}_2^N|\vect{B}}(\vect {Y}_2^N|\vect{B}) 
\leq \Gamma\Big|\;  \vect{B}=\left[1,1\right] \biggr\}.\label{case11-eps11-4}
\end{myequation}
The RHS of \eqref{case11-eps11-4} coincides with \eqref{case01-eps-3} conditioned on $\vect{B}=[1,1]$. The proof then follows the one in Appendix~\ref{case01-SZ}, with the probabilities and sup-entropy rates conditioned on $\vect{B}=[1,1]$ instead of $\vect{B}=[0,1]$.

\paragraph{Proof of Constraint \eqref{IC_1}}
We begin this proof by using \eqref{ii(b)-2} to write
 \begin{myequation}
\begin{aligned}\label{i11}
\infr_1+\infr_2 &= \log\frac{P_{\vect {X}_1^N|\vect {Y}_1^N,\vect{B}}(\vect {x}_1^N|\vect {y}_1^N,\vect{b})}{P_{\vect {X}_1^N|\vect{B}}(\vect {x}_1^N|\vect{b})}+\log\frac{P_{\vect {X}_2^N|\vect {Y}_2^N,\vect{B}}(\vect {x}_2^N|\vect {y}_2^N,\vect{b})}{P_{\vect {X}_2^N|\vect{B}}(\vect {x}_2^N|\vect{b})}\\
&\stackrel{(a)}{=}\log\frac{P_{\vect {X}_1^N|\vect {Y}_1^N,\mathsf{S}_{n_{c}}\vect {X}_1^N,\vect{B}}(\vect {x}_1^N|\vect {y}_1^N,\mathsf{S}_{n_{c}}\vect {x}_1^N,\vect{b})}{P_{\vect {X}_1^N|\mathsf{S}_{n_{c}}\vect {X}_1^N,\vect{B}}(\vect {x}_1^N|\mathsf{S}_{n_{c}}\vect {x}_1^N,\vect{b})}+\log\frac{P_{\vect {X}_1^N|\mathsf{S}_{n_{c}}\vect {X}_1^N,\vect{B}}(\vect {x}_1^N|\mathsf{S}_{n_{c}}\vect {x}_1^N,\vect{b})}{P_{\vect {X}_1^N|\vect{B}}(\vect {x}_1^N|\vect{b})}\\
&\quad\;{}+\log\frac{P_{\vect {X}_2^N|\vect {Y}_2^N,\mathsf{S}_{n_{c}}\vect {X}_2^N,\vect{B}}(\vect {x}_2^N|\vect {y}_2^N,\mathsf{S}_{n_{c}}\vect {x}_2^N,\vect{b})}{P_{\vect {X}_2^N|\mathsf{S}_{n_{c}}\vect {X}_2^N,\vect{B}}(\vect {x}_2^N|\mathsf{S}_{n_{c}}\vect {x}_2^N,\vect{b})} +\log\frac{P_{\vect {X}_2^N|\mathsf{S}_{n_{c}}\vect {X}_2^N,\vect{B}}(\vect {x}_2^N|\mathsf{S}_{n_{c}}\vect {x}_2^N,\vect{b})}{P_{\vect {X}_2^N|\vect{B}}(\vect {x}_2^N|\vect{b})}\\
&\quad\;{}
-\log\frac{P_{\vect {X}_1^N|\vect {Y}_1^N,\mathsf{S}_{n_{c}}\vect {X}_1^N,\vect{B}}(\vect {x}_1^N|\vect {y}_1^N,\mathsf{S}_{n_{c}}\vect {x}_1^N,\vect{b})}{P_{\vect {X}_1^N|\vect {Y}_1^N,\vect{B}}(\vect {x}_1^N|\vect {y}_1^N,\vect{b})}-\log\frac{P_{\vect {X}_2^N|\vect {Y}_2^N,\mathsf{S}_{n_{c}}\vect {X}_2^N,\vect{B}}(\vect {x}_2^N|\vect {y}_2^N,\mathsf{S}_{n_{c}}\vect {x}_2^N,\vect{b})}{P_{\vect {X}_2^N|\vect {Y}_2^N,\vect{B}}(\vect {x}_2^N|\vect {y}_2^N,\vect{b})} \\
&\stackrel{(b)}{=}\log{P_{\vect {Y}_1^N|\vect {X}_1^N,\mathsf{S}_{n_{c}}\vect {X}_1^N,\vect{B}}(\vect {y}_1^N|\vect {x}_1^N,\mathsf{S}_{n_{c}}\vect {x}_1^N,\vect{b})}-\log P_{\vect {Y}_1^N|\mathsf{S}_{n_{c}}\vect {X}_1^N,\vect{B}}(\vect {y}_1^N|\mathsf{S}_{n_{c}}\vect {x}_1^N,\vect{b})\\
&\quad\;{}+\log P_{\mathsf{S}_{n_{c}}\vect {X}_1^N|\vect {X}_1^N,\vect{B}}(\mathsf{S}_{n_{c}}\vect {x}_1^N|\vect {x}_1^N,\vect{b})-\log P_{\mathsf{S}_{n_{c}}\vect {X}_1^N|\vect{B}}(\mathsf{S}_{n_{c}}\vect {x}_1^N|\vect{b})\\
&\quad\;
{}+\log{P_{\vect {Y}_2^N|\vect {X}_2^N,\mathsf{S}_{n_{c}}\vect {X}_2^N,\vect{B}}(\vect {y}_2^N|\vect {x}_2^N,\mathsf{S}_{n_{c}}\vect {x}_2^N,\vect{b})}
-\log P_{\vect{Y}_2^N|\mathsf{S}_{n_{c}}\vect {X}_2^N,\vect{B}}(\vect {y}_2^N|\mathsf{S}_{n_{c}}\vect {x}_2^N,\vect{b})\\
&\quad\;{}+\log P_{\mathsf{S}_{n_{c}}\vect {X}_2^N|\vect {X}_2^N,\vect{B}}(\mathsf{S}_{n_{c}}\vect {x}_2^N|\vect {x}_2^N,\vect{b}) -\log P_{\mathsf{S}_{n_{c}}\vect {X}_2^N|\vect{B}}(\mathsf{S}_{n_{c}}\vect {x}_2^N|\vect{b})\\
&\quad\;
{}-\log\frac{P_{\dmat{S}_{n_c}\vect {X}_1^N|\vect {X}_1^N,\vect {Y}_1^N,\vect{B}}(\dmat{S}_{n_c}\vect {x}_1^N|\vect {x}_1^N,\vect {y}_1^N,\vect{b})}{P_{\dmat{S}_{n_c}\vect {X}_1^N|\vect {Y}_1^N,\vect{B}}(\dmat{S}_{n_c}\vect {x}_1^N|\vect {y}_1^N,\vect{b})}
-\log\frac{P_{\dmat{S}_{n_c}\vect {X}_2^N|\vect {X}_2^N,\vect {Y}_2^N,\vect{B}}(\dmat{S}_{n_c}\vect {x}_2^N|\vect {x}_2^N,\vect {y}_2^N,\vect{b})}{P_{\dmat{S}_{n_c}\vect {X}_2^N|\vect {Y}_2^N,\vect{B}}(\dmat{S}_{n_c}\vect {x}_2^N|\vect {y}_2^N,\vect{b})}
\end{aligned}
\end{myequation}
where (a) follows by adding and subtracting
\begin{equation*}
\frac{1}{N}\log \frac{P_{\vect {X}_1^N|\vect {Y}_1^N,\mathsf{S}_{n_{c}}\vect {X}_1^N,\vect{B}}(\vect {x}_1^N|\vect {y}_1^N,\mathsf{S}_{n_{c}}\vect {x}_1^N,\vect{b})}{P_{\vect {X}_1^N|\mathsf{S}_{n_{c}}\vect {X}_1^N,\vect{B}}(\vect {x}_1^N|\mathsf{S}_{n_{c}}\vect {x}_1^N,\vect{b})} \quad \text{and} \quad 
\frac{1}{N}\log \frac{P_{\vect {X}_2^N|\vect {Y}_2^N,\mathsf{S}_{n_{c}}\vect {X}_2^N,\vect{B}}(\vect {x}_2^N|\vect {y}_2^N,\mathsf{S}_{n_{c}}\vect {x}_2^N,\vect{b})}{P_{\vect {X}_2^N|\mathsf{S}_{n_{c}}\vect {X}_2^N,\vect{B}}(\vect {x}_2^N|\mathsf{S}_{n_{c}}\vect {x}_2^N,\vect{b})}
\end{equation*}
and by rearranging terms. Step (b) follows by applying the Bayes rule and by decomposing the logarithm terms.

We analyze the second and the seventh terms in \eqref{i11}. To this end, we define $n_{-}\eqdef\min\{(n_d-n_c)^+,n_c\}$ and $n_{+}\eqdef\max\{(n_d-n_c)^+,n_c\}$ and apply the chain rule of probability to obtain

 \begin{equation}
\begin{aligned}
&P_{\vect {Y}_1^N|\mathsf{S}_{n_{c}}\vect {X}_1^N,\vect{B}}(\vect {y}_1^N|\mathsf{S}_{n_{c}}\vect {x}_1^N,\vect{b})\\
&\qquad= P_{\mathsf{S}_{n_{-}}\vect {Y}_1^N|\mathsf{S}_{n_{c}}\vect {X}_1^N,\vect{B}}(\mathsf{S}_{n_{-}}\vect {y}_1^N|\mathsf{S}_{n_{c}}\vect {x}_1^N,\vect{b})P_{\mathsf{L}_{n_{+}}\vect {Y}_1^N|\mathsf{S}_{n_{c}}\vect {X}_1^N,\mathsf{S}_{n_{-}}\vect {Y}_1^N,\vect{B}}(\mathsf{L}_{n_{+}}\vect {y}_1^N|\mathsf{S}_{n_{c}}\vect {x}_1^N,\mathsf{S}_{n_{-}}\vect {y}_1^N,\vect{b})\\
&\qquad= P_{\mathsf{L}_{n_{+}}\vect {Y}_1^N|\mathsf{S}_{n_{c}}\vect {X}_1^N,\mathsf{S}_{n_{-}}\vect {Y}_1^N,\vect{B}}(\mathsf{L}_{n_{+}}\vect {y}_1^N|\mathsf{S}_{n_{c}}\vect {x}_1^N,\mathsf{S}_{n_{-}}\vect {y}_1^N,\vect{b})
\label{Px13}
\end{aligned}
\end{equation}
and
 \begin{equation}
\begin{aligned}
&P_{\vect {Y}_2^N|\mathsf{S}_{n_{c}}\vect {X}_2^N,\vect{B}}(\vect {y}_2^N|\mathsf{S}_{n_{c}}\vect {x}_2^N,\vect{b})\\
&\qquad = P_{\mathsf{S}_{n_{-}}\vect {Y}_2^N|\mathsf{S}_{n_{c}}\vect {X}_2^N,\vect{B}}(\mathsf{S}_{n_{-}}\vect {y}_2^N|\mathsf{S}_{n_{c}}\vect {x}_2^N,\vect{b})P_{\mathsf{L}_{n_{+}}\vect {Y}_2^N|\mathsf{S}_{n_{c}}\vect {X}_2^N,\mathsf{S}_{n_{-}}\vect {Y}_2^N,\vect{B}}(\mathsf{L}_{n_{+}}\vect {y}_2^N|\mathsf{S}_{n_{c}}\vect {x}_2^N,\mathsf{S}_{n_{-}}\vect {y}_2^N,\vect{b})\\
&\qquad = P_{\mathsf{L}_{n_{+}}\vect {Y}_2^N|\mathsf{S}_{n_{c}}\vect {X}_2^N,\mathsf{S}_{n_{-}}\vect {Y}_2^N,\vect{B}}(\mathsf{L}_{n_{+}}\vect {y}_2^N|\mathsf{S}_{n_{c}}\vect {x}_2^N,\mathsf{S}_{n_{-}}\vect {y}_2^N,\vect{b}).
\label{Px15}
\end{aligned}
\end{equation}
The probabilities \eqref{Px13} and \eqref{Px15} were simplified by recalling the underlying distribution \eqref{underlyingP}. Indeed, we have w.p.\ 1 that $P_{\mathsf{S}_{n_{-}}\vect {Y}_1^N|\mathsf{S}_{n_{c}}\vect {X}_1^N,\vect{B}}(\mathsf{S}_{n_{-}}\vect {Y}_1^N|\mathsf{S}_{n_{c}}\vect {X}_1^N,\vect{B})=1$ and  $P_{\mathsf{S}_{n_{-}}\vect {Y}_2^N|\mathsf{S}_{n_{c}}\vect {X}_2^N, \vect{B}}(\mathsf{S}_{n_{-}}\vect {Y}_2^N|\mathsf{S}_{n_{c}}\vect {X}_2^N,\vect{B})=1$, since $\dmat{S}_{n_{-}}\vect {Y}_i^N,\;  i=1,2$ is not affected by interference, so it is determined by $\dmat{S}_{n_c}\vect{X}_i^N,\; i=1,2$. Similarly, we have that 
$P_{\mathsf{S}_{n_{c}}\vect {X}_1^N|\vect {X}_1^N,\vect{B}}(\mathsf{S}_{n_{c}}\vect {X}_1^N|\vect {X}_1^N,\vect{B})=1$ and $P_{\mathsf{S}_{n_{c}}\vect {X}_2^N|\vect {X}_2^N,\vect{B}}(\dmat{S}_{n_{c}}\vect {X}_2^N|\vect {X}_2^N,\vect{B})=1$. 

We next note that, for the underlying distribution in \eqref{underlyingP}, the following identities hold w.p.\ 1: 
\begin{itemize}[leftmargin=21pt,labelsep=7pt]
\item [(i1)] $ \vect {Y}_1^N\oplus \mathsf{S}_{n_{d}} \vect {X}_1^{N}= \mathsf{S}_{n_{c}}\vect {X}_2^N$
\item [(i2)] $ \vect {Y}_2^N\oplus \mathsf{S}_{n_{d}}\vect {X}_2^{N}= \mathsf{S}_{n_{c}}\vect {X}_1^N$
\item [(i3)] $P_{\mathsf{S}_{n_{c}}\vect {X}_1^N|\vect {X}_1^N,\vect{B}}(\mathsf{S}_{n_{c}}\vect {X}_1^N|\vect {X}_1^N,\vect{B}=[1,1])=1$
\item [(i4)] $P_{\mathsf{S}_{n_{c}}\vect {X}_2^N|\vect {X}_2^N,\vect{B}}(\mathsf{S}_{n_{c}}\vect {X}_2^N|\vect {X}_2^N,\vect{B}=[1,1])=1$
\item [(i5)] $P_{\vect {Y}_1^N|\vect {X}_1^N,\dmat{S}_{n_{c}}\vect {X}_1^N,\vect{B}}(\vect {Y}_1^N|\vect {X}_1^N,\dmat{S}_{n_{c}}\vect {X}_1^N,\vect{B}=[1,1])=P_{\mathsf{S}_{n_{c}}\vect {X}_2^N|\vect{B}}(\vect {Y}_1^N\oplus \mathsf{S}_{n_{d}}\vect {X}_1^{N}|\vect{B}=[1,1])$
\item [(i6)] $P_{\vect {Y}_2^N|\vect {X}_2^N,\dmat{S}_{n_{c}}\vect {X}_2^N,\vect{B}}(\vect {Y}_2^N|\vect {X}_2^N,\dmat{S}_{n_{c}}\vect {X}_2^N,\vect{B}=[1,1])= P_{\mathsf{S}_{n_{c}}\vect {X}_1^N|\vect{B}}(\vect {Y}_2^N\oplus \mathsf{S}_{n_{d}}\vect {X}_2^{N}|\vect{B}=[1,1])$
\end{itemize}
We combine these identities with \eqref{Pi12}, \eqref{i11}--\eqref{Px15} to obtain
 \begin{equation}
\begin{aligned}
\epsilon_{11}  &= \Pr\left\{\frac{1}{N}(\infr_1+\infr_2)\leq \Gamma \Big|\;  \vect{B}=\left[1,1\right]\right\}\\
&=\Pr\left\{\frac{1}{N}\biggl(\log  P_{\mathsf{S}_{n_{c}}\vect {X}_2^N|\vect{B}}(\mathsf{S}_{n_{c}}\vect {X}_2^N|\vect{B})
-\log P_{\mathsf{\mathsf{L}}_{n_{+}}\vect {Y}_1^N|\mathsf{S}_{n_{c}}\vect {X}_1^N,\mathsf{S}_{n_{-}}\vect {Y}_1^N,\vect{B}}(\mathsf{L}_{n_{+}}\vect {Y}_1^N|\mathsf{S}_{n_{c}}\vect {X}_1^N,\mathsf{S}_{n_{-}}\vect {Y}_1^N,\vect{B})
\Biggr.\right.\\
&\qquad\quad\;{}-\log P_{\mathsf{S}_{n_{c}}\vect {X}_1^N|\vect{B}}(\mathsf{S}_{n_{c}}\vect {X}_1^N|\vect{B})+ \log P_{\mathsf{S}_{n_{c}}\vect {X}_1^N| \vect{Y}_1^N,\vect{B}}(\mathsf{S}_{n_{c}}\vect {X}_1^N | \vect{Y}_1^N,\vect{B})\\
&\qquad\quad\;{}+\log P_{\mathsf{S}_{n_{c}}\vect {X}_1^N|\vect{B}}(\mathsf{S}_{n_{c}}\vect {X}_1^N|\vect{B}) -\log P_{\mathsf{L}_{n_{+}}\vect {Y}_2^N|\mathsf{S}_{n_{c}}\vect {X}_2^N,\mathsf{S}_{n_{-}}\vect {Y}_2^N,\vect{B}}(\mathsf{L}_{n_{+}}\vect {Y}_2^N|\mathsf{S}_{n_{c}}\vect {X}_2^N,\mathsf{S}_{n_{-}}\vect {Y}_2^N,\vect{B}) \\
&\qquad\quad\;\Bigg.\Bigg.{}-\log P_{\mathsf{S}_{n_{c}}\vect {X}_2^N|\vect{B}}(\mathsf{S}_{n_{c}}\vect {X}_2^N|\vect{B})+\log P_{\mathsf{S}_{n_{c}}\vect {X}_2^N|\vect{Y}_2^N,\vect{B}}(\mathsf{S}_{n_{c}}\vect {X}_2^N|\vect{Y}_2^N,\vect{B})\bigg)\leq \Gamma \Big|\;  \vect{B}=\left[1,1\right] \Bigg\}\\
&\geq \Pr\Bigg\{-\frac{1}{N}\log P_{\mathsf{\mathsf{L}}_{n_{+}}\vect {Y}_1^N|\mathsf{S}_{n_{c}}\vect {X}_1^N,\mathsf{S}_{n_{-}}\vect {Y}_1^N,\vect{B}}(\mathsf{L}_{n_{+}}\vect {Y}_1^N|\mathsf{S}_{n_{c}}\vect {X}_1^N,\mathsf{S}_{n_{-}}\vect {Y}_1^N,\vect{B})\\
&\qquad\quad\;\Bigg.{}-\frac{1}{N}\log P_{\mathsf{L}_{n_{+}}\vect {Y}_2^N|\mathsf{S}_{n_{c}}\vect {X}_2^N,\mathsf{S}_{n_{-}}\vect {Y}_2^N,\vect{B}}(\mathsf{L}_{n_{+}}\vect {Y}_2^N|\mathsf{S}_{n_{c}}\vect {X}_2^N,\mathsf{S}_{n_{-}}\vect {Y}_2^N,\vect{B})\leq \Gamma\Big|\;  \vect{B}=\left[1,1\right] \Bigg\}\label{e11_3}
\end{aligned}
\end{equation}
where in the last step we canceled the terms $\log P_{\mathsf{S}_{n_{c}}\vect {X}_1^N|\vect{B}}(\mathsf{S}_{n_{c}}\vect {X}_1^N|\vect{B})$ and $\log P_{\mathsf{S}_{n_{c}}\vect {X}_2^N|\vect{B}}(\mathsf{S}_{n_{c}}\vect {X}_2^N|\vect{B})$ and we used that, w.p.\ 1, $\log P_{\mathsf{S}_{n_{c}}\vect {X}_1^N|\vect{Y}_1^N,\vect{B}}(\mathsf{S}_{n_{c}}\vect {X}_1^N|\vect{Y}_1^N,\vect{B})\leq 0$ and $\log P_{\mathsf{S}_{n_{c}}\vect {X}_2^N|\vect{Y}_2^N,\vect{B}}(\mathsf{S}_{n_{c}}\vect {X}_2^N|\vect{Y}_2^N,\vect{B})\leq 0$.

By \eqref{lemma 10} and \eqref{lemma 5} in Lemma~\ref{Lemma Verdu-Han}, the conditional sup-entropy rates satisfy
\begin{eqnarray}
\overline{H}(\mathsf{L}_{n_{+}}\vect {Y}_i^N|\mathsf{S}_{n_{c}}\vect {X}_i^N,\mathsf{S}_{n_{-}}\vect {Y}_i^N,\vect{B})\leq\overline{H}(\mathsf{L}_{n_{+}}\vect {Y}_i^N|\vect{B})<\max\{(n_d-n_c)^+,n_c\}, \quad i=1,2.
\label{He11_3gvv}
 \end{eqnarray}
 Then, setting $\Gamma=2\max\{(n_d-n_c)^+,n_c\}+2\delta$ for some arbitrary $\delta>0$, we obtain from \eqref{Pe_LB} that \eqref{e11_3} can be lower-bounded by
 \begin{equation}
\begin{aligned}
\epsilon_{11} 
&\geq   {\Pr}\Biggl\{-\frac{1}{N}\log P_{\mathsf{\mathsf{L}}_{n_{+}}\vect {Y}_1^N|\mathsf{S}_{n_{c}}\vect {X}_1^N,\mathsf{S}_{n_{-}}\vect {Y}_1^N,\vect{B}}(\mathsf{L}_{n_{+}}\vect {Y}_1^N|\mathsf{S}_{n_{c}}\vect {X}_1^N,\mathsf{S}_{n_{-}}\vect {Y}_1^N,\vect{B})\Biggr.\\
&\qquad\quad\;\Biggl.< \overline{H}(\mathsf{L}_{n_{+}}\vect {Y}_1^N|\mathsf{S}_{n_{c}}\vect {X}_1^N,\mathsf{S}_{n_{-}}\vect {Y}_1^N,\vect{B}) +\delta\;\Big|\;  \vect{B}=\left[1,1\right]\Biggr\}\\
&\quad\;{}-\Pr\Biggl\{ \frac{1}{N}\log P_{\mathsf{L}_{n_{+}}\vect {Y}_2^N|\mathsf{S}_{n_{c}}\vect {X}_2^N,\mathsf{S}_{n_{-}}\vect {Y}_2^N,\vect{B}}(\mathsf{L}_{n_{+}}\vect {Y}_2^N|\mathsf{S}_{n_{-}}\vect {X}_2^N,\mathsf{S}_{n_{-}}\vect {Y}_2^N,\vect{B})\Biggr.\\
&\qquad\qquad\;\:\Biggl.\geq \overline{H}(\mathsf{L}_{n_{+}}\vect {Y}_2^N|\mathsf{S}_{n_{c}}\vect {X}_2^N,\mathsf{S}_{n_{-}}\vect {Y}_2^N,\vect{B})+ \delta\;\Big|\;  \vect{B}=\left[1,1\right]\Biggr\}.\label{epsilon_3}
\end{aligned}
\end{equation}

By the definition of $\overline{H}(\mathsf{L}_{n_{+}}\vect {Y}_1^N|\mathsf{S}_{n_{c}}\vect {X}_1^N,\mathsf{S}_{n_{-}}\vect {Y}_1^N,\vect{B})$, the fist probability on the RHS of \eqref{epsilon_3} tends to 1 as $N\to\infty$. Similarly, by the definition of $\overline{H}(\mathsf{L}_{n_{+}}\vect {Y}_2^N|\mathsf{S}_{n_{c}}\vect {X}_2^N,\mathsf{S}_{n_{-}}\vect {Y}_2^N,\vect{B})$, the second probability on the RHS of \eqref{epsilon_3} tends to 0 as $N\to\infty$. This demonstrates that if $\Gamma >2\max\{(n_d-n_c)^+,n_c\}$, then the lower bound in \eqref{e11_3} tends to 1 as $N\to\infty$. Thus, $\epsilon_{11}\to0$ as $N\to\infty$ only if
\begin{equation}
\Gamma \leq2\max\{(n_d-n_c)^+,n_c\}.
\end{equation}

\section{Achievability for Local CSIRT}\label{App: Local_CSIRT}

\makeatletter 
\setcounter{figure}{6} 
\@addtoreset{figure}{section}
\renewcommand{\thefigure}{D\arabic{figure}}
\makeatletter

\makeatletter 
\setcounter{equation}{0} 
\@addtoreset{equation}{section}
\renewcommand{\theequation}{D\arabic{equation}}
\makeatletter

In this appendix we present the achievability schemes for local CSIRT.

\subsubsection{Very Weak Interference}\label{Local_CSIRT_VWI} 
The sum rate \eqref{VWI_local_CSIRT} coincides with that of local CSIR, which in this interference region is equal to the sum rate of global CSIRT. The achievability scheme presented in Section~\ref{Ap:Ach-NoCSI-1} is thus optimal for local CSIRT and VWI.

\subsubsection{Weak Interference}\label{Local_CSIRT_WI} 
We follow a random-coding argument where the codebooks of Tx$_1$ and Tx$_2$ are drawn i.i.d.\ at random according to the distribution depicted in Figure~\ref{Fig: WI_CSIRT}. Specifically,  we divide the transmitted signal by Tx$_1$ into three regions. For each symbol (corresponding to a coherence block) we denote the bits in regions \framebox[0.3cm][c]{\footnotesize $A$}, \framebox[0.3cm][c]{\footnotesize $B$} and \framebox[0.3cm][c]{\footnotesize $C$} by $\mat{X}_1^A$, $\mat{X}_1^B$ and $\mat{X}_1^C$, respectively.
In each region the bits are i.i.d.\, but they follow a different distribution. 
\begin{itemize}
\item Regions \framebox[0.3cm][c]{\footnotesize $A$} and \framebox[0.3cm][c]{\footnotesize $C$}: The bits $\mat{X}_1^A$ and $\mat{X}_1^C$ are i.i.d.\ with marginal probability mass function (pmf)
\begin{eqnarray}
P_{X_1|B_1}(1|0)=P_{X_1|B_1}(1|1)=\tfrac{1}{2}.\label{p12_1}
\end{eqnarray}

\item Region \framebox[0.3cm][c]{\footnotesize $B$}: The bits $\mat{X}_1^B$ are i.i.d.\ with marginal pmf
\begin{eqnarray}
P_{X_1|B_1}(1|0)&=&{p}_1\label{p31_1}\\
P_{X_1|B_1}(1|1)&=& {p}_2\label{p32_1}\\
\qquad \qquad\quad 
P_{X_1}(1)&=&{p}_3=(1-p){p}_1+p{p}_2.\label{p33_1}
\end{eqnarray}

\end{itemize}
We further assume that $\mat{X}_1^A,\mat{X}_1^B$ and $\mat{X}_1^C$ are mutually independent. For Tx$_2$, the input distributions coincide with that of Tx$_1$ in the corresponding regions but with probabilities $q_i$ instead of $p_i$, with $i=1,2$. Evaluating $I(\mat{X}_1;\mat{Y}_1|B_1)$ for these distributions,
it follows that user 1 achieves the rate 
 \begin{equation}
\begin{aligned}
R_1=&(1-p)[(n_d-n_c)H_b(\tfrac{1}{2})+(2n_c-n_d)H_b({p}_1)+(n_d-n_c)]+p(n_d-n_c)H_b(\tfrac{1}{2})\\
& {}+p(2n_c-nd)\left[\Hsum({p}_2,\tfrac{1}{2})-H_b(q_3)\right]+p(2n_d-3n_c)(\Hsum(\tfrac{1}{2},\tfrac{1}{2})-H_b(\tfrac{1}{2}))\\
& {}+p(2n_c-n_d)(\Hsum(\tfrac{1}{2},q_3)-H_b({q}_3))\\
=&(n_d-n_c)+(1-p)[(n_d-n_c)+(2n_c-n_d)H_b(p_1)]+p(2n_c-n_d)(1-H_b(q_3)).
\end{aligned}
\end{equation}

Similarly, for user 2, we obtain \eqref{WI_local_CSIRT_R2}.

\begin{figure}[tbp]
		\centering
		\includegraphics[width=0.3\linewidth]{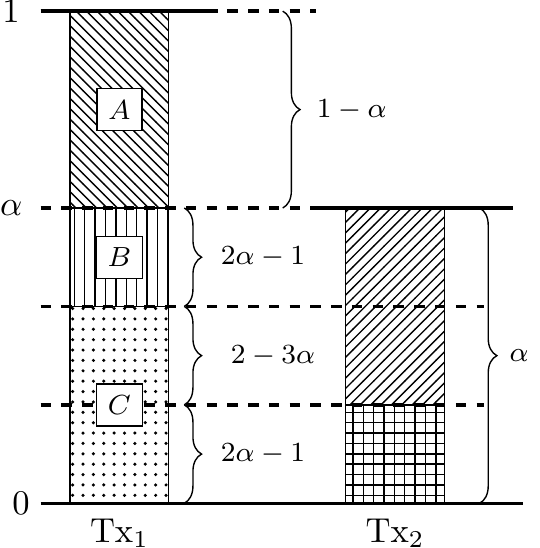}
		\caption{Normalized signal levels at Rx$_1$ (WI).}
		\label{Fig: WI_CSIRT}
\end{figure}

\paragraph{Moderate Interference}
We follow along similar lines to obtain the achievable rates for MI. However, in contrast to WI, for MI we need to consider different input distributions, depending on the value of $\alpha$.
In the proofs, we shall make use of the following auxiliary results, which can be proven by direct evaluation of the entropies considered.
\begin{lemma}\label{Lemma:XX}
Let $X$ and $\tilde{X}$ be two binary random variables  with joint pmf $P_{X\tilde{X}}(0,0)=P_{X\tilde{X}}(1,1)=\frac{\eta}{2}$, and $P_{X\tilde{X}}(0,1)=P_{X\tilde{X}}(1,0)=\tfrac{1-\eta}{2}$.  Then, 
\begin{eqnarray}
H(X|\tilde{X})=H(\tilde{X}|X)=H_b(\eta).
\end{eqnarray}
\end{lemma}

\begin{lemma}\label{Lemma:XXB}
Let $X,\tilde{X}$ and $B$ be  binary random variables with joint pmf $P_{X\tilde{X}B}(0,0,0)=P_{X\tilde{X}B}(1,1,0)=\frac{\eta_1}{2}(1-p)$, $P_{X\tilde{X}B}(0,1,0)=P_{X\tilde{X}B}(1,0,0)=\tfrac{1-\eta_1}{2}(1-p)$, $P_{X\tilde{X}B}(0,0,1)=P_{X\tilde{X}B}(1,1,1)=\frac{\eta_2}{2}p$, and $P_{X\tilde{X}B}(0,1,1)=P_{X\tilde{X}B}(1,0,1)=\tfrac{1-\eta_2}{2}p$. Then, 
\begin{eqnarray}
H(\tilde{X}|X,B)=(1-p)H_b(\eta_1)+pH_b(\eta_2)
\end{eqnarray}
and 
\begin{eqnarray}
H(\tilde{X}|X)=H_b\bigl((1-p)\eta_1+p\eta_2\bigr).
\label{eqn:XXB-Hxx}
\end{eqnarray}
\end{lemma}

\begin{lemma}\label{Lemma:X+Wsum}
Let $X_1$ and $\tilde{X}_1$ be two binary random variables with joint pmf $P_{X_1\tilde{X}_1}(0,0)=P_{X_1\tilde{X}_1}(1,1)=\frac{\eta_1}{2}$ and $P_{X_1\tilde{X}_1}(0,1)=P_{X_1\tilde{X}_1}(1,0)=\tfrac{1-\eta_1}{2}$.
Similarly, let the pair of binary random variables $X_2$ and $\tilde{X}_2$ be independent of $X_1$ and $\tilde{X}_1$ have the same joint pmf but with parameter $\eta_2$. Further let $Z \sim Ber(p_z)$. Then,
\begin{eqnarray}\label{Hsum1}
H(X_1 | \tilde{X}_1 \oplus \tilde{X}_2, X_2) =
H(\tilde{X}_1 \oplus \tilde{X}_2 | X_1, X_2) = 
H_{\text{sum}}(\eta_1,\eta_2)
\end{eqnarray}
and
\begin{eqnarray}\label{Hsum2}
H(X_1\oplus Z | \tilde{X}_1 \oplus \tilde{X}_2, X_2) =
H_{\text{sum}}(p_z,\eta_1(1-\eta_2)+\eta_2(1-\eta_1)).
\end{eqnarray}
\end{lemma}

To derive the achievable rates for MI, we again follow a random-coding argument where the codebooks are drawn i.i.d.\ at random. We next describe the input distributions for different values of~$\alpha$:
\paragraph{MI, $\frac{2}{3}<\alpha\leq \frac{3}{4}$}\label{Local_CSIRT_MI_1}

Consider the regions shown in Figure~\ref{Fig: MI_1_CSIRT} for the received signal at Rx$_1$.
For the transmitted signal $X_1$, we denote the bits in region \framebox[0.3cm][c]{\footnotesize $j$}
by $\mat{X}_1^j$, $j=\{A,\ldots,F\}$. In each of these regions we consider the following input distributions:
\begin{itemize}

\item Regions \framebox[0.3cm][c]{\footnotesize $A$} and \framebox[0.3cm][c]{\footnotesize $\tilde{A}$}:
We group the bits $\mat{X}_1^A$ and $\mat{X}_1^{\tilde{A}}$ in pairs, and we let each of these pairs $(X_1, \tilde{X}_1)$ be i.i.d.
and have the distribution from Lemma~\ref{Lemma:XXB} with $\eta_2=1$, i.e., their marginal pmf is
\begin{eqnarray} \label{Dist_MI_1}
P_{X_1 \tilde{X}_1 | B_1}(0,0|0)=P_{X_1 \tilde{X}_1 | B_1}(1,1|0) &=&\frac{\eta_1}{2}\\
P_{X_1 \tilde{X}_1 | B_1}(0,1|0)=P_{X_1 \tilde{X}_1 | B_1}(1,0|0) &=&\frac{1-\eta_1}{2}\\
P_{X_1 \tilde{X}_1 | B_1}(0,0|1)=P_{X_1 \tilde{X}_1 | B_1}(1,1|1) &=&\frac{1}{2}\\
P_{X_1 \tilde{X}_1 | B_1}(0,1|1)=P_{X_1 \tilde{X}_1 | B_1}(1,0|1)&=&0\\
\qquad\qquad \qquad \qquad\qquad \quad P_{\tilde{X}_1 | X_1}(1|1)&=&\tilde{\eta}=p+\eta_1(1-p) \label{Dist_MI_2}.
\end{eqnarray}
where $\tfrac{1}{2}\leq\eta_1\leq 1$. 
\item Regions \framebox[0.3cm][c]{\footnotesize $B$} and \framebox[0.3cm][c]{\footnotesize $F$}: The bits $\mat{X}_1^B$ and $\mat{X}_1^F$ are i.i.d.\ with marginal pmf
\begin{eqnarray}
P_{X_1|B_1}(1|0)=
P_{X_1|B_1}(1|1)&=& \tfrac{1}{2}.\label{p12_2}
\end{eqnarray}

\item Region \framebox[0.3cm][c]{\footnotesize $C$}: The bits $\mat{X}_1^C$  are i.i.d.\ with marginal pmf
\begin{eqnarray}
P_{X_1|B_1}(1|0)&=&{p}_1\label{p21_2}\\
P_{X_1|B_1}(1|1)&=&{p}_2\label{p22_2}\\
\qquad\qquad \quad P_{X_1}(1)&=&{p}_3=(1-p){p}_1+p{p}_2.\label{p33_3}
\end{eqnarray}

\item Region \framebox[0.3cm][c]{\footnotesize $D$}: The bits $\mat{X}_1^D$ are i.i.d.\ with marginal pmf
\begin{eqnarray}
P_{X_1|B_1}(1|0)&=&\tilde{p}_1\label{p41_2}\\
P_{X_1|B_1}(1|1)&=&\tilde{p}_2\label{p42_2}\\
\qquad\qquad \quad P_{X_1}(1)&=&\tilde{p}_3=(1-p)\tilde{p}_1+p\tilde{p}_2.\label{p43_3}
\end{eqnarray}

\item Region \framebox[0.3cm][c]{\footnotesize $E$}: The bits $\mat{X}_1^E$ are i.i.d.\ with marginal pmf
\begin{eqnarray}
P_{X_1|B_1}(1|0)&=&\hat{p}_1\label{p51_2}\\
P_{X_1|B_1}(1|1)&=& 0\label{p52_2}\\
\qquad \qquad\quad P_{X_1}(1)&=&\hat{p}_3=(1-p)\hat{p}_1.\label{p 53_3}
\end{eqnarray}

\end{itemize}
Furthermore, we assume that $\mat{X}_1^j$, $j=\{A,\ldots,F\}$ are independent.
For user 2, the input distributions coincide with that of user 1 in the corresponding regions, but with parameters $q_i$ instead of $p_i$, $\tilde{q}_i$ instead of $\tilde{p}_i$, $\hat{q}_1$ instead of $\hat{p}_1$, and $\gamma_i$ instead of $\eta_i$.

\begin{figure}[tbp]
		\centering
		\includegraphics[width=0.48\linewidth]{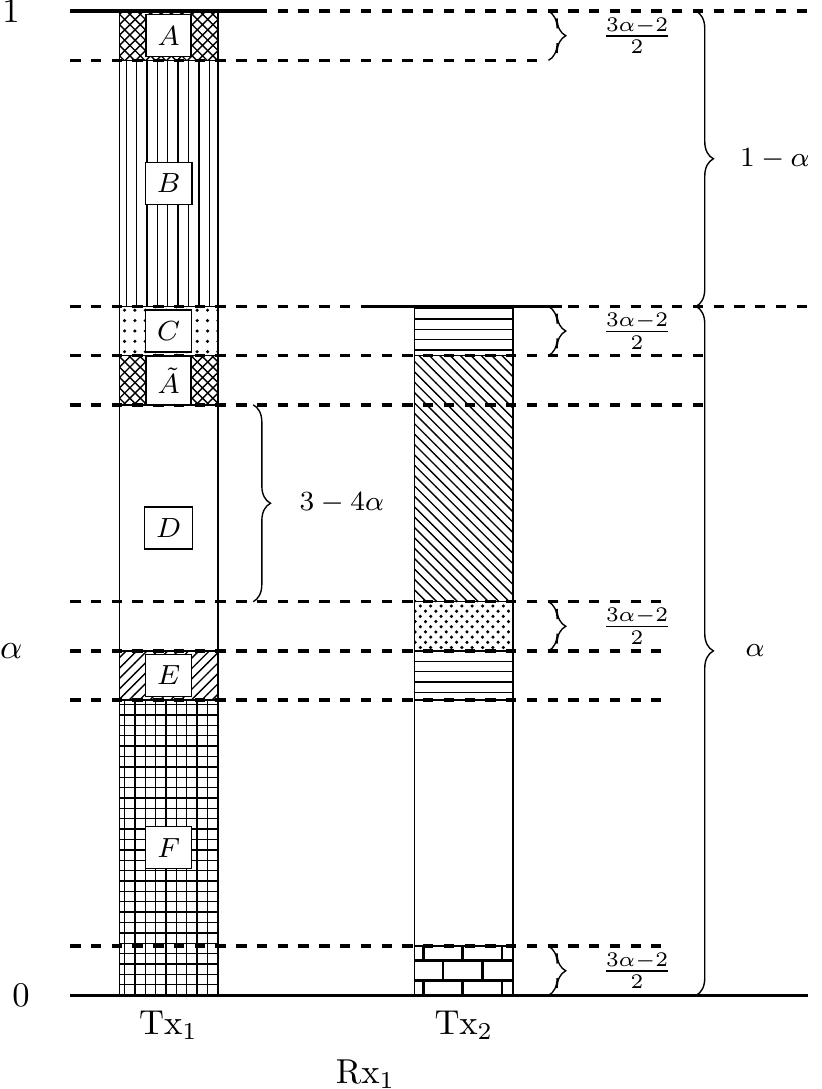}
		\caption{Normalized signal levels at Rx$_1$ (MI) for $\tfrac{2}{3}<\alpha\leq \tfrac{3}{4}$.}
		\label{Fig: MI_1_CSIRT}
\end{figure}

From the random-coding argument, we know that the rate $R_1 = \frac{1}{N} I(\mat{X}_1^{K};\mat{Y}_1^{K}|B_1)$ is achievable.
Since the distributions considered are temporally i.i.d., it suffices to evaluate $I(\mat{X}_1;\mat{Y}_1|B_1)$
for one coherence block, obtaining
\begin{equation}
\begin{aligned}
\label{Eq: MI_1}
T R_1 = & I(\mat{X}_1;\mat{Y}_1^{A}|B_1)+I(\mat{X}_1;\mat{Y}_1^{\tilde{A}}|\mat{Y}_1^{A},B_1)+I(\mat{X}_1;\mat{Y}_1^{E}|\mat{Y}_1^{A},\mat{Y}_1^{\tilde{A}},B_1)+I(\mat{X}_1;\mat{Y}_1^{C}|\mat{Y}_1^{A},\mat{Y}_1^{\tilde{A}},\mat{Y}_1^E,B_1)\\
&{}+I(\mat{X}_1;\mat{Y}_1^{B}|\mat{Y}_1^{A},\mat{Y}_1^{\tilde{A}},\mat{Y}_1^E,\mat{Y}_1^C,B_1)+I(\mat{X}_1;\mat{Y}_1^{D}|\mat{Y}_1^{A},\mat{Y}_1^{\tilde{A}},\mat{Y}_1^E,\mat{Y}_1^C,\mat{Y}_1^B,B_1)\\
&{}+I(\mat{X}_1;\mat{Y}_1^{F}|\mat{Y}_1^{A},\mat{Y}_1^{\tilde{A}},\mat{Y}_1^E,\mat{Y}_1^C,\mat{Y}_1^B,\mat{Y}_1^D,B_1)\\
=&(1-p)[H(\mat{X}_1^{A}|B_1=0)+H(\mat{X}_1^{\tilde{A}}|\mat{X}_1^A,B_1=0)+H(\mat{X}_1^{E}|B_1=0)+H(\mat{X}_1^{C}|B_1=0)\\
& {}+H(\mat{X}_1^{B}|B_1=0)+H(\mat{X}_1^{D}|B_1=0)+H(\mat{X}_1^{F}|B_1=0)]\\
&{}+p[H(\mat{X}_1^{A}|B_1=1)+H(\mat{X}_1^{\tilde{A}}\oplus \mat{X}_2^{\tilde{A}}|\mat{X}_1^A,B_1=1)-H(\mat{X}_2^{\tilde{A}})+H(\mat{X}_1^{{E}}\oplus \mat{X}_2^{{E}}|B_1=1)\\
&{}-H(\mat{X}_2^E)+H(\mat{X}_1^{{C}}\oplus \mat{X}_2^{{C}}|\mat{X}_1^{{E}}\oplus \mat{X}_2^{{E}},B_1=1)-H(\mat{X}_2^C|\mat{X}_2^E)+H(\mat{X}_1^{{B}}|B_1=1)\big.\\
&{}+H(\mat{X}_1^{{D}}\oplus \mat{X}_2^{{D}}|B_1=1)-H(\mat{X}_2^D)+H(\mat{X}_1^{{F}}\oplus \mat{X}_2^{{F}}|B_1=1)-H(\mat{X}_2^F)].
\end{aligned}
\end{equation}

By Lemma~\ref{Lemma:XXB}, we have that 
\begin{equation}
H(\mat{X}_1^{\tilde{A}}|\mat{X}_1^A,B_1=0)=T\tfrac{3n_c-2n_d}{2} H_b(\eta_1).
\end{equation}
Furthermore, by  Lemma~\ref{Lemma:X+Wsum}, we have that
\begin{eqnarray}
H(\mat{X}_1^C\oplus \mat{X}_2^C|\mat{X}_1^E\oplus \mat{X}_2^E,B_1=1) &=& H(\mat{X}_1^C\oplus \mat{X}_2^C|\mat{X}_2^E,B_1=1) \nonumber\\
& = & T \tfrac{3n_c-2n_d}{2}\Hsum(p_2,\tilde{\gamma})
\end{eqnarray}
 because for the bits $\mat{X}_1^C$, $P_{X_1^C|B_1}(1|1)=0$. Similarly, we have
 \begin{equation}
 \Hsum(\mat{X}_1^{\tilde{A}}\oplus \mat{X}_2^{\tilde{A}}|\mat{X}_1^A,B_1=1)= T \tfrac{3n_c-2n_d}{2}\Hsum(1,\tfrac{1}{2})=T\tfrac{3n_c-2n_d}{2}.
 \end{equation}
 The terms in the other regions follow analogously. Therefore, using \eqref{Eq: MI_1} we obtain the rate
 
\begin{myequation}
\begin{aligned}
R_1
=&(n_d-n_c)+(1-p)\left[\left(\tfrac{3n_c-2n_d}{2}\right)\left(H_b(\eta_1)+H_b(\hat{p}_1)+H_b({p}_1)\right)+\left(\tfrac{4n_d-5n_c}{2}\right)H_b(\tilde{p}_1)+(n_d-n_c)\right]\\
&{}+ p\left[\left(\tfrac{3n_c-2n_d}{2}\right)\left(1+\Hsum(p_2,\tilde{\gamma})-H_b(\tilde{\gamma})+\Hsum(\tilde{p}_2,q_3)-H_b(q_3)-H_b(\hat{q}_3)\right)\right.\\
&{}+\left(\tfrac{4n_d-5n_c}{2}\right)\left(1-H_b(\tilde{q}_3)\right)\big].
\end{aligned}
\end{myequation}
Similarly, user 2  achieves the rate \eqref{MI_local_CSIRT_sr1_R2}.

\paragraph{MI, $\frac{3}{4}\leq\alpha\leq\frac{4}{5}$ }

We use a similar transmission strategy as for the case where $\tfrac{2}{3}\leq\alpha\leq\tfrac{3}{4}$ (Section~\ref{Local_CSIRT_MI_1}), but where the regions have different sizes; see Figure~\ref{Fig: A9}a. Following the same steps as in Section~\ref{Local_CSIRT_MI_1}, we obtain the achievable rates \eqref{MI_local_CSIRT_sr2_R1} for $R_1$ and  \eqref{MI_local_CSIRT_sr2_R2} for $R_2$.

  \begin{figure}[tbp]
  	\centering
  	\subfigure[\label{Fig: MI_2_CSIRT}]{\includegraphics[width=0.44\linewidth]{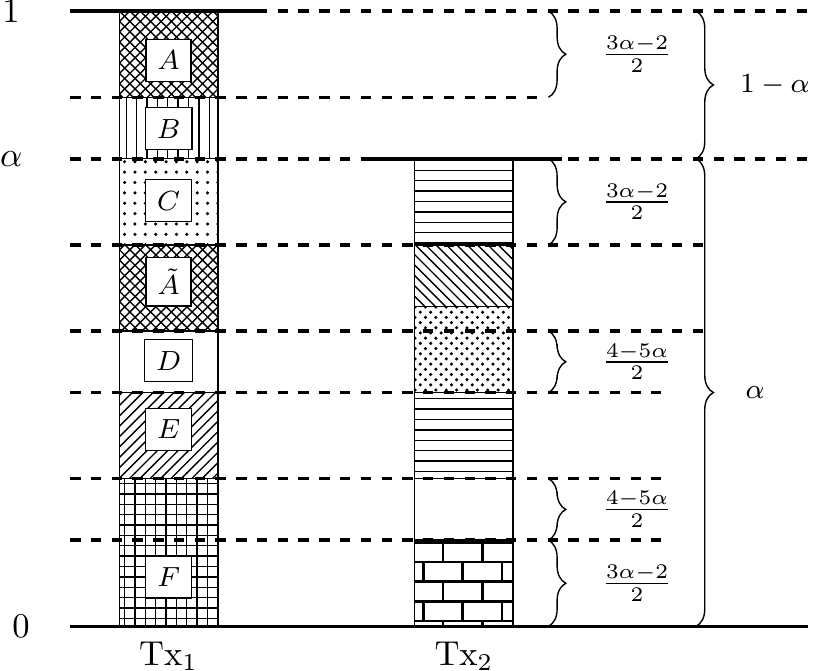}}\quad
    \subfigure[
     		\label{Fig: MI_3_CSIRT}]{\includegraphics[width=0.37\linewidth]{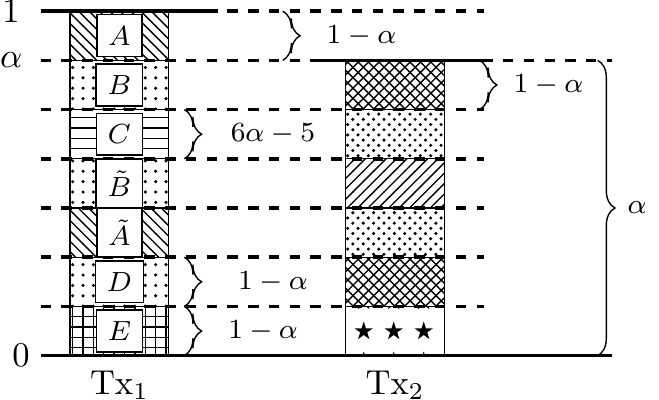}}\\	
  	\caption{Normalized signal levels at Rx$_1$. (\textbf{a}) (MI) for $\tfrac{3}{4}\leq\alpha\leq\tfrac{4}{5}$; (\textbf{b}) (MI) for $\alpha=\tfrac{6}{7}$.} \label{Fig: A9}
  \end{figure}

\paragraph{MI, $\alpha=\frac{6}{7}$}\label{Local_CSIRT_MI_6/7}
In this subsection we consider the particular case $\alpha=\tfrac{6}{7}$. The proposed achievability scheme features two nested regions with a certain correlation. In particular, we consider the division of the bit-pipes for the transmitted signal Tx$_1$ in the subregions shown in Figure~\ref{Fig: A9}b. The input distributions considered in each of these regions are described next (for Tx$_2$, we shall consider the same input distributions parametrized by $q_i$, $\hat{q}_1$, $\gamma_1$ and $\gamma'$, instead of $p_i$, $\hat{p}_1$, $\eta_1$ and $\eta'$):

\begin{itemize}
\item Regions \framebox[0.3cm][c]{\footnotesize $A$} and \framebox[0.3cm][c]{\footnotesize $\tilde{A}$}: The bits $\mat{X}_1^A$ and $\mat{X}_1^{\tilde{A}}$ are grouped in i.i.d. pairs with the marginal pmf given by \eqref{Dist_MI_1}--\eqref{Dist_MI_2}.
\item Regions \framebox[0.3cm][c]{\footnotesize $B$} and \framebox[0.3cm][c]{\footnotesize $\tilde{B}$}: The bits $\mat{X}_1^B$ and $\mat{X}_1^{\tilde{B}}$ are grouped in  i.i.d. pairs with marginal pmf
\begin{eqnarray}\label{Dist_MI_3}
P_{X_1 \tilde{X}_1 | B_1}(0,0|0)=P_{X_1 \tilde{X}_1 | B_1}(1,1|0) &=&\frac{\eta'}{2}\\
P_{X_1 \tilde{X}_1 | B_1}(0,1|0)=P_{X_1 \tilde{X}_1 | B_1}(1,0|0) &=&\frac{1-\eta'}{2}\\
P_{X_1 \tilde{X}_1 | B_1}(0,0|1)=P_{X_1 \tilde{X}_1 | B_1}(1,1|1) &=&\frac{\eta'}{2}\\
P_{X_1 \tilde{X}_1 | B_1}(0,1|1)=P_{X_1 \tilde{X}_1 | B_1}(1,0|1) &=&\frac{1-\eta'}{2}\\
\qquad \qquad \qquad \qquad \qquad \quad P_{\tilde{X}_1 | X_1}(1|1) &=& \eta'\label{Dist_MI_4}
\end{eqnarray}
where $\tfrac{1}{2}\leq\eta'\leq 1$. 

\item Region \framebox[0.3cm][c]{\footnotesize $C$}: The bits $\mat{X}_1^C$ are i.i.d.\ with marginal pmf
\begin{eqnarray}
P_{X_1|B_1}(1|0)&=&{p}_1\\
P_{X_1|B_1}(1|1)&=&{p}_2\\
\qquad\qquad\ P_{X_1}(1)&=&{p}_3=(1-p){p}_1+p{p}_2.
\end{eqnarray}

\item Region \framebox[0.3cm][c]{\footnotesize $D$}: The bits $\mat{X}_1^D$ are i.i.d.\ with marginal pmf
\begin{eqnarray}
P_{X_1|B_1}(1|0)&=&\hat{p}_1\\
P_{X_1|B_1}(1|1)&=&0\\
\qquad\qquad \  P_{X_1}(1)&=&\hat{p}_3=(1-p)\hat{p}_1.
\end{eqnarray}

\item Region \framebox[0.3cm][c]{\footnotesize $E$}: The bits $\mat{X}_1^E$ are i.i.d.\ with marginal pmf
\begin{eqnarray}
P_{X_1|B_1}(1|0)=
P_{X_1|B_1}(1|1)=\tfrac{1}{2}.
\end{eqnarray}
\end{itemize}
Furthermore, we assume that $\mat{X}_i^j$, i=1,2, $ j=\{A,B,C,D,E\}$ are mutually independent.
For the input distributions described above, we obtain for user 1 that
\begin{myequation}
\label{Eq: MI_2}
\begin{aligned}
T R_1
=& I(\mat{X}_1;\mat{Y}_1^{A}|B_1)+I(\mat{X}_1;\mat{Y}_1^{\tilde{A}}|\mat{Y}_1^{A},B_1)+I(\mat{X}_1;\mat{Y}_1^{D}|\mat{Y}_1^{A},\mat{Y}_1^{\tilde{A}},B_1)+I(\mat{X}_1;\mat{Y}_1^{B}|\mat{Y}_1^{A},\mat{Y}_1^{\tilde{A}},\mat{Y}_1^D,B_1)\big.\\
&{}+I(\mat{X}_1;\mat{Y}_1^{\tilde{B}}|\mat{Y}_1^{A},\mat{Y}_1^{\tilde{A}},\mat{Y}_1^D,\mat{Y}_1^B,B_1)+I(\mat{X}_1;\mat{Y}_1^{C}|\mat{Y}_1^{A},\mat{Y}_1^{\tilde{A}},\mat{Y}_1^D,\mat{Y}_1^B,\mat{Y}_1^{\tilde{B}},B_1)\\
&{}\big.+I(\mat{X}_1;\mat{Y}_1^{E}|\mat{Y}_1^{A},\mat{Y}_1^{\tilde{A}},\mat{Y}_1^D,\mat{Y}_1^B,\mat{Y}_1^{\tilde{B}},\mat{Y}_1^C,B_1)\\
=&(1-p)[H(\mat{X}_1^{A}|B_1=0)+H(\mat{X}_1^{\tilde{A}}|\mat{X}_1^A,B_1=0)+H(\mat{X}_1^{D}|B_1=0)\big.\\
&  {}+H(\mat{X}_1^{B}|B_1=0)+H(\mat{X}_1^{\tilde{B}}|\mat{X}_1^B,B_1=0)+H(\mat{X}_1^{C}|B_1=0)+H(\mat{X}_1^{E}|B_1=0)]\\
&\big.{}+ p[H(\mat{X}_1^{A}|B_1=1)+H(\mat{X}_1^{\tilde{A}}\oplus \mat{X}_2^{\tilde{A}}|\mat{X}_1^A,B_1=1)-H(\mat{X}_2^{\tilde{A}})+H(\mat{X}_1^{{D}}\oplus \mat{X}_2^{{D}}|B_1=1)\big.\\
&\big.{}-H(\mat{X}_2^D)+H(\mat{X}_1^{{B}}\oplus \mat{X}_2^{{B}}|\mat{X}_1^{{D}}\oplus \mat{X}_2^{{D}},B_1=1)-H(\mat{X}_2^B|\mat{X}_2^D)\\
&\big.{}+H(\mat{X}_1^{\tilde{B}}\oplus \mat{X}_2^{\tilde{B}}|\mat{X}_1^B\oplus \mat{X}_2^B, \mat{X}_1^{{D}}\oplus \mat{X}_2^{{D}}, B_1=1)-H(\mat{X}_2^{\tilde{B}})\big.\\
&\big.{}+H(\mat{X}_1^{{C}}\oplus \mat{X}_2^{{C}}|\mat{X}_1^{\tilde{A}}\oplus \mat{X}_2^{\tilde{A}}, \mat{X}_1^A,B_1=1)-H(\mat{X}_2^C|\mat{X}_2^{\tilde{A}})+H(\mat{X}_1^{{E}}\oplus \mat{X}_2^{{E}}|B_1=1)\\
&{}-H(\mat{X}_2^E)].
\end{aligned}
\end{myequation}
We next evaluate the different terms in \eqref{Eq: MI_2} by applying Lemmas~\ref{Lemma:XXB} and \ref{Lemma:X+Wsum} to obtain 
\begin{eqnarray}
H(\mat{X}_1^{\tilde{A}}|\mat{X}_1^A,B_1=0)&=&T(n_d-n_c)H_b(\eta_1) \label{Eq1}\\ 
H(\mat{X}_1^{\tilde{B}}|\mat{X}_1^B,B_1=0)&=&T(n_d-n_c)H_b(\eta')\\ 
H(\mat{X}_1^{\tilde{A}}\oplus \mat{X}_2^{\tilde{A}}|\mat{X}_1^A,B_1=1)&=&T(n_d-n_c)\Hsum(1,\tfrac{1}{2})=T(n_d-n_c).
\end{eqnarray}
Similarly, using Lemma~\ref{Lemma:X+Wsum}, and since for $\mat{X}_1^D$ we have that $P_{X_1|B_1}(0|1)=1$, we obtain
\begin{equation}
\label{Eq2}
\begin{aligned}
H(\mat{X}_1^{\tilde{B}}\oplus \mat{X}_2^{\tilde{B}}|\mat{X}_1^B\oplus \mat{X}_2^B, \mat{X}_1^{{D}}\oplus \mat{X}_2^{{D}}, B_1=1) =&
H(\mat{X}_1^{\tilde{B}}\oplus \mat{X}_2^{\tilde{B}}|\mat{X}_1^B\oplus \mat{X}_2^B,  \mat{X}_2^{{D}}, B_1=1) \\
=& T(n_d-n_c)\Hsum(q_3,\eta'(1-\tilde{\gamma})+(1-\eta')\tilde{\gamma}). 
\end{aligned}
\end{equation}
Combining \eqref{Eq1}--\eqref{Eq2} with \eqref{Eq: MI_2} yields
\begin{equation}
\begin{aligned}
R_1 
=&(n_d-n_c)+(1-p)\big[(6n_c-5n_d)H_b(p_1)+(n_d-n_c)\left(2+H_b(\eta_1)+H_b(\eta')+H_b(\hat{p}_1)\right)\big]\\
&{}+ p\big[(n_d-n_c)\left(2-H_b(\tilde{\gamma})-H_b(\hat{q}_3)+\Hsum(\eta'(1-\tilde{\gamma})+(1-\eta')\tilde{\gamma},q_3)-H_b(q_3)\right)\\
&{}+(6n_c-5n_d)\left(\Hsum(p_2,\gamma')-H_b(\gamma')\right)\big].
\end{aligned}
\end{equation}
Following along similar lines, it can be shown that user 2 achieves the rate
\begin{equation}
\begin{aligned}
R_2=&(n_d-n_c)+(1-p)\big[(6n_c-5n_d)H_b(q_1)+(n_d-n_c)\left(2+H_b(\gamma_1)+H_b(\gamma')+H_b(\hat{q}_1)\right)\big]\\
&{}+ p\big[(n_d-n_c)\left(2-H_b(\tilde{\eta})-H_b(\hat{p}_3)+\Hsum(\gamma'(1-\tilde{\eta})+(1-\gamma')\tilde{\eta},p_3)-H_b(p_3)\right)\\
&{}+(6n_c-5n_d)\left(\Hsum(q_2,\eta')-H_b(\eta')\right)\big].
\end{aligned}
\end{equation}

\paragraph{MI, $\frac{4}{5}<\alpha<\frac{6}{7}$}\label{Local_CSIRT_MI_2}

We consider the input distribution depicted in  Figure~\ref{Fig: A10}a with:
\begin{itemize}
\item Regions \framebox[0.3cm][c]{\footnotesize $A$} and \framebox[0.3cm][c]{\footnotesize $\tilde{A}$}: The bits $(\mat{X}_1^A,\mat{X}_1^{\tilde{A}})$ are i.i.d.\, with marginal pmf given by \eqref{Dist_MI_1}-\eqref{Dist_MI_2}.
\item Regions \framebox[0.3cm][c]{\footnotesize $B$} and \framebox[0.3cm][c]{\footnotesize $\tilde{B}$}: The bits $(\mat{X}_1^B,\mat{X}_1^{\tilde{B}})$ are i.i.d.\, with marginal pmf given by \eqref{Dist_MI_3}-\eqref{Dist_MI_4}.

\item Region \framebox[0.3cm][c]{\footnotesize $C$}: The bits $\mat{X}_1^C$ are i.i.d.\ with marginal pmf
\begin{eqnarray}
P_{X_1|B_1}(1|0)&=&{p}_1\\
P_{X_1|B_1}(1|1)&=&{p}_2\\
\qquad \qquad \  P_{X_1}(1)&=&{p}_3=(1-p){p}_1+p{p}_2.
\end{eqnarray}

\item Region \framebox[0.3cm][c]{\footnotesize $D$}: The bits $\mat{X}_1^D$ are i.i.d.\ with marginal pmf
\begin{eqnarray}
P_{X_1|B_1}(1|0)&=&\hat{p}_1\\
P_{X_1|B_1}(1|1)&=&0\\
\qquad \qquad \  P_{X_1}(1)&=&\hat{p}_3=(1-p)\hat{p}_1.
\end{eqnarray}

\item Region \framebox[0.3cm][c]{\footnotesize $E$}: The bits $\mat{X}_1^E$ are i.i.d.\ with marginal pmf
\begin{eqnarray}
P_{X_1|B_1}(1|0)=
P_{X_1|B_1}(1|1)=\tfrac{1}{2}
\end{eqnarray}
\end{itemize}
Furthermore, we assume that $\mat{X}_1^j$, $ j=\{A,B,C,D,E\}$ are independent. For Tx$_2$, the input distributions coincide with that of Tx$_1$ in the corresponding regions, but with parameters $q_i$ instead of $p_i$, $\hat{q}_1$ instead of $\hat{p}_1$, $\gamma_i$ instead of $\eta_i$ and $\gamma'$ instead of $\eta'$. Following similar steps as in the previous sections, we obtain \eqref{MI_local_CSIRT_sr3_R1} for $R_1$ and  \eqref{MI_local_CSIRT_sr3_R2} for $R_2$.

\paragraph{MI, $\frac{6}{7}<\alpha< 1$}
The transmission strategy is similar to the one for $\tfrac{4}{5}< \alpha<\tfrac{6}{7}$ (Section~\ref{Local_CSIRT_MI_2}), but with different sizes for the regions \framebox[0.3cm][c]{\footnotesize $A$} - \framebox[0.3cm][c]{\footnotesize $E$}, see  Figure~\ref{Fig: A10}b.
Following similar steps as in previous sections, we  obtain~\eqref{MI_local_CSIRT_sr4_R1} for $R_1$ and  \eqref{MI_local_CSIRT_sr4_R2} for $R_2$.

\begin{figure}[tbp]
  	\centering
  	\subfigure[\label{Fig: MI_4_CSIRT}]{\includegraphics[width=0.4\linewidth]{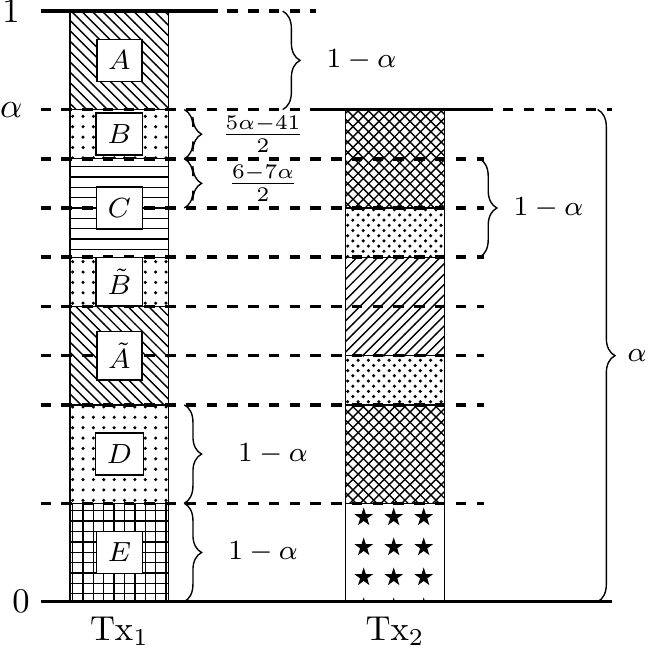}}\quad
    \subfigure[
     	\label{Fig: MI_6_CSIRT}]{\includegraphics[width=0.42\linewidth]{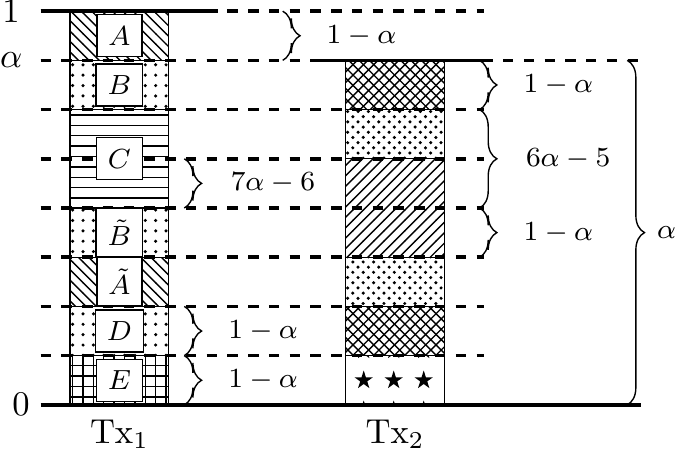}}\\	
  	\caption{Normalized signal levels at Rx$_1$. (\textbf{a}) (MI) for $\tfrac{4}{5}\leq\alpha\leq\tfrac{6}{7}$; (\textbf{b}) (MI) for $\tfrac{6}{7}\leq\alpha\leq 1$.} \label{Fig: A10}
  \end{figure}

\subsubsection{Strong Interference}
To obtain the achievable rates for SI, we again need to consider different input distributions, depending on the value of $\alpha$.

\paragraph{SI, $1\leq\alpha\leq\frac{6}{5}$}\label{Local_CSIRT_SI_1}
 
We consider the input distribution depicted in Figure~\ref{Fig: A11}a with:
\begin{itemize}
\item Regions \framebox[0.3cm][c]{\footnotesize $A$} and \framebox[0.3cm][c]{\footnotesize $\tilde{A}$}: The bits $(\mat{X}_1^A,\mat{X}_1^{\tilde{A}})$ are i.i.d.\, with marginal pmf given by \eqref{Dist_MI_1}--\eqref{Dist_MI_2}.
\item Regions \framebox[0.3cm][c]{\footnotesize $B$} and \framebox[0.3cm][c]{\footnotesize $\tilde{B}$}: The bits $(\mat{X}_1^B,\mat{X}_1^{\tilde{B}})$ are i.i.d.\, with marginal pmf given by \eqref{Dist_MI_3}--\eqref{Dist_MI_4}.

\item Region \framebox[0.3cm][c]{\footnotesize $C$}: The bits $\mat{X}_1^C$ are i.i.d.\ with marginal pmf
\begin{eqnarray}
P_{X_1|B_1}(1|0)&=&{p}_1\\
P_{X_1|B_1}(1|1)&=&{p}_2\\
\qquad \qquad \ P_{X_1}(1)&=&{p}_3=(1-p){p}_1+p{p}_2.
\end{eqnarray}
\end{itemize}
Furthermore, we assume that $\mat{X}_1^j$, $j=\{A,B,C\}$ are independent. For Tx$_2$, the input distributions coincide with that of Tx$_1$ in the corresponding regions, but with parameters $q_i$ instead of $p_i$, $\gamma_1$ instead of $\eta_1$ and $\gamma'$ instead of $\eta'$. Following similar steps as in previous sections, we obtain the achievable rate pair \eqref{SI_local_CSIRT_sr1_R1} and \eqref{SI_local_CSIRT_sr1_R2}.

\begin{figure}[tbp]
  	\centering
  	\subfigure[\label{Fig: SI_0_CSIRT}]{\includegraphics[width=0.42\linewidth]{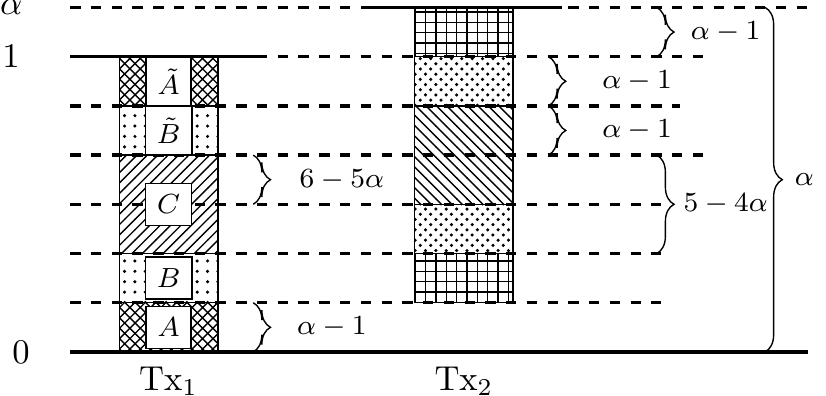}}\quad
    \subfigure[	\label{Fig: SI_1_CSIRT}]{\includegraphics[width=0.42\linewidth]{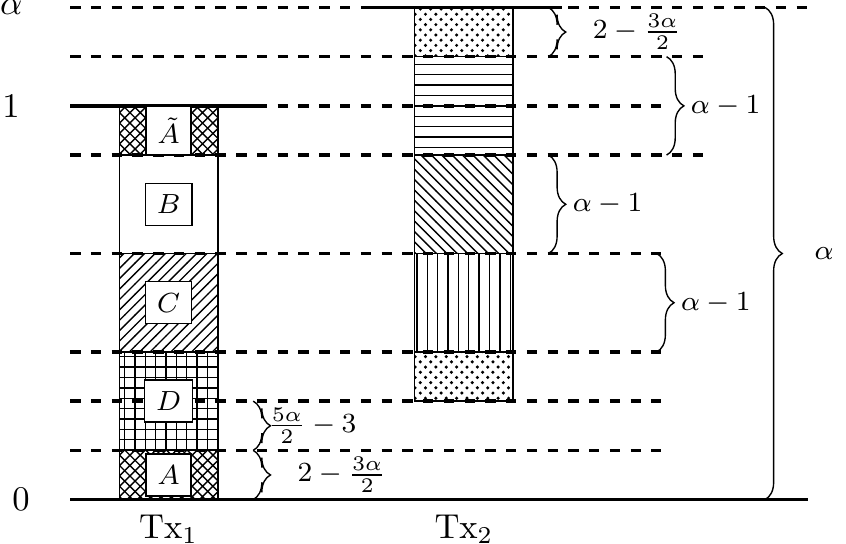}}\\	
  	\caption{Normalized signal levels at Rx$_1$. (\textbf{a}) (SI) for $1\leq\alpha\leq\tfrac{6}{5}$.; (\textbf{b}) (SI) for $\tfrac{6}{5}\leq\alpha\leq\tfrac{4}{3}$.} \label{Fig: A11}
  \end{figure}

\paragraph{SI, $\frac{6}{5}\leq\alpha\leq\frac{4}{3}$}\label{Local_CSIRT_SI_2}
We consider the input distribution depicted in Figure~\ref{Fig: A11}b with the following distributions:
\begin{itemize}
\item Regions \framebox[0.3cm][c]{\footnotesize $A$} and \framebox[0.3cm][c]{\footnotesize $\tilde{A}$}: The bits $(\mat{X}_1^A,\mat{X}_1^{\tilde{A}})$ are i.i.d.\, with marginal pmf given by \eqref{Dist_MI_1}--\eqref{Dist_MI_2}.
\item Regions \framebox[0.3cm][c]{\footnotesize $B$} and  \framebox[0.3cm][c]{\footnotesize $D$}: The bits $\mat{X}_1^B$ and $\mat{X}_1^D$ are independent and temporally i.i.d.\ with marginal pmf
\begin{eqnarray}
P_{X_1|B_1}(1|0)=
P_{X_1|B_1}(1|1)=\tfrac{1}{2}.
\end{eqnarray}

\item Region \framebox[0.3cm][c]{\footnotesize $C$}: The bits $\mat{X}_1^C$ are i.i.d.\ with marginal pmf
\begin{eqnarray}
P_{X_1|B_1}(1|0)&=&{p}_1\label{p31_}\\
P_{X_1|B_1}(1|1)&=&{p}_2\label{p32_}\\
P_{X_1}(1)&=&{p}_3=(1-p){p}_1+p{p}_2.\label{p33_}
\end{eqnarray}
\end{itemize}
Furthermore, we assume that $\mat{X}_1^j$, $j=\{A,B,C,D\}$ are independent. For Tx$_2$, the input distributions coincide with that of Tx$_1$ in the corresponding regions, but with parameters $q_i$ instead of $p_i$, $\hat{q}_1$ instead of $\hat{p}_1$ and $\gamma_1$ instead of $\eta_1$. Following similar steps as in previous sections,  we obtain the achievable rate pair \eqref{SI_local_CSIRT_sr2_R1} and \eqref{SI_local_CSIRT_sr2_R2}.

\paragraph{SI, $\frac{4}{3}\leq\alpha\leq\frac{3}{2}$}

We consider the input distribution depicted in Figure~\ref{figurea12}a with the following distributions:
\begin{itemize}
\item Regions \framebox[0.3cm][c]{\footnotesize $A$} and \framebox[0.3cm][c]{\footnotesize $\tilde{A}$}: The bits $(\mat{X}_1^A,\mat{X}_1^{\tilde{A}})$ are i.i.d.\, with marginal pmf given by \eqref{Dist_MI_1}--\eqref{Dist_MI_2}.
\item Regions \framebox[0.3cm][c]{\footnotesize $B$} , \framebox[0.3cm][c]{\footnotesize $C$} ,  \framebox[0.3cm][c]{\footnotesize $E$} and \framebox[0.3cm][c]{\footnotesize $F$}: The bits $\mat{X}_1^B$, $\mat{X}_1^C$, $\mat{X}_1^E$ and $\mat{X}_1^F$ are independent and temporally i.i.d.\ with marginal pmf
\begin{eqnarray}
P_{X_1|B_1}(1|0)=
P_{X_1|B_1}(1|1)=\tfrac{1}{2}.
\end{eqnarray}

\item Region \framebox[0.3cm][c]{\footnotesize $D$}: The bits $\mat{X}_1^D$ are i.i.d.\ with marginal pmf
\begin{eqnarray}
P_{X_1|B_1}(1|0)&=&\hat{p}_1\\
P_{X_1|B_1}(1|1)&=&\hat{p}_1\\
\qquad \qquad \ P_{X_1}(1)&=&\hat{p}_3=\hat{p}_1.
\end{eqnarray}
\end{itemize}
Furthermore, we assume that $\mat{X}_1^j$, $j=\{A,B,C,D,E,F\}$ are independent. For Tx$_2$, the input distributions coincide with that of Tx$_1$ in the corresponding regions, but with parameters $q_i$ instead of $p_i$ and $\gamma_1$ instead of $\eta_1$.
Following similar steps as in previous sections, we obtain an achievable rate pair for $\tfrac{4}{3}<\alpha\leq \tfrac{3}{2}$ which is given by \eqref{SI_local_CSIRT_sr3_R1} and \eqref{SI_local_CSIRT_sr3_R2}.

\paragraph{SI, $\frac{3}{2}\leq\alpha\leq 2$}
We consider the input distribution depicted in Figure~\ref{figurea12}b with the following distributions:
\begin{itemize}
\item Regions \framebox[0.3cm][c]{\footnotesize $A$} and \framebox[0.3cm][c]{\footnotesize $\tilde{A}$}: The bits $(\mat{X}_1^A,\mat{X}_1^{\tilde{A}})$ are i.i.d.\, with marginal pmf given by \eqref{Dist_MI_1}--\eqref{Dist_MI_2}.
\item Region \framebox[0.3cm][c]{\footnotesize $B$}: The bits are i.i.d.\ with marginal pmf
\begin{eqnarray}
P_{X_1|B_1}(1|0)=
P_{X_1|B_1}(1|1)=\tfrac{1}{2}.
\end{eqnarray}
\end{itemize}
Furthermore, we assume that $\mat{X}_1^j$, $j=\{A,B\}$ are independent. For Tx$_2$, the input distributions coincide with that of Tx$_1$ in the corresponding regions, but with parameters $q_i$ instead of $p_i$, $\hat{q}_1$ instead of $\hat{p}_1$ and $\gamma_1$ instead of $\eta_1$. Proceeding as in the previous sections we obtain the achievable rate pair~\eqref{SI_local_CSIRT_sr4_R1} and \eqref{SI_local_CSIRT_sr4_R2}.

\begin{figure}[tbp]
  	\centering
  	\subfigure[\label{Fig: SI_2_CSIRT}]{\includegraphics[width=0.45\linewidth]{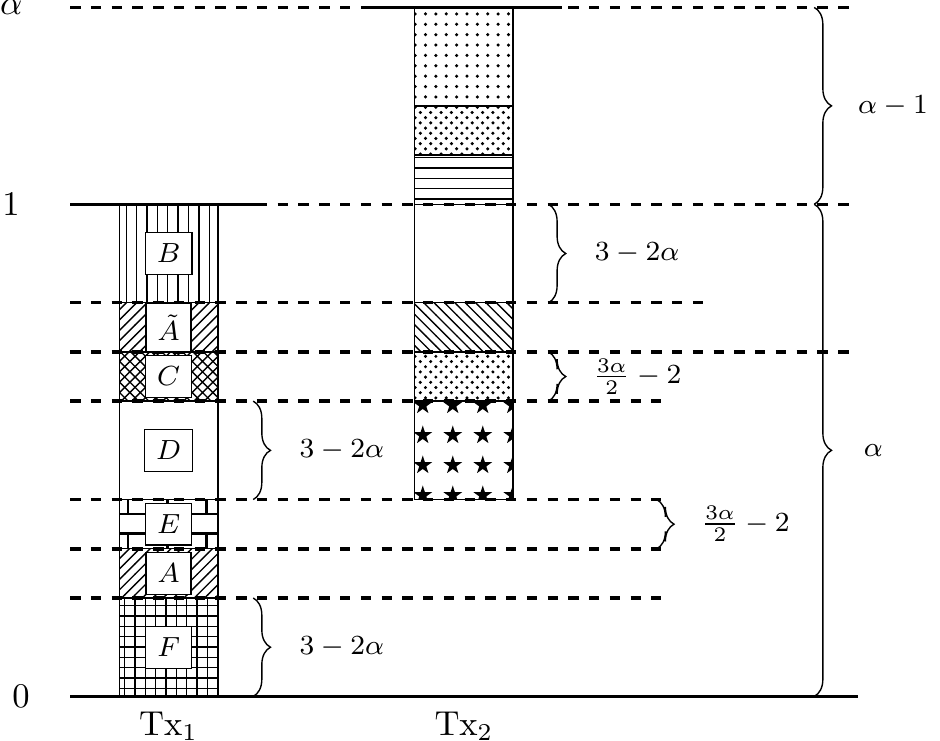}}\quad
    \subfigure[
     	\label{Fig: SI_3_CSIRT}]{\includegraphics[width=0.45\linewidth]{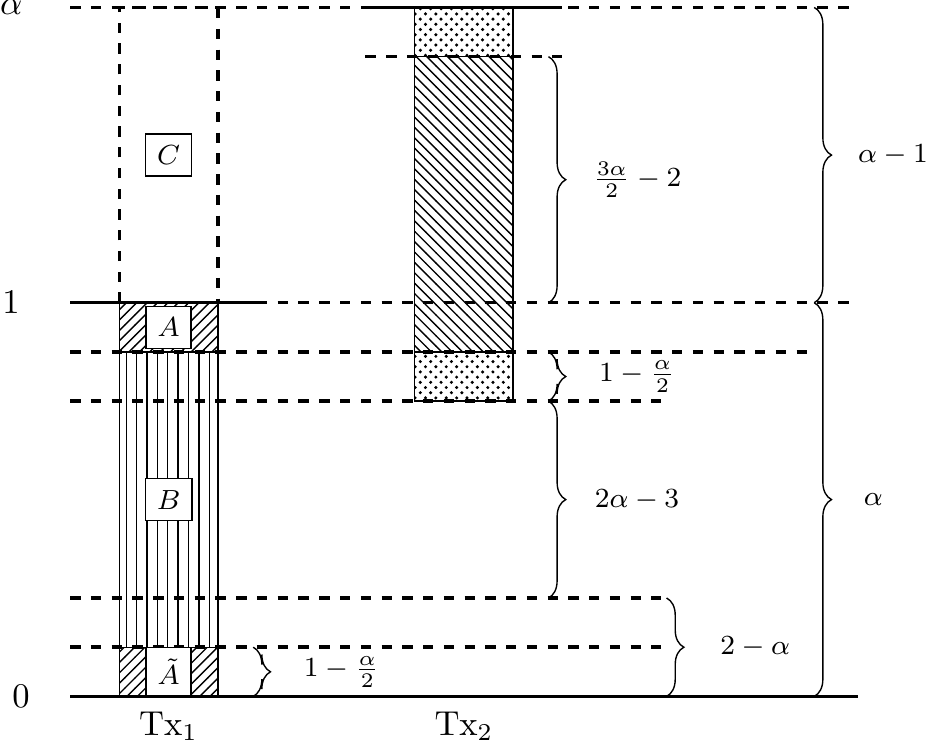}}\\	
  	\caption{Normalized signal levels at Rx$_1$. (\textbf{a}) (SI) for $\tfrac{4}{3}\leq\alpha\leq\tfrac{3}{2}$; (\textbf{b}) (SI) for $\tfrac{3}{2}\leq\alpha\leq 2$.}  \label{figurea12}
  \end{figure}



\bibliographystyle{IEEEtran}


\end{document}